\catcode`\@=11
\newif\if@fewtab\@fewtabtrue
{\count255=\time\divide\count255 by 60
\xdef\hourmin{\number\count255}
\multiply\count255 by-60\advance\count255 by\time
\xdef\hourmin{\hourmin:\ifnum\count255<10 0\fi\the\count255}}
\def\ps@draft{\let\@mkboth\@gobbletwo
    \def\@oddfoot{\hbox to 7 cm{\tiny \versionno
       \hfil}\hskip -7cm\hfil\rm\thepage \hfil {\tiny\draftdate}}
    \def\@oddhead{}
    \def\@evenhead{}\let\@evenfoot\@oddfoot}
\def\draftdate{\number\month/\number\day/\number\year\ \ \ \hourmin }

\def\citen#1{\if@filesw \immediate\write \@auxout {\string\citation{#1}}\fi%
\@tempcntb\m@ne \let\@h@ld\relax \def\@citea{}%
\@for \@citeb:=#1\do {\@ifundefined {b@\@citeb}%
    {\@h@ld\@citea\@tempcntb\m@ne{\bf ?}%
    \@warning {Citation `\@citeb ' on page \thepage \space undefined}}%
    {\@tempcnta\@tempcntb \advance\@tempcnta\@ne
    \setbox\z@\hbox\bgroup\ifcat0\csname b@\@citeb \endcsname \relax
    \egroup \@tempcntb\number\csname b@\@citeb \endcsname \relax
    \else \egroup \@tempcntb\m@ne \fi \ifnum\@tempcnta=\@tempcntb
    \ifx\@h@ld\relax \edef \@h@ld{\@citea\csname b@\@citeb\endcsname}%
    \else \edef\@h@ld{\hbox{--}\penalty\@highpenalty
    \csname b@\@citeb\endcsname}\fi
    \else \@h@ld\@citea\csname b@\@citeb \endcsname \let\@h@ld\relax \fi}%
\def\@citea{,\penalty\@highpenalty\hskip.13em plus.13em minus.13em}}\@h@ld}
\def\@citex[#1]#2{\@cite{\citen{#2}}{#1}}%
\def\@cite#1#2{\leavevmode\unskip\ifnum\lastpenalty=\z@\penalty\@highpenalty\fi%
  \ [{\multiply\@highpenalty 3 #1%
  \if@tempswa,\penalty\@highpenalty\ #2\fi}]}   %
\makeatother 
\catcode`\@=12

\def\Alpha         {{A}}
\def\alphd         {{\alpha_{\!2}^{}}}
\def\alphe         {{\alpha_{\!1}^{}}}
\def\alphv         {{\alpha_{\!4}^{}}}
\def\alphz         {{\alpha_{\!3}^{}}}

\newcommand\bb[2]  {\BB(#1#2)}
\def\BB            {{\mathscr B}}
\def\BBc           {{\mathscr B}^-}
\newcommand\bbc[2] {\BBc(#1#2)}
\newcommand\bcoef[3]{b^{#1,#2}_{#3}}
\newcommand\BCS[1] {B(#1)}
\newcommand\BCSc[1]{B^-(#1)}
\def\be            {\begin{equation}}
\def\bearl         {\begin{array}{l}}
\def\bearll        {\begin{array}{ll}}
\def\bet           {{\!\beta}}
\def\Beta          {{B}}
\def\blue          {\color{mydarkblue}}

\def\boxA          {{\color{Acol}\framebox(8,8){$\sss \!A$}}}
\def\boxB          {{\color{Bcol}\framebox(8,8){$\sss \!B$}}}
\def\bP            {\begin{picture}}
\def\btcs          {\small \begin{tabular}c}
\def\budef         {defect-crossing}
\def\budefa        {defect-crossing factorization}
\def\Budefa        {Defect-crossing factorization}
\def\C             {\ensuremath{\mathcal C}}
\def\CA            {\ensuremath{\mathscr B_{\!A}}}
\def\CAB           {\ensuremath{\mathcal C_{\!A|B}}}

\newcommand\CAC[6] {C_{#1#2,#3#4}^{~~~#5#6}}

\def\cbulk         {c^{\text{bulk}}}
\def\cbulki        {c^{\text{bulk}^{\,\scriptstyle-1}}}
\def\ccA           {{\sss (A)}}
\def\ccB           {{\sss (B)}}
\newcommand\clc[9] {C^{#1,#2#3;#4,#5#6}_{#7,#8#9}}
\newcommand\Clc[9] {C^{#1;#4}_{#7}}
\def\CD            {\ensuremath{\mathscr D_{\!A|B}}}
\newcommand\cdef[3]{c^{\text{def}}_{#1,#2#3}}
\newcommand\cdefinv[5]{(c^{\text{def}\;-1}_{#1,#2#3})_{#4#5}^{}}

\def\chii          {\raisebox{.15em}{$\chi$}}

\def\complex       {{\ensuremath{\mathbbm C}}}
\def\Cong          {\,{\cong}\,}
\newcommand\coronebd[2]{c(#1;#2)}

\def\CX            {\ensuremath{C_{\!X}}}
\def\cXo           {\ensuremath{c_{\!X;0}^{}}}
\newcommand\dcoef[5]{d^{#1#2,#3#4}_{#5}}
\newcommand\Dcoef[5]{d^{#1#2}_{#5}}
\def\dim           {\mathrm{dim}}
\def\dimc          {\dim_\complex}
\def\dsty          {\displaystyle }
\def\dtc           {defect transmission coefficient}
\def\ee            {\end{equation}}
\def\eE            {{\rm e}}
\def\eear          {\end{array}}
\def\End           {\mathrm{End}}
\def\eP            {\end{picture}}

\def\eq            {\,{=}\,}
\newcommand\erf[1] {(\ref{#1})}

\newcommand\FA[9]  {\mbox{\sf F}[A]^{(#1#2#3#4)}_{#5#6#7,#8#9}}

\newcommand\FB[9]  {\mbox{\sf F}[B]^{(#1#2#3#4)}_{#5#6#7,#8#9}}
\newcommand\FC[9]  {\mbox{\sf F}[C]^{(#1#2#3#4)}_{#5#6#7,#8#9}}
\newcommand\fbY[3] {{\color{Brcl}\fbox{$\scriptstyle\mathrm Y_{\!#1;#2}^{#3}$}}}

\def\findim        {fi\-ni\-te-di\-men\-si\-o\-nal}

\def\gc            {\Psi}
\newcommand\gcoef[5]{\Psi^{#1#2,#3#4}_{#5}}

\def\glutorus      {factorization torus}
\def\glutori       {factorization tori}
\def\gree          {\color{ForestGreen}}
\def\hom           {\mathrm{Hom}}

\def\HomA          {\hom_{\!A}}
\newcommand\Homaa[2]{\ensuremath{\HomAA(#1,#2)}}
\def\HomAA         {\hom_{\!A|A}}
\newcommand\Hombb[2]{\ensuremath{\HomBB(#1,#2)}}
\def\HomBB         {\hom_{\!B|B}}
\def\HomCC         {\hom_{\!C|C}}
\newcommand\hsp[1] {\mbox{\hspace{#1 em}}}
\def\I             {\ensuremath{\mathcal I}}
\def\ia            {{\ensuremath{\imath}}}
\def\ib            {{\ensuremath{{\bar\imath}}}}
\def\id            {\mbox{\sl id}}

\def\ii            {{\rm i}}
\def\iN            {\,{\in}\,}

\def\ja            {{\ensuremath{\jmath}}}
\def\jb            {{\ensuremath{{\bar\jmath}}}}
\newcommand\K[8]   {\mathcal K_{#1#2#3;#4#5#6}^{#7;#8}}
\def\kb            {\ensuremath{{\bar k}}}
\newcommand\labl[1]{\label{#1}\ee}
\def\lb            {\ensuremath{{\bar l}}}

\def\M             {{\ensuremath{\mathscr M}}}

\def\MN            {{\ensuremath{\mathscr N}}}
\def\MrmE          {\ensuremath{\M^{AB}}} 
\def\MrmN          {\ensuremath{\M^A}}    
\def\MrmS          {\ensuremath{\M^B}}    

\newcommand\N[3]   {{N_{#1#2}}^{\!\!#3}}
\def\nl            {{\circ}}
\newcommand\nxl[1] {\\[#1mm]}
\newcommand\Nxl[1] {\\[-1.3em]\\[#1mm]}
\def\nxt           {\raisebox{.08em}{\rule{.41em}{.41em}}~\,}
\newcommand\nxtp[1]{\def\leftmargini{1.54em}~\\[-1.45em]\begin{itemize}%
                   \item[\nxt]{#1}\end{itemize}}
\def\nxu           {~\\[-1.52em]} 
\def\nxv           {\nxl{1.5}\indent} 
\def\nxx           {\nxl{-2.3}} 
\def\nxy           {\nxl{-1.6}} 

\def\one           {{\ensuremath{\mathbf 1}}}
\def\onedim        {one-dimen\-sio\-nal}

\def\ota           {\,{\otimes_{\!A}}\,}
\def\Ota           {{\otimes_{\!A}}}
\def\oti           {\,{\otimes}\,}
\def\otic          {\,{\otimes_\complex}\,}

\def\otim          {\,{\otimes^{\!-}}\,}
\def\otip          {\,{\otimes^{\!+}}}
\newcommand\pA[1]  {\sse$\color{Acol} #1$}
\def\pb            {{\bar p}}
\newcommand\pB[1]  {\sse$\color{Bcol} #1$}
\newcommand\pg[1]  {\sse$\gree #1$}
\def\Phi           {\varPhi}
\newcommand\PHi[1] {\Phi_{\!#1}}
\newcommand\pl[1]  {\sse$#1$}

\def\Psi           {\varPsi}
\newcommand\pX[1]  {\sse$\color{defcol} #1$}
\def\qb            {{\bar q}}
\def\qd            {{r}}
\def\qdb           {{\bar r}}
\def\qe            {{p}}
\def\qeb           {{\bar p}}
\def\qf            {{s}}
\def\qfb           {{\bar s}}

\def\qz            {{q}}
\def\qzb           {{\bar q}}
\newcommand\R[5]   {{\sf R}^{(#1\,#2)#3}_{#4\,#5}}
\def\reals         {{\ensuremath{\mathbb R}}}
\def\rep           {representation}

\newcommand\Rm[5]  {{\sf R}^{-\,(#1\,#2)#3}_{\;#4\,#5}}
\def\rmY           {{\mathrm Y}}
\newcommand\SCIT[8]{{\mathcal L}_{#1#2#3;#4#5#6}^{#7;#8}}
\newcommand\SCmorph[5] {\omega^{#1#2#3}_{#4#5}}
\newcommand\SCmorphb[5]{\varpi^{#1#2#3}_{#4#5}}
\newcommand\setulen[2]{\setlength\unitlength{.#1#2pt}}
\def\sse           {\scriptsize}
\def\sss           {\scriptscriptstyle}

\newcommand\STg[6] {\mathcal N_{#1#2,#3#4}^{#5,#6}}
\def\SzI           {\ensuremath{S^2\Times [-1,1]}}
\def\SzSe          {\ensuremath{S^2\Times S^1}}
\def\T             {{\mathscr T}}
\def\torus         {{\mathrm T}}
\newcommand\Tfact[6] {\M^{\torus,#1#2}_{#3#5,#4#6}}

\def\Times         {\,{\times}\,}
\def\To            {\,{\to}\,} 
\def\tovacuum      {\btcs projection\nxx \vlrightarrow46 \nxx to vacuum
                    \nxy channel \end{tabular}}
\def\tovacuuw      {\btcs projection\nxx \vlleftarrow46  \nxx to vacuum
                    \nxy channel \end{tabular}}
\def\twodim        {two-di\-men\-sio\-nal}
\def\TXY           {_{\torus}^{X|Y}}
\def\TXYij         {_{\torus,\,ij}^{X|Y}}

\newcommand\vleq[2]{\ensuremath{\stackrel{\mbox{\rule{#1.#2em}{.03em}}}
                                         {\mbox{\rule{#1.#2em}{.03em}}}}}
\newcommand\vlleftarrow[2]{\ensuremath{\longleftarrow\!\!\!\raisebox{.28em}
                   {\rule{#1.#2em}{.03em}}}}
\newcommand\vlrightarrow[2]{\ensuremath{\raisebox{.28em}
                   {\rule{#1.#2em}{.03em}}\!\!\!\longrightarrow}}
\def\wsemb         {\iota}
\def\X             {{\rm X}}

\def\Z             {{\mathrm Z}}
\def\zet           {\ensuremath{\mathbb Z}}

\newcommand\includepicclax[3] {{\begin{picture}(0,0)(0,0) \scalebox{.#1#2}
                   {\includegraphics{imgs/pic_clal_#3.eps}}\end{picture}}}
\newcommand\Includepicclal[1] {{\begin{picture}(0,0)(0,0) \scalebox{.38}
                   {\includegraphics{imgs/pic_clal_#1.eps}}\end{picture}}}
\newcommand\eqpc[4]{$$\begin{picture}(#2,#3) #4 \end{picture}\label{#1} $$}
\newcommand\eqpic[4]{\begin{eqnarray}
                   \begin{picture}(#2,#3){}\end{picture}\nonumber\\
                   \raisebox{-#3pt}{ \begin{picture}(#2,#3) #4 \end{picture} }
                   \label{#1} \\~\nonumber \end{eqnarray} }
\newcommand\Eqpic[4]{\begin{eqnarray}
                   \begin{picture}(#2,#3){}\end{picture}\nonumber\\
                   \raisebox{-#3pt}{ \begin{picture}(#2,#3) #4 \end{picture} }
                   \nonumber \\[3pt]~\label{#1} \end{eqnarray} }

\documentclass[12pt]{article}
\usepackage{amsthm,latexsym,amsmath,amssymb,amsfonts,color,colordvi,bbm}
\usepackage[mathscr]{eucal}
\usepackage{graphics}
\usepackage{rotating}
\numberwithin{equation}{section}

\definecolor{Acol}       {rgb}  {0.494118,0.015686,0.015686}
\definecolor{Bcol}       {rgb}  {0.098039,0.070588,0.552941}
\definecolor{Brcl}       {rgb}  {0.600000,0.200000,0.200000}
\definecolor{DarkGreen}  {rgb}  {0.000000,0.392156,0.000000}
\definecolor{defcol}     {rgb}  {0.623529,0.062745,1.000000}
\definecolor{DarkViolet} {rgb}  {0.580392,0.000000,0.827450}
\definecolor{ForestGreen}{rgb}  {0.100000,0.408823,0.100000}
\definecolor{green3}     {rgb}  {0.000000,0.803921,0.000000}
\definecolor{mydarkblue} {rgb}  {0.282352,0.239215,0.803921}
\definecolor{OrangeRed}  {rgb}  {1.000000,0.270588,0.000000}
\definecolor{red3}       {rgb}  {0.803921,0.000000,0.000000}
\definecolor{SP1}        {rgb}  {0.4,0.16,0.4}
\definecolor{SP2}        {rgb}  {0.39,0.23,0.47}
\def\hom           {\mathrm{Hom}} 
\begin{document}
                  
\begin{flushright}
   {\sf ZMP-HH/10-17}\\
   {\sf Hamburger$\;$Beitr\"age$\;$zur$\;$Mathematik$\;$Nr.$\;$384}\\[2mm]
   July 2010
\end{flushright}
\vskip 3.5em
\begin{center}\Large
THE CLASSIFYING ALGEBRA FOR DEFECTS
\end{center}\vskip 2.1em
\begin{center}
  ~J\"urgen Fuchs\,$^{\,a}$,~
  ~Christoph Schweigert\,$^{\,b}$,~
  ~Carl Stigner\,$^{\,a}$
\end{center}

\vskip 9mm

\begin{center}\it$^a$
  Teoretisk fysik, \ Karlstads Universitet\\
  Universitetsgatan 21, \ S\,--\,651\,88\, Karlstad
\end{center}
\begin{center}\it$^b$
  Organisationseinheit Mathematik, \ Universit\"at Hamburg\\
  Bereich Algebra und Zahlentheorie\\
  Bundesstra\ss e 55, \ D\,--\,20\,146\, Hamburg
\end{center}
\vskip 3.5em
\noindent{\sc Abstract}
\\[3pt]
We demonstrate that topological defects in a rational conformal field theory 
can be described by a classifying algebra for defects -- a finite-dimensional
semisimple unital commutative associative
algebra whose irreducible representations give the defect transmission 
coefficients. We show in particular that the structure constants
of the classifying algebra are traces of operators on spaces of conformal 
blocks and that the defect transmission coefficients determine the
defect partition functions.

\vskip3em
\newpage

\section{Introduction}

In \twodim\ quantum field theory, in particular in conformal quantum field 
theory, the \onedim\ structures of boundaries and defect lines have attracted 
much interest. Boundaries require the notion of boundary conditions. The study 
of these has provided much structural insight into conformal field theory. 
Boundary conditions are also of considerable interest in applications,
ranging from percolation problems in statistical mechanics and
impurity problems in condensed matter systems to D-branes in string theory.

A defect line separates two regions of the world sheet on which the
theory is defined. In fact, on the two sides of a defect line there can be 
two different conformal field theories. Hence defects relate different 
conformal field theories; this forms the ground for their structural importance.
Indeed, defects allow one to determine non-chiral symmetries of a conformal
field theory and dualities between different theories. For
further applications see e.g.\ \cite{kaWi,kaTi,drgg,dakr}.

Among the conformal field theories those with `large' symmetry algebras are 
particularly accessible for an explicit treatment. Technically, these are 
\emph{rational} conformal field theories, for which the representation category 
of the chiral symmetry algebra is a modular tensor category. Similarly, those 
boundary conditions and defect lines are particularly tractable which preserve 
rational chiral symmetries. In fact,
rational CFTs are amenable to a precise mathematical treatment \cite{fuRs},
in which boundary conditions and defects can be analyzed with the help of
certain algebras internal to the \rep\ category of the chiral symmetry algebra.

An earlier approach \cite{prss3} to boundary conditions can be understood with 
the help of a classifying algebra \cite{fuSc5}, 
the presence of which was established rigorously in \cite{fuSs} for all rational 
CFTs. The classifying algebra is an algebra over the complex numbers spanned 
by certain bulk fields. It is semisimple, associative and commutative, and its
irreducible representations are in bijection with the elementary boundary 
conditions. (Any boundary condition of a rational CFT is a superposition, 
with suitable Chan-Paton multiplicities, of elementary boundary conditions.)

In fact, the classifying algebra allows one to obtain also the corresponding 
reflection coefficients for bulk fields: the homomorphisms given by its 
irreducible representations give the bulk field reflection coefficients in the 
presence of elementary boundary conditions. The latter coefficients determine
not only boundary states and boundary entropies \cite{aflu}, but also annulus 
partition functions. The classifying algebra approach is also of interest in the
study of boundary conditions and defects in non-rational theories, like e.g.\ 
in Liouville theory,
for which a description in terms of algebras in the \rep\ category of the chiral
symmetry algebra is not yet available \cite{gaRec,gaRw,sark5,petk4,caRun2}.

\medskip

Obviously, it is desirable to have similar tools at one's disposal for the 
study of defects. The basic result of this paper is that such tools indeed
exist. In the case of defects the role of reflection coefficients for boundary 
conditions is taken over by \emph{\dtc s}, which we carefully discuss. We show 
in particular that these coefficients determine the defect partition functions. 
As discussed in \cite{doug15}, such quantities are also expected to be relevant 
for the structure of the moduli space of conformal field theories.
Again there is a classifying algebra for defects, spanned by certain pairs
of bulk fields, which determines these coefficients,
namely through the homomorphisms given by its irreducible representations.

One way to capture aspects of defect lines is to think of them as boundary 
conditions for a doubled theory, the so-called folding trick. From this point
of view it is reasonable to expect that a classifying algebra for defects indeed
exists. In fact, this algebra has already been used successfully on a heuristic 
basis in the study of Liouville theory \cite{sark5}. As we will see, the folding 
trick does not render the problem trivial, though, since when resorting to this 
mechanism one loses essential information about defects, like e.g.\ the fusion 
of defect lines, a structure that is crucial for the relation between defects 
and non-chiral symmetries and dualities. Indeed, general \dtc s cannot be
understood as boundary reflection coefficients of the doubled theory, see the
discussion at the end of section \ref{sec.d-dtc} below.

\medskip

This paper is organized as follows. In section 2 we review basic properties of 
defects and explain the fundamental role played by the \dtc s. Section 3 
summarizes our strategy for the derivation of the classifying algebra for 
defects. It involves the comparison of two different ways of factorizing 
correlators with defect fields, which we carry out in Sections 4 and 5, 
respectively (two technical steps are relegated to appendices). In section 6 
we establish the main properties of the classifying algebra: it is a 
semisimple unital commutative associative algebra over the complex numbers, 
whose irreducible representations furnish the \dtc s. At the end of section 6 
we also discuss the relation between the classifying algebra and an algebra 
introduced in \cite{pezu6}. Section 7 finally shows 
how defect partition functions can be expressed in terms of the \dtc s.


\section{Defects and \dtc s}\label{sec.d-dtc}

Let us start by discussing a few general aspects of codimension-one defects 
that separate different quantum field theories. Our treatment is adapted to
the case of \twodim\ conformal quantum field theories, but part of the
discussion is relevant to non-conformal and to higher-di\-men\-si\-onal
theories as well. 

In the case of \emph{rational} conformal field theories, 
our statements have in fact the status of mathematical theorems.
The understanding of rational conformal field theory is based on the use of 
`large' chiral symmetry structures. Technically, this means that the
representation category of the chiral symmetry structure, say of a conformal 
vertex algebra, is a modular tensor category. To keep this technically important 
tool available also in the discussion of defects that separate different
quantum field theories, we assume that all theories in question share the 
same chiral symmetry structure and that the defects preserve all those
symmetries. But chiral symmetries do not determine the local quantum
field theory completely -- a fact that sometimes is, slightly inappropriately, 
expressed by saying that for a given (chiral) CFT there can be different
modular invariant partition functions. As a consequence we can still realize
non-trivial situations in which two different rational CFTs are separated by 
a defect. In addition, as we will see, even defects adjacent to one and the 
same CFT carry important structural information.

We will say that different local conformal field theories based on the same 
chiral symmetry realize different \emph{phases} of the theory. An instructive 
example one may wish to keep in mind is given by the conformal sigma models 
based on the group manifolds SU$(2)$ and SO$(3)$, respectively. It makes sense 
to consider world sheets that are composed of several regions in which different 
phases of the theory live and which are separated by \onedim\ phase boundaries. 
We refer to these phase boundaries as \emph{defect lines}; more specifically, 
the interface between regions in phases $A$ and $B$ constitutes an
$A$-$B$-\emph{defect}. For instance, in the sigma model example, in each region
one deals with maps to either SU$(2)$ or SO$(3)$, depending on the phase of the 
region. At the defect line a transition condition for these maps must be 
specified; it determines the type of defect assigned to the defect line.

Defects can be characterized by the behavior of the energy-momentum tensor in 
the vicinity of the defect line. In particular, for a \emph{conformal} defect 
the difference $T\,{-}\,\bar T$ between holomorphic and antiholomorphic 
components is continuous across the defect line (see e.g.\ \cite{bddo}). Among 
the conformal defects there are the totally reflective defects, for which 
$T \eq \bar T$ on either side of the defect line, and the totally transmissive 
ones, for which $T$ and $\bar T$ individually are continuous across the defect 
line. Totally transmissive defects are tensionless, i.e.\ they can be deformed 
on the world sheet without affecting the value of a correlator, as long as 
they are not taken through any field insertion or another defect line. To allow 
us to attain a maximal mathematical control of the situation, a defect should 
preserve even more chiral symmetries than the conformal symmetries, namely the 
ones of a rational chiral algebra. We refer to totally transmissive defects with
this stronger property as \emph{topological defects}. Such topological defects 
have been studied in e.g.\ \cite{pezu5,pezu6,grwa,ffrs3,ffrs5,runk7,ffrs6}.
All the defects considered in the present paper are topological; accordingly 
we will usually suppress the qualification `topological' in the sequel.  

\medskip

Let us explain how properties of defects are reflected in the mathematical 
description of rational CFTs. As has been shown in \cite{fuRs4}, for a rational 
CFT the phases $A$ are in bijection with certain Frobenius algebras internal to 
the representation category \C\ of the chiral symmetry algebra, and for given
phases $A$ and $B$ the (topological) $A$-$B$-defects are the objects of the 
category \CAB\ of $A$-$B$-bimodules in \C. Morita equivalent Frobenius algebras
give equivalent local conformal field theories, so that with regard to all 
observable aspects a phase can be identified with a Morita class of Frobenius 
algebras. Similarly, different defects can be physically equivalent. Equivalence
classes of $A$-$B$-defects correspond to isomorphism classes of objects of the 
bimodule category \CAB; we refer to such classes of defects as 
\emph{defect types}.

The category of bimodules over Frobenius algebras in the category of chiral
symmetries can be analyzed explicitly. One derives the
following finiteness statements, which are specific to rational theories:
\nxtp
{The number of inequivalent phases is finite.
\\[-1.38em]~\item[\nxt]
The number of elementary defect types is finite.}
\nxu
Indeed, for any two phases $A$ and $B$ there is a finite number of inequivalent 
elementary, or \emph{simple}, $A$-$B$-defects. Every
defect is a finite superposition (finite direct sum) of such simple defects.

\subsubsection*{Fusion of defects}

A central feature of all defects that are transparent to both chiralities of the
stress-energy tensor, and thus in particular of all topological defects, is that
there is an operation of \emph{fusion} of defects \cite{pezu5,pezu6,fuRs4,grwa}:
\nxtp
{Defects with matching adjacent phases can be \emph{fused}.}
\nxu
Specifically, when an $A$-$B$-defect $X \,{\equiv}\, X_{\!A|B}$ and a 
$B$-$C$-defect $Y \,{\equiv}\, Y_{\!B|C}$ are running parallel to each other, 
then due to the transparency property, we can take a smooth limit of vanishing 
distance in which the two defects constitute an $A$-$C$-defect $X \,{\otimes_B}
\, Y$. This fusion operation is associative up to equivalence of defects. The
fused defect $X \,{\otimes_B}\, Y$ is, in general, not simple, even if both $X$
and $Y$ are simple.
\nxtp
{For every defect $X$ there is a \emph{dual defect} $X^\vee\!$.}
\nxu
Namely, another operation that we can perform on a defect line is to change its
orientation, and this results in a new defect type, the dual defect. More 
precisely, for an $A$-$B$-defect $X$ the dual defect $X^\vee$ corresponding
to the defect line with opposite orientation is a $B$-$A$-defect. 
\nxl1
These two operations have again a clear-cut representation theoretic meaning:
Fusion is the tensor product over $B$ of the bimodules $X$ and $Y$, and the 
categories of bimodules have a duality which implements orientation reversal. 
\nxtp
{For each phase there is an \emph{invisible defect}, the fusion with which does 
not change any defect type.}
\nxu
The presence of this distinguished defect signifies that the tensor category 
$\C_{\!A|A}$ of $A$-$A$-bimodules has a tensor unit. The tensor unit of 
$\C_{\!A|A}$, and thus the invisible $A$-$A$-defect, is in fact just the algebra 
$A$, seen as a bimodule over itself. For any $A$-$B$-defect $X$ the fused defects 
$X \,{\otimes_B}\, X^\vee$
and $X^\vee \,{\otimes_A}\, X$ contain the invisible defects $A$ and $B$, 
respectively, as sub-bimodules (with multiplicity one if $X$ is simple).
\nxtp
{For each phase $A$ there is a \emph{fusion ring} $\mathscr F_{\!A}$ of $A$-$A$ 
defect types. 
\\
The $A$-$B$-defect types span a left module over 
$\mathscr F_{\!A}$ and a right module over $\mathscr F_{\!B}$.}
\nxu
Indeed, the fusion of $A$-$A$ defects in a rational CFT is sufficiently 
well-be\-haved such that it induces the structure of a fusion ring on the set 
of $A$-$A$ defect types. This defect fusion ring has a distinguished basis 
consisting of simple defect types. (But there does not exist a braiding of
defect lines, hence the fusion ring of defects is not, in general, commutative).
\nxv
The defect fusion ring contains a surprising amount of information. To
elucidate this, consider the subset of those $A$-$A$-defects $X$ which when 
fused with their dual defect just give the invisible defect or, in other words, 
for which $X^\vee{\otimes_A}\, X$ is isomorphic to $A$ as an $A$-$A$-bimodule.
Such defects, which we call \emph{invertible} (or group-like) defects, 
turn out to be particularly interesting: 
\nxtp
{Non-chiral \emph{internal symmetries} of a full CFT in phase $A$ are in 
bijection with the equivalence classes of invertible $A$-$A$-defects.}
\nxu
This insight \cite[Sect.\,3.1]{ffrs5} has important consequences in applications;
we will return to this effect and to its generalization to Kramers-Wannier
dualities in a moment.
The types of invertible 
defects form a group, the \emph{Picard group} of the phase $A$. The group 
ring of the Picard group is a subring of the fusion ring $\mathscr F_{\!A}$.

\subsubsection*{Defect fields}

In the language of bimodules, another aspect of defects becomes obvious:
Defects can also be joined. At such a junction, a coupling needs to
be chosen. The possible couplings of an $A$-$B$-de\-fect $X$
and a $B$-$C$-defect $Y$ to an $A$-$C$-defect $Z$ are given by the space
$\hom_{A|C}(X{\otimes_B}Y,Z)$ of bi\-mo\-du\-le morphisms. Moreover,
defect lines can start and end at field insertion points -- at insertions of
so-called \emph{disorder fields}. A field insertion on a defect line -- a
\emph{defect field} -- can also change the type of a defect. 
\nxtp
{Disorder fields are special instances of defect fields, namely those 
which turn the invisible defect into a non-trivial defect or inversely. Bulk 
fields are special instances of disorder fields, turning the 
invisible defect to the invisible defect. 
}
\nxu\indent
Just like bulk fields, defect fields carry two chiral
labels $U$, $V$ which correspond to \rep s of the chiral symmetry algebra.
In representation theoretic terms, the space of defect fields changing an
$A$-$B$-defect $X$ to another $A$-$B$-defect $Y$ is given by the space
$\hom_{A|B}(U\otip X\otim V,Y)$ of bimodule morphisms, where $U\otip X\otim V$
carries a specific structure of $A$-$B$-bimodule that is determined by the
(left, respectively right) actions of the algebras $A$ and $B$ on $X$ and by 
the braiding of \C. 
\nxv
For \emph{rational} conformal field theories one can classify defect fields. 
The corresponding partition functions obey remarkable consistency relations:
\nxtp
{The expansion coefficients, in the basis of characters, of the partition 
functions of a torus with two parallel defect lines inserted 
furnish a NIM-rep of the double fusion algebra \cite{pezu5,fuRs4}.}

\subsubsection*{Acting with defects}

We next turn to another important consequence of the fusion structure
on defects: their action on various quantities of a rational CFT.
\nxtp
{Defects can be fused to boundary conditions.}
\nxu
Specifically, when an $A$-$B$-defect $X \,{\equiv}\, X_{\!A|B}$ and a boundary 
condition $M \,{\equiv}\, M_B$ adjacent to phase $B$ are running parallel to 
each other, then in the case of a topological defect one can take the limit of 
vanishing distance between defect line and boundary, after which defect and 
boundary condition together constitute a boundary condition $X\,{\otimes_{B\,}}M$
adjacent to phase $A$. Representation theoretically one deals again with the 
tensor product over $B$, now between a bimodule and a left module. 
Especially, the space of boundary conditions in phase $A$ carries an action 
of the $A$-$A$-type defects and in particular of their Picard group.
\nxtp
{Defects act on bulk fields.}
\nxu
Namely, consider a bulk field insertion in phase $B$ that is encircled 
by an $A$-$B$-defect. Invoking again the transparency property we can shrink the 
circular defect line to zero size, whereby we obtain another bulk field, now in 
phase $A$. This way the defect gives rise to a linear map on the space of bulk 
fields, which was considered in \cite{pezu6}.
The action of defects on bulk fields is sufficient to distinguish 
inequivalent defects (i.e.\ non-isomorphic bimodules) \cite[Prop.\,2.8]{ffrs5}. 
As a consequence, any two different defect-induced internal symmetries can 
already be distinguished by their action on bulk fields \cite[Sect.\,3.1]{ffrs5}.
\nxv
Combining the various actions of defects on other structures on the world sheet,
one arrives at a notion of \emph{inflating an $A$-$B$-defect in phase $A$}, by
which one can relate correlators on arbitrary world sheets in phase $A$ to
correlators in phase $B$ \cite[Sect.\,2.3]{ffrs5}.

\subsubsection*{Defects and dualities}

An $A$-$B$-defect $X$ is called a \emph{duality defect} iff there exists a 
$B$-$A$-defect $X'$ such that the fused defect $X \,{\otimes_B}\, X'$ is a 
superposition of only invertible $A$-$A$-defects \cite[Thm.\,3.9]{ffrs5}.
These generalize the invertible defects:
\nxtp
{\emph{Order-disorder dualities} of Kramers-Wannier type can be deduced from 
the existence of duality defects.}
\nxu
In short, the relationship between defects and internal symmetries generalizes 
to order-disorder dualities.
\nxtp
{\emph{T-duality} between free boson CFTs is a variant of order-disorder 
duality.} 
\nxu
More precisely, T-duality is generated by a special kind of duality defect,
namely one in the \emph{twisted} sector of the $\zet_2$-orbifold of the free
boson theory, corresponding to a $\zet_2$-twisted representation of the U$(1)$
current algebra (\hsp{-.3}\cite[Sect.\,5.4]{fGrs}, compare also 
\cite{shWi,akosy}). While ordinary order-disorder dualities in general relate 
correlators of bulk fields to correlators of genuine disorder fields, in the 
case of T-duality the resulting disorder fields are in fact again local bulk 
fields.
\nxv
If two phases $A$ and $B$ can be separated by a duality defect $X$, then the 
torus partition function for phase $A$ can be written in terms of partition 
functions with defect lines of phase $B$, in a manner reminiscent of the way the
partition function of an orbifold is expressed as a sum over twisted sectors.
\cite[Prop.\,3.13]{ffrs5}. (For $A\eq B$, one thus deals with an `auto-orbifold 
property', as first observed in \cite{ruel'5}.) The relevant orbifold group 
consists of the types of invertible defects that appear in the fusion of $X$ 
with $X^\vee$.
\nxv
This orbifold construction can be generalized to arbitrary pairs of phases $A$ 
and $B$: correlators in phase $B$ can be obtained from those in phase $A$ as
a generalized orbifold corresponding to an $A$-$A$-defect which (as an object
of \C) is given by $A\oti B\oti A$ \cite{ffrs6}.

\subsubsection*{Transmission through a defect}

To further investigate defects, and in particular to derive a classifying 
algebra for them, we need to consider yet another phenomenon arising from the 
presence of defects, namely the transmission of bulk fields through a defect. 
Let us describe this phenomenon in some detail. When a bulk field passes 
through a defect, it actually turns into a superposition of \emph{disorder} 
fields. This process is similar to the excitation of boundary fields that 
results when a bulk field approaches the boundary of the world sheet. 

Consider a bulk field $\PHi\alpha$ in phase $A$ close to an $A$-$B$-defect 
$X$. Since the defect is topological, we can deform the defect line around 
$\PHi\alpha$. Fusing the resulting two parallel pieces of defect line (with 
opposite orientation) then amounts to the creation of disorder fields in phase 
$B$ with the same chiral labels as $\PHi\alpha$, connected by defects $Y$ to 
the original defect $X$. In the case of rational CFTs these can be expanded in a
basis of elementary disorder fields $\varTheta_{Y,B;\gamma}$. Pictorially, this 
process is described schematically as follows (compare \cite[(2.28)]{ffrs5}):
  \Eqpic{def_trans:99}{420}{95}{ \setulen 90
  \put(-25,115) { \includepicclax3{42}{99a}
  \put(56,44)  {$\color{DarkGreen} \PHi\alpha $}
  \put(109,80) {\pX X }
  \put(72,90)  {\pA {\fbox{$ A $}}}
  \put(124,60) {\pB {\fbox{$ B $}}}
  }
  \put(164,167) {$ = $}
  \put(205,115) { \includepicclax3{42}{99b}
  \put(109,80) {\pX X }
  \put(72,90)  {\pA {\fbox{$ A $}}}
  \put(124,60) {\pB {\fbox{$ B $}}}
  }
  \put(-10,42) {$=~ \dsty \sum_Y\sum_\tau$}
  \put(59,-10) { \includepicclax3{42}{99c}
  \put(109,80) {\pX X }
  \put(74,40)  {\pX X }
  \put(56,57) {\begin{rotate}{51}{\pX Y}\end{rotate}}
  \put(72,90)  {\pA {\fbox{$ A $}}}
  \put(124,60) {\pB {\fbox{$ B $}}}
  }
  \put(235,42) {$=~ \dsty \sum_{Y,\tau}\sum_{\gamma}~
                  d^{\alpha\gamma}_{A,X,B;Y,\tau} $}
  \put(358,-10) { \includepicclax3{42}{99d}
  \put(109,80) {\pX X }
  \put(58,54.5){\begin{rotate}{51}{\pX Y}\end{rotate}}
  \put(75,43)  {$\blue \varTheta_{\gamma} $}
  \put(72,90)  {\pA {\fbox{$ A $}}}
  \put(124,60) {\pB {\fbox{$ B $}}}
  } }
Here the $Y$-summation is over isomorphism classes of simple $B$-$B$-defects,
while the $\tau$-summation is over a basis of the space of $B$-bimodule 
morphisms from $X^\vee\Ota X$ to $Y$. (The notations for defect 
fields will be explicated after \erf{def_b:40,def_op:44} below.) 
 
Now take the bulk field $\PHi\alpha^{}\,{\equiv}\,\PHi\alpha^{\ia\ja}$ to have
chiral labels $\ia,\ja$ (corresponding to simple objects $U_\ia$ and $U_\ja$ of 
the \rep\ category of the chiral symmetry algebra, so that the label $\alpha$
takes values in the bimodule morphism space $\Homaa{U_\ia\otip A\otim U_\ja}A$).
Then the expansion \erf{def_trans:99} of the bulk field $\PHi\alpha$ in phase 
$A$ into disorder fields $\varTheta_{Y,B;\gamma}$ in phase $B$ reads more 
explicitly
  \be
  \PHi\alpha^{\ia\ja}(z) = \sum_Y\sum_\tau\sum_{\gamma}
  d^{\ia\ja,\alpha\gamma}_{A,X,B;Y,\tau}\, \varTheta^{\ia\ja}_{Y,B;\gamma}(z)
  \labl{def-dtc}
(as an equality valid inside correlators with suitable bulk fields). Note that 
the coefficients of this 
expansion do not depend on the position of the insertion point of the bulk, 
respectively disorder, field, simply because we can freely move a topological 
defect line and keep the position of the insertion point fixed while 
deforming the defect.

Next consider the situation that we perform the manipulations in 
\erf{def_trans:99} on a sphere on which besides the bulk field $\PHi\alpha$ in 
phase $A$ there is one further field insertion, another bulk field $\PHi\beta$ 
in phase $B$, so that we are dealing with the correlator $C_X$ of two bulk 
fields $\PHi\alpha^{}$ and $\PHi\beta^{}$ separated by the defect $X$ on a 
sphere. Then in the sum over $Y$ in \erf{def-dtc} only a single summand gives a 
non-zero contribution to the correlator, namely the one with $Y\eq B$, and in 
this case the label $\tau$ takes only a single value, which we denote by 
`$\nl$'. The corresponding disorder fields $\varTheta_{B,B;\gamma}$ are just 
ordinary bulk fields in phase $B$, and thus the resulting contribution to $C_X$ 
involves the coefficient $(\cbulk_{B;\ia,\ja})^{}_{\gamma\beta}$ (in a standard 
basis, compare \cite[App.\,C.2]{fjfrs}) of an ordinary two-point function 
$C(\PHi\alpha^{\ia\ja\,\ccB},\PHi\gamma^{\ib\jb\,\ccB})$ of bulk fields on 
the sphere. Furthermore, if $Y\eq B$, then the circular defect line 
carrying the defect $X$, which no longer encloses any field insertion in phase 
$A$, can be shrunk to zero size and thus be completely removed from the world 
sheet at the expense of multiplying the correlator with $\dim(X)\,/\dim(B)$.
The contribution of $Y\eq B$ to the expansion \erf{def-dtc},
and thus to the correlator $C_X$, is therefore given by
  \be
  \frac{\dim(X)}{\dim(B)} \sum_\gamma d^{\ia\ja,\alpha,\gamma}_{A,X,B;B,\nl}\,
  (\cbulk_{B;\ia,\ja})^{}_{\gamma\beta}
  =: \frac1{S_{0,0}^{}} \dim(X)\, \dcoef \ia\ja\alpha\beta X
  \labl{newdef-dtc}
(keeping the factor $\dim(X){/}S_{0,0}$ on the right hand side will prove to be 
convenient).

\medskip 

The so defined numbers $\dcoef \ia\ja\alpha\beta X$ thus encode the contribution
of a bulk field (in the guise of a particular disorder field) to the 
expansion \erf{def-dtc}. The presence of a \emph{bulk} field in that expansion
may be described as the effect of \emph{transmitting} the original bulk field
$\PHi\alpha$ from phase $A$ to phase $B$ through the defect $X$; accordingly
we will refer to the numbers $\dcoef \ia\ja\alpha\beta X$ as \emph{\dtc s}.
Equivalently one may also think of $\dcoef \ia\ja\alpha\beta X$ as a particular 
operator product coefficient, namely for the operator product of the two bulk 
fields $\PHi\alpha$ and $\PHi\beta$ in different phases separated by the defect 
$X$. Another way of interpreting the \dtc s is as the matrix elements of the 
action of the defect $X$ on bulk fields \cite{pezu6,fGrs} (i.e.\ for the 
operators that implement the shrinking of a defect line around a bulk field), 
and yet another way to view the situation is as a scattering 
of bulk fields in the background of the defect, as studied in \cite{fusW}.
The \dtc s also appear naturally in the expansion of the partition function
on a torus with defect lines into characters, see formula \erf{ZTXYij} below.

Since, as noted above, defect types are completely characterized by
their action on bulk fields, the set of two-point correlators of bulk fields 
separated by a simple defect on a sphere, and thus the collection of \dtc s,
carries essential information about simple defect types.

\medskip

It is instructive to compare the \dtc s introduced above with the analogous
quantities in the case of boundaries, which are the \emph{reflection 
coefficients} $\bcoef\ia\alpha M \,{\equiv}\, \bcoef\ia{\alpha;0\nl}M\!$. These 
are the numbers which appear as the coefficient 
of the boundary identity field $\Psi^{MM;0}$ in the short-distance expansion 
  \be
  \PHi\alpha^{\ia\ja}(r\eE^{\ii\sigma}) \,\sim\, \sum_{j\in\I} \sum_\beta
  (r^2{-}1)^{-2\Delta_\ia+\Delta_j}_{}\, \bcoef\ia{\alpha;j\beta} M \,
  \Psi^{MM;j}_\beta(\eE^{\ii\sigma}) \qquad {\rm for}\;\ r\To 1 
  \labl{Phi-to-Psi}
of the bulk field $\Phi^{\ia\ib}_\alpha$ into boundary fields when it 
approaches the boundary of the (unit) disk with boundary condition $M$ 
(see \cite{cale,lewe3,bppz}
or \cite[Sect.\,2.7]{fuSc6}). In contrast to the expansion \erf{def-dtc},
the coefficients in \erf{Phi-to-Psi} involve a non-trivial dependence
$(r^2{-}1)^{-2\Delta_\ia+\Delta_j}_{}$ on the position of the bulk field,
in agreement with the fact that the boundary of the world sheet, unlike a 
topological defect, in general cannot be deformed without changing the
value of a correlator.

\medskip

When formulating rational CFT with the help of the TFT construction, correlators
are described as invariants of three-manifolds \cite{fuRs}. The third dimension 
allows for a geometric separation of left- and right-movers, whereby the two 
chiral labels of a bulk field $\PHi\alpha^{\ia\ja}$ label ribbons which run
through the regions above and below the world sheet, respectively. When 
applied to the transmission of a bulk field through a defect as shown in 
\erf{def_trans:99}, this amounts to the equalities
  \eqpc{def_trans_1,2:89,90}{420}{100}{ \setulen 90
  \put(-43,0)    {\includepicclax3{42}{89}
  \put(145.5,106){\pg i}
  \put(149,3)    {\pg j}
  \put(141,71)   {\pl{\phi_{\!\alpha}^{}}}
  \put(164,63.7) {\pA A}
  \put(209,68)   {\pB B}
  \put(90,50.8)  {\pX X}
  }
  \put(192,50) {$=\,\dsty \sum_Y\sum_\tau$}
  \put(255,0)    {\includepicclax3{42}{90}
  \put(145.5,106){\pg i}
  \put(149,4)    {\pg j}
  \put(143,66)   {\pl{\phi_{\!\alpha}^{}}}
  \put(157,60)   {\pA A}
  \put(208,70)   {\pB B}
  \put(102.4,50.5) {\pX X}
  \put(90.5,60.5){\pl {\scriptstyle\tau}}
  \put(111.6,63) {\pl {\scriptstyle\bar\tau}}
  \put(111,71)   {\pX Y}
  \put(72,71)    {\pX X}
  } }
  \Eqpic{def_trans_3:91}{420}{26}{ \setulen 90
  \put(133,31) {$=~ \dsty \sum_{Y,\tau}\sum_{\gamma}~
                  d^{ij,\alpha\gamma}_{A,X,B;Y,\tau} $}
  \put(255,-19)  {\includepicclax3{42}{91}
  \put(145.5,106){\pg i}
  \put(149,3)    {\pg j}
  \put(140.7,70.5) {\pl{\theta_\gamma}}
  \put(209,73)   {\pB B}
  \put(103,66.5) {\pl {\scriptstyle\tau}}
  \put(121,55.5) {\pX Y}
  \put(74,73.9)  {\pX X}
  } }
which translate into mathematical identities for corresponding correlators. It 
follows in particular that the \dtc s, and likewise the reflection coefficients,
can be expressed as invariants of ribbon graphs in the three-sphere $S^3$, 
and thus as morphisms in $\End(\one) \,{\cong}\, \complex$. One finds
  \eqpic{def_b:40,def_op:44}{420}{54} {
  \put(0,63) {$  \dim(M)\; \bcoef\ia\gamma M ~= $}
  \put(89,20) { {\Includepicclal{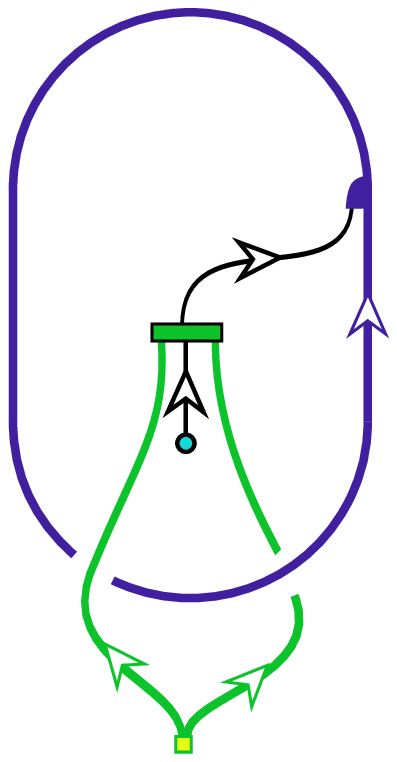}} \setlength\unitlength{1.2pt}
  \put(4.7,37)   {\pl{\phi_{\!\gamma}^{}}}
  \put(3.8,8)    {\pl \ia}
  \put(27.9,8)   {\pl \ib}
  \put(33,22.5)  {\pX M}
  \put(23.9,48.3){\pA A}
  } 
  \put(172,63)   {and}
  \put(221,63)   {$ \dim(X)\; \dcoef \ia\ja\alpha\beta X ~= $}
  \put(317,-6){ {\Includepicclal{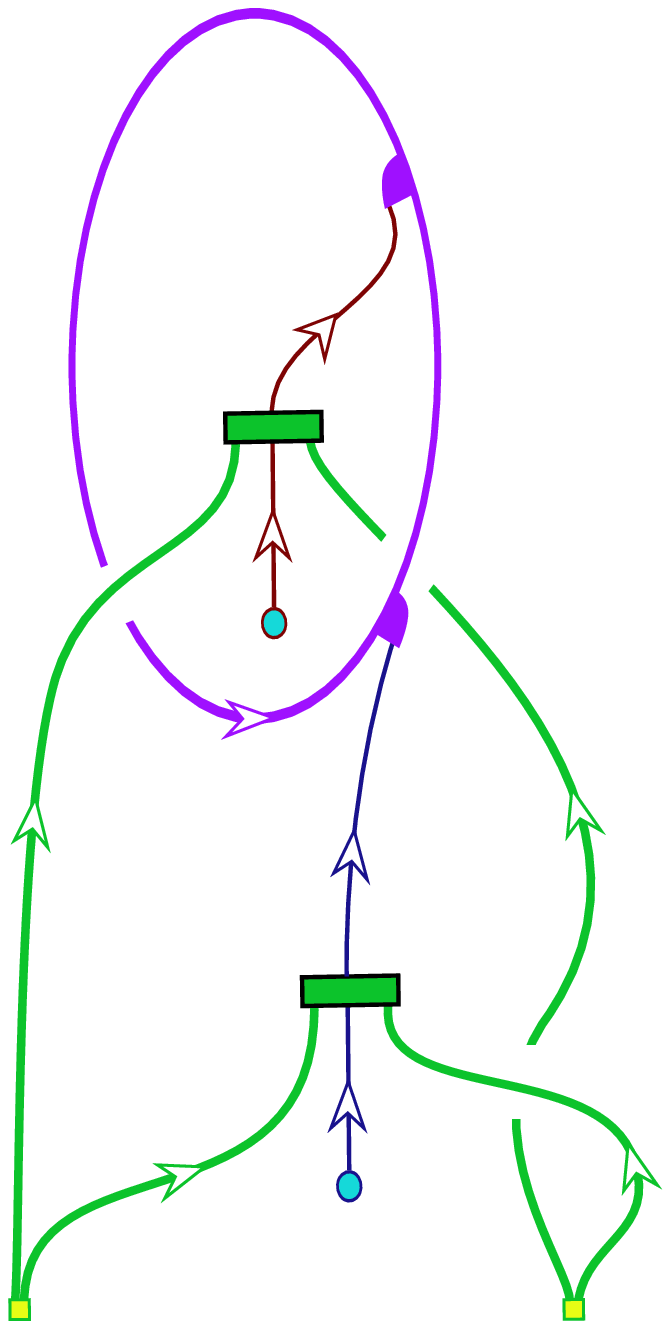}}
  \put(47,138)   {\pX X}
  \put(22,103)   {\pl{\phi_{\!\alpha}^{}}}
  \put(50,39)    {\pl{\phi_{\!\beta}^{}}}
  \put(39.5,116) {\pA A}
  \put(46.8,58)  {\pB B}
  \put(2.5,15)   {\pl{\ia}}
  \put(17,7)     {\pl{\ib}}
  \put(58.8,8)   {\pl{\ja}}
  \put(76.7,8)   {\pl{\jb}}
  } }

As for the notations used here and in the sequel, we follow the conventions 
listed in the appendix of \cite{fuSs}. In particular, in a rational CFT the set 
\I\ of chiral sectors, i.e.\ isomorphism classes of simple objects of the \rep\ 
category \C\ of the chiral symmetry algebra, is finite. We denote representatives 
of these isomorphism classes by $U_i$ with $i\iN\I$, reserving the label $i\eq0$ 
for the vacuum sector, i.e.\ for the tensor unit of \C, $U_0 \eq \one$. Boundary 
fields adjacent to phase $A$ which change the boundary condition from $M$ to $N$ 
(both of which are left $A$-modules) are denoted by $\Psi_{\!\alpha}^{} 
\,{\equiv}\, \Psi_{\!\alpha}^{MN;j}$ with $\alpha\,{\equiv}\,
\psi_\alpha \iN \HomA(M\oti U_i,N)$ a morphism of $A$-modules, 
bulk fields in phase $A$ by $\PHi\alpha^{} \,{\equiv}\, \PHi\alpha^{ij}$ with 
$\alpha \,{\equiv}\,\phi_\alpha \iN \Homaa{U_i\otip A\otim U_j}A$ a morphism 
of $A$-$A$-bimodules, and general defect fields changing an $A$-$B$-defect $X$ 
to $Y$ by $\varTheta_{\!\alpha}^{} \,{\equiv}\, \varTheta_{\!X,Y;\alpha}^{ij}$
with $\alpha \,{\equiv}\, \theta_\alpha \iN \hom_{A|B}(U_i\otip X\otim U_j,Y)$; 
see appendix A.5 of \cite{fuSs} for more details. The object $U^\vee$ of \C\ is 
the one dual to $U$, and $\ib\iN\I$ is the label such that the simple object 
$U_\ib$ is isomorphic to $U_\ia^\vee$. For any $i\iN\I$ the intertwiner space 
$\hom(\one, U_i^{}\oti U_i^\vee)$ is \onedim\ and we choose a basis $\Upsilon^i$
of that space. In pictures like in \erf{def_b:40,def_op:44} the tensor 
unit $\one$ is invisible, and the basis $\Upsilon^i$ is depicted by a piece of 
ribbon graph that looks as $\bP(12,0)\put(0,-4){{\includepicclax29{87}}}\eP$. 
 \\ 
The morphisms corresponding to bulk fields are depicted as 
$\phi_\alpha \eq \bP(20,0)\put(1.3,-17) {{\includepicclax29{96}}}\eP$. When
displaying ribbon graphs in three-manifolds, we use blackboard framing, whereby 
ribbons are depicted as lines (with an arrow indicating the orientation of the
ribbon core). For other aspects of the graphical calculus
see e.g.\ the appendix of \cite{ffrs} and the references given there.

The graphical calculus will be a crucial tool in our analysis of the 
classifying algebra for defects. Let us give two simple applications of this 
calculus which are easy consequences of \erf{def_b:40,def_op:44}.
First, the equality
  \eqpic{TC_chiral:92,TC_RC:93}{172}{63} {
  \put(0,0)   {\Includepicclal{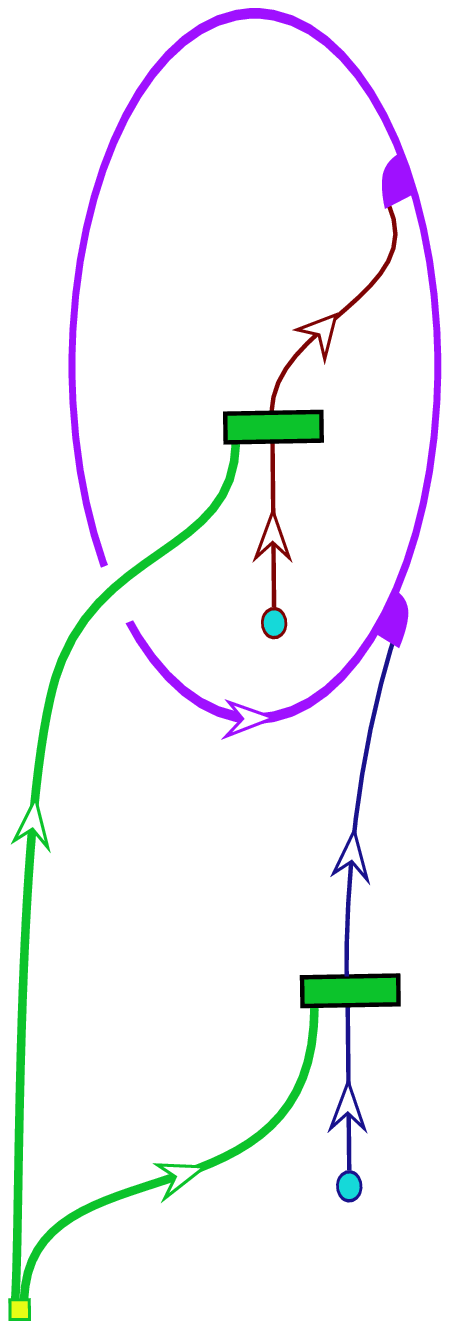}}
  \put(47,138)   {\pX X}
  \put(22,103)   {\pl{\phi_{\!\alpha}^{}}}
  \put(50,39)    {\pl{\phi_{\!\beta}^{}}}
  \put(39.5,116) {\pA A}
  \put(46.8,58)  {\pB B}
  \put(3,14)     {\pl{\ia}}
  \put(18,7)     {\pl{\ib}}
  \put(87,70)    {$=$}   
  \put(110,9) { {\Includepicclal{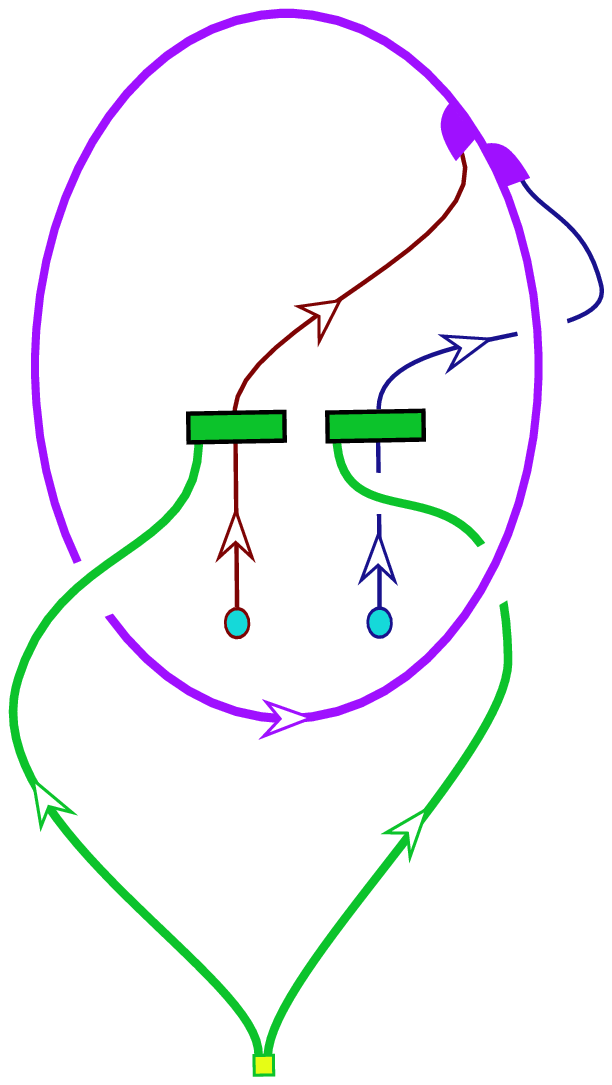}}
  \put(58,116)   {\pX X}
  \put(22,81)    {\pl{\phi_{\!\alpha}^{}}}
  \put(57,75)    {\pl{\phi_{\!\beta}^{}}}
  \put(47,96)    {\pA A}
  \put(42.3,58)  {\pB B}
  \put(21,18)    {\pl{\ia}}
  \put(50,18)    {\pl{\ib}}
  } } 
demonstrates that those \dtc s for which one of the chiral labels is 0 coincide
with specific reflection coefficients, namely
  \be
  \dcoef \ia0\alpha\beta{X_{\!A|B}^{}}
  = \bcoef \ia{\alpha\otimes\widehat\beta}{X_{\!A\otimes B^-}} \,,
  \ee
where $B^-$ denotes the algebra opposite to $B$ (i.e., the one obtained from $B$ 
by replacing the product $m_B$ of $B$ with the composition of $m_B$ and an 
inverse self-braiding of $B$), the boundary condition $X_{\!A\otimes B^-}$ is 
the $A$-$B$-bimodule $X_{\!A|B}^{}$ regarded as a left $A{\otimes}B^-$-module, 
and the morphism $\phi_{\widehat\beta}$ is obtained from $\phi_\beta$ by 
composition with the inverse braiding of $U_\ib$ and $B$. In contrast, for 
generic \dtc s a similar relation with reflection coefficients does not exist. 

Second, in the so-called Cardy case, which is the phase for which the Frobenius 
algebra $A$ is (Morita equivalent to) the tensor unit $\one$, both the
elementary boundary conditions and the simple defects are labeled by the same
set $\I$ as the chiral sectors, whereby the reflection coefficients and \dtc s 
reduce to
  \eqpic{40c,44c}{390}{44} {
  \put(-14,52) {$  \dim(U_m)\, \bcoef\ia\nl m ~= $}
  \put(75,-4) { {\Includepicclal{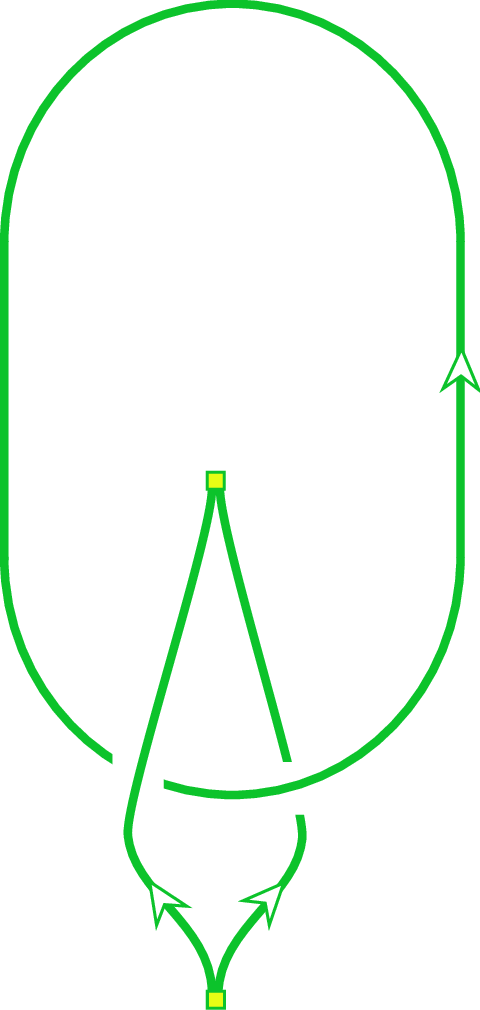}}
  \put(12.2,8)  {\pl \ia}
  \put(32.2,7)  {\pl \ib}
  \put(52,40)   {\pl m}
  }
  \put(172,52)  {and}
  \put(221,52)  {$ \dim(U_x)\, \dcoef \ia\ib\nl\nl x ~= $}
  \put(313,0){
     {\Includepicclal{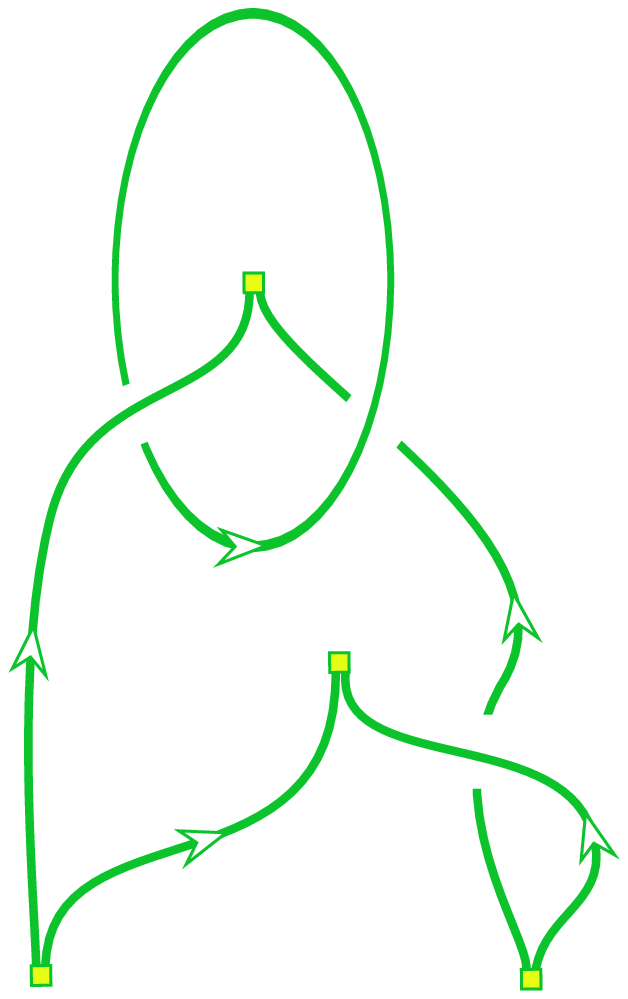}}
  \put(50.7,70) {\pl x}
  \put(4,15)    {\pl{\ia}}
  \put(19,6.2)  {\pl{\ib}}
  \put(55.5,6)  {\pl{\ib}}
  \put(70.8,5)  {\pl{\ia}}
  } }
With the help of the identity $\dim(U_j) \eq S_{j,0}/S_{0,0}$ and elementary 
braiding and fusing moves (see e.g.\ the formulas (2.45)\,--\,(2.48) of 
\cite{fuRs4}), one sees that these are nothing but simple multiples of 
entries of the modular matrix $S$:
  \be
  \bcoef\ia\nl m = \frac1{\dim(U_\ia)}\, \frac{S_{\ib,m}^{}}{S_{0,m}^{}} 
  \qquad{\rm and}\qquad
  \dcoef \ia\ib\nl\nl x
  = \frac{\theta_\ia^{}}{{\dim(U_\ia)}^2}\, \frac{S_{\ib,x}^{}}{S_{0,x}^{}}
  \labl{cardy-case}
($\theta_j\eq\exp(-2\pi\ii\Delta_j)$ is the twist of $U_j$).
This illustrates the multifaceted role played in rational CFT by the matrix $S$:
Besides representing the modular transformation $\tau\,{\mapsto}\,{-}1{/}\tau$ 
on the characters of the \emph{chiral} CFT, in the Cardy case it also gives the  
reflection coefficients as well as the \dtc s of the \emph{full} CFT.\,%
  \footnote{~The equality between the topological defect operators of a full
  CFT and the so-called Verlinde loop operators of the corresponding chiral CFT,
  which holds \cite{petk4,drgg} for the Cardy case, is another consequence of
  this versatility of $S$. There is thus in particular no reason to expect that
  this equality will survive beyond the Cardy case.}

\medskip

It is also worth mentioning that the folding trick does not seem to provide
any simple relation between the classifying algebra $\mathscr D$ (and \dtc s)
for the topological defects of a given CFT and the classifying algebra
$\tilde{\mathscr B}$ (and boundary reflection coefficients) for the boundary
conditions of the doubled theory.  What the folding trick \emph{can}
do is to give information about certain defects that are \emph{conformal}
(rather than even topological) in terms of boundary conditions of the
doubled theory. To the best of our knowledge, this has not been spelt out
in the literature in sufficient detail; for instance, it is unclear what
Frobenius algebra is appropriate for describing $A$-$B$-defects with the
folding trick. Moreover, the analysis of the symmetries preserved by
topological defects reveals that one would have to relax the conditions on
preserved chiral quantities and to work with the $\zet_2$-permutation
orbifold of $\C\,{\boxtimes}\,\C$, see \cite[App.\,A.2]{brRo2}. In contrast, 
our analysis only makes use of the chiral data of the original theory
and does not require to construct the orbifold category.


\section{The classifying algebra for defects -- synopsis}\label{syno}

Let us briefly present the basic ingredients of our construction. The 
derivation of the classifying algebra for boundary conditions employs the
fact that there are two types of factorization operations by which one can 
relate the correlator on any world sheet without defect lines to more 
fundamental correlators: on the one hand, boundary factorization, which 
involves a cutting of the world sheet along an interval that connects two 
points on its boundary, and on the other hand, bulk factorization, for which 
one cuts along a circle in the interior of the world sheet. Comparing the 
two types of factorization for the correlator of two bulk fields on the disk, 
one arrives at a quadratic identity for the reflection coefficients 
$\bcoef\ia\alpha M$. To obtain the classifying algebra for defects in a rational
CFT we proceed along similar lines. We select an appropriate correlation 
function \CX\ involving a circular defect line separating phases $A$ and $B$. 
By comparing two different factorizations of \CX\ we derive a 
quadratic identity for the \dtc s $\dcoef \ia\ja\alpha\beta X$.

\subsubsection*{The algebra structure on {\boldmath \CD}}

Suppressing multiplicity labels, the quadratic identity for the \dtc s that we
are going to derive in this paper reads
  \be
  \Dcoef ij\alpha\beta X\, \Dcoef kl\gamma\delta X = \sum_{p,q}
  \Clc{ij}\alpha\beta{kl}\gamma\delta{pq}\mu\nu\, \Dcoef pq\mu\nu X
  \labl{clc00}
with complex numbers $\Clc{ij}\..{kl}..{pq}..$ that do not depend on the simple
defect $X$. It is natural to interpret the latter coefficients as the 
structure constants of a multiplication on the vector space
  \be
  \CD := \bigoplus_{\ia,\ja\in\I} \Homaa{U_\ia\otip A\otim U_\ja}{A}
  \otic \Hombb{U_\ib\otip B\otim U_{\jb}}{B} \,,
  \labl{CD-as-vs}
and doing so indeed endows the space \CD\ with the structure of an associative 
algebra over \complex. Each summand in the expression \erf{CD-as-vs} is the 
space of a \emph{pair of bulk fields}, namely a bulk field with chiral labels 
$\ia,\ja$ in phase $A$ and one with conjugate chiral labels $\ib,\jb$ in phase 
$B$, and the structure constants $\Clc{ij}\..{kl}..{pq}..$ describe the product
on \CD\ in a natural basis of such pairs of bulk fields. Further, by making use
of the TFT approach to RCFT correlators, the derivation of this result at the
same time supplies a description of the structure constants 
$\Clc{ij}\..{kl}..{pq}..$ as invariants of ribbon graphs in the three-manifold 
\SzSe\ and thus as traces on spaces of conformal blocks. This description 
allows us to establish that the product on \CD\ obtained this way is indeed 
associative, as well as commutative, and moreover that there is a unit, that 
the algebra \CD\ with this product is semisimple, and that the irreducible 
\CD-\rep s are in bijection with the types of simple defects separating the 
phases $A$ and $B$.

\subsubsection*{The strategy for obtaining the boundary classifying algebra 
                (revisited)}

Our procedure is in fact largely parallel to the derivation of the classifying 
algebra \CA\ for boundary conditions in \cite{fuSs}. It is therefore instructive
to recapitulate the main steps of that derivation.  The relevant correlation 
function $C\eq C_M$ is in this case the correlator of two bulk fields on a disk 
in phase $A$ with elementary boundary condition $M$. Correlators of a full 
rational CFT are elements of an appropriate space of conformal blocks; in the 
TFT approach, the coefficients of a correlator in an expansion in a basis of 
the space of conformal blocks are expressed as invariants of ribbon graphs in 
a closed three-manifold; in the case of $C_M$, this three-manifold is the 
three-sphere $S^3$.

By boundary factorization of $C_M$ one obtains an expression in which two 
additional boundary fields are inserted, and which can be written as a linear 
combination of products of two factors, with each factor corresponding to the 
correlator of one bulk field and one boundary field $\Psi$ on a disk with 
boundary condition $M$. Moreover, as one deals with a space of conformal blocks 
at genus zero, this space has a distinguished subspace, corresponding to the 
\emph{vacuum channel}. By restriction to this subspace one can extract the term 
in the linear combination for which the boundary fields $\Psi$ are identity 
fields, so that one deals just with correlators $C(\PHi\alpha^{\ia\ib};M)$ for 
one left-right symmetric bulk field $\PHi\alpha^{\ia\ib}$ on a disk with 
boundary condition $M$. Now a correlator of bulk fields on a disk is naturally 
separated in a product of a normalized correlator and of the boundary vacuum 
two-point function $c^{\text{bnd}}_{M,0}$, which by formula (C.3) of 
\cite{fjfrs} equals the (quantum) dimension of $M$. When doing so for the 
correlators $C(\PHi\alpha^{\ia\ib};M)$, the resulting expansion coefficients 
$\coronebd{\PHi\alpha^{\ia\ib}}M$ of these correlators in a standard basis of 
conformal two-point blocks on the sphere become precisely the reflection 
coefficients, i.e.
  \be
  \coronebd{\PHi\alpha^{\ia\ib}}M = \dim(M)\, \bcoef\ia\alpha M .
  \ee

Bulk factorization of $C_M$ yields a correlator with two additional bulk field 
insertions. Relating this new factorized correlator to the unfactorized one 
involves a surgery along a solid torus embedded in the relevant 
three-manifold. This amounts to a modular S-transformation (see e.g.\ 
\cite[Sects.\,5.2,5.3]{fjfrs}), and accordingly now the coefficients are 
described by ribbon graphs in \SzSe\ rather than in the three-sphere; this 
implies that each such coefficient can be interpreted as the trace of an 
endomorphism of the space of three-point conformal blocks on $S^2$. After 
again specializing to the vacuum channel (as well as simplifying the result 
with the help of the \emph{dominance} property of the category of chiral 
sectors of a rational CFT, i.e.\ the fact that any morphism between objects $V$ 
and $W$ of the category can be written as a sum of morphisms between simple 
subquotients of $V$ and of $W$) one can separate the expression for the
coefficients into the product of a reflection coefficient with the trace of
an endomorphism that involves the insertion of three left-right symmetric
bulk fields on $S^2$.

Comparing the results of boundary and bulk factorization of $C_M$ in the vacuum 
channel one thus finds that the reflection coefficients $\bcoef\ia\alpha M$
satisfy
  \be
  \bcoef\ia\alpha M\, \bcoef k\beta M = \sum_{q\in\I}\sum_{\gamma=1}^{\Z_{q\qb}}
  \CAC \ia\alpha k\beta q\gamma\, \bcoef q\gamma M
  \labl{CA}
with complex numbers $\CAC \ia\alpha k\beta q\gamma$ that do not depend on the 
elementary boundary condition $M$. The latter numbers, which are obtained as 
traces on spaces of conformal blocks \cite{scfu2}, can be used to define an 
associative multiplication on the space $\CA \,{:=}\, \bigoplus_{\ia\in\I} 
\Homaa{U_\ia\otip A\otim U_\ib}A$ of left-right symmetric bulk fields, i.e.\ 
they play the role of the structure constants of the algebra \CA.

The derivation of \erf{CA} is summarized schematically in the following diagram,
where some simple prefactors and all summations are suppressed:
   \be
   \bP(400,310) \put(0,-6){
   \put(161,242)   { {\includepicclax24{86}} }
   \put(70,225){\begin{turn}{35}  \btcs boundary \nxx \vlleftarrow58
                 \nxx factorization \end{tabular}\end{turn}} 
   \put(250,275){\begin{turn}{-35} \btcs bulk    \nxx \vlrightarrow58
                 \nxx factorization \end{tabular}\end{turn}} 
   \put(-5,145)    { {\includepicclax24{08}} }
   \put(37,145)    { {\includepicclax24{08}} }
   \put(340,140)   { {\includepicclax12{41t}} }
   \put(25,130){\begin{turn}{-40} \tovacuum \end{turn}}
   \put(312,79){\begin{turn}{40}  \tovacuuw \end{turn}}
   \put(90,18)     { {\includepicclax29{40}} \put(29,18){\pX{\scriptstyle M}}}
   \put(130,18)    { {\includepicclax29{40}} \put(29,18){\pX{\scriptstyle M}}}
   \put(175,51)    { \vleq28 }
   \put(222,3)     { {\includepicclax15{23t}} }
   \put(295,18)    { {\includepicclax29{40}} \put(29,18){\pX{\scriptstyle M}}}
   } \eP
   \labl{pic-bdy}

\vskip1.2em

\subsubsection*{The strategy for obtaining the defect classifying algebra \CD}

As we will demonstrate, each of the steps in the description above has a 
counterpart in the derivation of the classifying algebra for defects. More 
specifically, we have already pointed out that just like the collection of 
one-point functions of bulk fields on the disk characterizes an elementary 
boundary condition, for characterizing a simple defect one needs the two-point 
functions of bulk fields on a world sheet that is a two-sphere containing a 
circular defect line, with the two bulk insertions on opposite sides of the 
defect line. Accordingly, the relevant correlator to start from is now the one 
for four bulk fields on a two-sphere, with a circular simple $A$-$B$-defect 
along the equator that separates the bulk fields into two pairs, one pair on the
Northern hemisphere in phase $A$ and the other pair on the Southern hemisphere
in phase $B$. We label the defect line by $X$ and denote this correlator by \CX.

Again we consider two different factorizations of \CX; the role of bulk 
factorization is taken over by a \emph{double bulk factorization}, while 
instead of boundary factorization we now must consider \emph{bulk factorization 
across the defect line}, to which for brevity we refer to as \emph{\budefa}.
This is indicated schematically in the following picture:
  \Eqpic{schematicview}{420}{117}{ \put(27,0){ \setulen90
  \put(157,198)   {\includepicclax35{501}
                   \put(23,33){\pX X} \put(57,57){\boxA} \put(57,16){\boxB} }
  \put(-20,58)  {\begin{turn}{50} \btcs \nxx\vlleftarrow{17}1
                  \nxx \budef\hspace*{7.1em}factoriz.\end{tabular}\end{turn}}
  \put(19,132)    {\includepicclax3{42}{502}}
  \put(-61,-5)    {\includepicclax4{42}{503}
                   \put(12,14){\pX X} \put(38,42){\boxA} \put(38,10){\boxB} }
  \put(11,-5)     {\includepicclax4{42}{503}
                   \put(12,14){\pX X} \put(38,42){\boxA} \put(38,10){\boxB} }
  \put(430,130)   {\includepicclax4{42}{506} \put(32,30){\boxA} }
  \put(430,65)    {\includepicclax4{42}{503}
                   \put(12,14){\pX X} \put(38,42){\boxA} \put(38,10){\boxB} }
  \put(430,0)     {\includepicclax4{42}{505} \put(32,30){\boxB} }
  \put(240,239) {\begin{turn}{-40} \btcs \nxx\vlrightarrow{16}6 \nxx double
                   bulk\hspace*{6.5em}factorization \end{tabular}\end{turn}}
  \put(307,122)   {\includepicclax3{42}{504}}
   } }
More explicitly, together with the projection to the vacuum channel we proceed 
according to the following picture, which is the analogue of \erf{pic-bdy}
(suppressing simple prefactors and summations):
   \be
   \bP(400,350)
   \put(146,270)   { {\includepicclax12{42}} }
   \put(65,259) {\begin{turn}{35} \btcs \budef     \nxx \vlleftarrow57 
                  \nxx factorization \end{tabular}\end{turn}} 
   \put(266,313){\begin{turn}{-35}\btcs double bulk\nxx \vlrightarrow57
                  \nxx factorization \end{tabular}\end{turn}} 
   \put(-35,144)   { \includepicclax15{46t} }
   \put(349,144)   { \includepicclax0{87}{62t} }
   \put(-3,125) {\begin{turn}{-35} \tovacuum \end{turn}} 
   \put(340,82) {\begin{turn}{35}  \tovacuuw \end{turn}} 
   \put(72,17)     { {\includepicclax20{44}}\put(30,51){\pX{\scriptstyle X}}}
   \put(117,17)    { {\includepicclax20{44}}\put(30,51){\pX{\scriptstyle X}}}
   \put(168,51)    { \vleq34 }
   \put(177,59)    { {\footnotesize \erf{clc00}} }
   \put(221,-5)    { {\includepicclax10{70t}} }
   \put(311,17)    { {\includepicclax20{44}}\put(30,51){\pX{\scriptstyle X}}}
   \eP
   \labl{scheme_CD}
   
\noindent
When comparing this description with the corresponding picture \erf{pic-bdy} for
the boundary case, one must bear in mind that what is depicted in the middle row
of \erf{pic-bdy} (i.e.\ after implementing factorization, but before projecting 
to the vacuum channel) are the coefficients of the correlator in a chosen basis 
of four-point blocks on the sphere, whereas the middle row of \erf{scheme_CD} 
gives results for the entire correlator. Accordingly the ribbon graphs shown 
in the middle row of \erf{scheme_CD} are embedded in a manifold with boundary 
(namely \SzSe\ with two four-punctured three-balls removed, see 
\erf{cob_4p_df_glue:46} and \erf{fact_MF_S2S1:62} for details) rather than in 
a closed three-manifold. We have chosen this alternative description because, 
unlike in the boundary case, the complexity of the ribbon graphs does not get 
significantly reduced when one expands the correlator in a basis 
(and subsequently invokes dominance).

\medskip

Our task in sections \ref{s3} and \ref{s4} of this paper will be to explain the 
various ingredients of the picture \erf{scheme_CD} in appropriate detail. In 
section \ref{s3} we introduce the correlator \CX\ and perform the \budefa, 
followed by projection to the vacuum channel. Section \ref{s4} is devoted to the
double bulk factorization and ensuing projection to the vacuum channel, with 
some of the details deferred to appendices \ref{app.glue4} and \ref{app.dubuvacpr}. 
Afterwards we are in a position to compare the two factorizations, whereby we 
arrive, in section \ref{sec:CD}, at the precise form \erf{clc0} of the equality
in the bottom line of the picture \erf{scheme_CD}; this allows us to define the 
classifying algebra \CD\ and establish its various properties.

For the case that the two phases $A$ and $B$ to the left and right of the defect
line coincide, a commutative associative algebra $\mathscr D_{\!A|A}^{\rm PZ}$
related to defects has already been obtained in \cite[Sect.\,7.4]{pezu6}.
As we will explain at the end of section \ref{sec:CD}, this algebra is
isomorphic to the classifying algebra $\mathscr D_{\!A|A}$
and indeed contains the same physical information, provided that the \dtc s 
possess additional properties (namely, a specific behavior under complex 
conjugation and unitarity of a certain matrix obtained from them), which we 
cannot, however, obtain from our results.

Let us finally note that, in accordance with \erf{cardy-case}, in the Cardy
case the classifying algebra \CD, as well as the classifying algebra for
boundary conditions, just coincides with the chiral fusion rules. In a
suitable basis (corresponding to re-normalizing the defect transmission, 
respectively reflection, coefficients by the fractions of dimensions and twist
eigenvalues that appear on the right hand side of \erf{cardy-case}) the 
structure constants are nothing but the fusion rule multiplicities.
For \CD\ this is demonstrated explicitly in \erf{CD-caca}.


\section{\Budefa}\label{s3}

As already pointed out in section \ref{sec.d-dtc}, in order to characterize 
a simple defect $X\,{\equiv}\,X_{\!A|B}$ separating phases $A$ and $B$, one
needs the correlators $C(\PHi\alpha;X;\PHi\beta)$ of two bulk fields 
$\PHi\alpha^{} \,{\equiv}\, \PHi\alpha^{\ia\ja\,\ccA}$ and 
$\PHi\beta^{} \,{\equiv}\, \PHi\beta^{kl\,\ccB}$ on the Northern and
Southern hemisphere which live in phases $A$ and $B$, respectively, and 
are separated by a circular defect line labeled by $X$ and running along 
the equator. Such a correlator can be non-zero only if
$k\eq\ib$ and $l\eq\jb$, and in that case it can be written as
  \be
  C(\PHi\alpha^{\ia\ja\,\ccA};X;\PHi\beta^{\ib\jb\,\ccB})
  = \frac1{S_{0,0}^{}} \dim(X)\,
  \dcoef \ia\ja\alpha\beta X\, \bb\ia\ib \oti \bbc\ja\jb
  \labl{CaXb}
with $\bb\ia\ib$ a standard basis for the (\onedim) space of two-point
conformal blocks on a sphere with standard (outward) orientation and 
$\bbc\ja\jb$ a basis for the corresponding blocks on a sphere with opposite 
orientation. Via the TFT construction, the numbers $\dcoef \ia\ja
\alpha\beta X$ appearing in \erf{CaXb} are easily seen to be precisely those 
defined in \erf{def-dtc}, i.e.\ they are \dtc s. (This is again in analogy 
with the situation for boundary conditions: For the one-point functions of 
bulk fields on a disk with elementary boundary condition $M$ one has an 
expression analogous to \erf{CaXb} with a single factor of $\bb\ia\ib$, 
and the corresponding coefficients $\bcoef\ia\alpha M$ coincide with the 
expansion coefficients $\bcoef\ia{\alpha;0\nl} M$ in \erf{Phi-to-Psi}, 
i.e.\ with the reflection coefficients.)

To gain information about properties of \dtc s with the help of factorization, 
the obvious starting point is the correlator \CX\ for four bulk fields 
$\PHi{\alpha_r^{}}^{} \,{\equiv}\, \PHi{\alpha_r^{}}^{\ia_r\ja_r}$ on a 
two-sphere, separated by the defect $X$ into two pairs, as already described 
above. We label the bulk fields $\PHi{\alpha_r^{}}^{}$ such that the ones with 
$r\eq 3,4$ are in phase $A$ and those with $r\eq 1,2$ are in phase $B$. The 
TFT construction provides an expression 
  \be
  \CX \equiv C(\PHi{\alphz}^{\ia_3\ja_3\,\ccA},\PHi{\alphv}^{\ia_4\ja_4\,\ccA};
  X;\PHi{\alphe}^{\ia_1\ja_1\,\ccB},\PHi{\alphd}^{\ia_2\ja_2\,\ccB})
  = Z(\M_\X)
  \labl{CX-equiv}
for \CX\ as the invariant of the \emph{connecting manifold} $\M_\X$, a 
three-manifold with embedded ribbon graph that is associated to \X. In the case
at hand, the connecting manifold $\M_\X$ is \SzI\ (a two-sphere times an 
interval) as a three-manifold, and including the ribbon graph it looks as 
follows \cite{fuRs4}:
  \eqpic{cob_4p:42}{380}{100}{ \setulen 90
  \put(0,115) {$\M_\X ~=$} \put(50,0){
  \put(5,0)   {\includepicclax3{42}{42}}
  \put(69,121.5) {\pl{\phi_{\!\alphe}^{}}}
  \put(140,129.9){\pl{\phi_{\!\alphz}^{}}}
  \put(237.8,96.1){\pl{\phi_{\!\alphd}^{}}}
  \put(253,139.7){\pl{\phi_{\!\alphv}^{}}}
  \put(168,117)  {\pA A}
  \put(275,129.6){\pA A}
  \put(80,104)   {\pB B}
  \put(265.5,108){\pB B}
  \put(83,219)   {\pg {\ia_1}}
  \put(146,219)  {\pg {\ia_3}}
  \put(202,219)  {\pg {\ia_2}}
  \put(262,219)  {\pg {\ia_4}}
  \put(81,26)    {\pg {\ja_1}}
  \put(144,26)   {\pg {\ja_3}}
  \put(201.4,26) {\pg {\ja_2}}
  \put(260,26)   {\pg {\ja_4}}
  \put(305.5,134){\pX X}
  } }
(the three accented horizontal pieces of $\M_\X$ are parts of the spheres 
$S^2\Times\{-1\}$, $S^2\Times\{0\}$, and $S^2\Times\{1\}$, respectively).

We now want to perform a factorization of this correlator along a circle in
the bulk that separates the bulk fields with labels $r\eq 1,3$ from those
with $r\eq 2,4$. When applying this procedure directly to the correlator as
displayed in \erf{cob_4p:42}, the cutting circle would intersect the defect line
twice, with the defect line oriented in opposite directions and thus 
corresponding once to the defect line labeled by $X \eq X_{\!A|B}^{}$ and once to
a defect line labeled by the dual $X_{}^\vee \eq X_{\!B|A}^\vee$. To circumvent 
the resulting complications, we take advantage of the fact that defects have a 
fusion structure.  We deform the defect line in such a way that the relevant 
pieces labeled by $X$ and $X^\vee$ are close to each other and thus form the 
fused defect $X^\vee\ota X$, and then use dominance in the tensor category of 
defects to expand $X^\vee\ota X$ into a direct sum of simple 
$B$-$B$-defects $Y$. Hereby \erf{cob_4p:42} is rewritten as
  \eqpic{cob_4p_df:43}{430}{100}{ \setulen 90
  \put(0,105) {$\M_\X ~=~ \dsty \sum_Y\sum_\tau$}
  \put(105,-2){
  \put(0,0)   {\includepicclax3{42}{43}}
  \put(65,124)   {\pl{\phi_{\alphe}}}
  \put(134,131)  {\pl{\phi_{\alphz}}}
  \put(220,98)   {\pl{\phi_{\alphd}}}
  \put(248.2,144){\pl{\phi_{\alphv}}}
  \put(152,117)  {\pA A}
  \put(269,134)  {\pA A}
  \put(76,107)   {\pB B}
  \put(261,111)  {\pB B}
  \put(139,103)  {\pX X}
  \put(179,136)  {\pX Y}
  \put(176,101)  {\begin{rotate}{51}\pl {\wsemb(S)}\end{rotate}}
  \put(166,138.1){\pl \tau}
  \put(209.1,136){\pl {\bar\tau}}
  \put(79,222)   {\pg {\ia_1}}
  \put(141,222)  {\pg {\ia_3}}
  \put(197,222)  {\pg {\ia_2}}
  \put(256,222)  {\pg {\ia_4}}
  \put(76.5,27)  {\pg {\ja_1}}
  \put(140,27)   {\pg {\ja_3}}
  \put(196.4,27) {\pg {\ja_2}}
  \put(255,27)   {\pg {\ja_4}}
  \put(300,136)  {\pX X}
  } }
where the $Y$-summation is over isomorphism classes of simple $B$-$B$-defects,
while the $\tau$-summation is over a basis of the space $\Hombb{X^\vee\Ota X}Y$
of $B$-bimodule morphisms and $\bar\tau$ is an element of the dual basis of
$\Hombb Y{X^\vee\Ota X}$.

In \erf{cob_4p_df:43} also the embedded cutting circle $\wsemb(S)$ is indicated,
as a dashed-dotted line. Note that, in agreement with the expectation from the 
folding trick, the situation displayed in \erf{cob_4p_df:43} bears some 
resemblance with the one for a boundary factorization for a doubled theory; 
still we deal with a genuine \emph{bulk} factorization, albeit one for which the
cutting circle crosses a defect line. Such a factorization works just like in 
the case without defect lines, as discussed in Section 5 of \cite{fjfrs}, up to 
the following modifications (details will be given elsewhere): First, some 
ribbons labeled by $B$ -- i.e.\ by the tensor unit of the category of 
$B$-$B$-bimodules -- namely those appearing in picture 
\erf{glue_tor_A:52,glue_tor_B:53} below, are exchanged with ribbons labeled by 
the defect $Y$ -- i.e.\ by another $B$-$B$-bimodule (compare 
\erf{glue_tor_A:52,glue_tor_B:53} to \erf{glue_tor_inv:45}). Second, the bulk 
two-point function on the sphere gets replaced by the correlator $\cdef Ypq$ 
for the sphere with an $Y$-defect line running between two disorder fields.

More explicitly, the factorization results in the expression
  \be
  C_X = \sum_{Y}\sum_{\tau}
  \sum_{\qe,\qz\in\I}\sum_{\gamma,\delta}\dim(U_{\qe})\, \dim(U_{\qz})\,
  \cdefinv Y{\qe}{\qz}\delta\gamma\, Z(\M_{\qe\qz\gamma\delta}^{Y,\tau})
  \ee
for the correlator, with appropriate cobordisms $\M_{\qe\qz\gamma\delta}
^{Y,\tau}$. To obtain these cobordisms one has to cut the three-ma\-ni\-fold in 
\erf{cob_4p_df:43} along the connecting intervals over the cutting circle 
$\wsemb(S)$, which yields the disconnected sum of two three-balls, and glue the 
so obtained manifold to a specific manifold $\T_{\!\qe\qz\gamma\delta}^{B,Y}$.\,%
  \footnote{~Both the three-balls and $\T_{\!\qe\qz\gamma\delta}^{B,Y}$ are
  actually manifolds with corners. Their corners separate the annular parts
  along which the gluing is performed from the rest of the manifold; for brevity
  we refer to such annuli as the \emph{sticky} parts of a manifold with corners.
  Also note that $\T_{\!\qe\qz\gamma\delta}^{B,Y}$ is the analogue of the solid
  torus with corners that appears in picture (3.4) of \cite{fuSs}.
  \label{fn-sticky}}
The following picture describes $\T_{\!\qe\qz\gamma\delta}^{B,Y}$ as the 
exterior of a solid torus embedded in the closed three-manifold \SzSe\ 
(with the $S^1$-factor running vertically, i.e.\ top and bottom are identified,
and with the boundary of the excised solid torus oriented inwards), to which 
we will refer as the \emph{\glutorus}; the sticky\,$^{\ref{fn-sticky}}$ 
annular parts which are to be identified with corresponding sticky 
parts of the two three-balls are indicated by a darker shading:
  \eqpic{glue_tor_inv:45} {300} {110} { \setulen 90 \put(0,-17){
  \put(0,144)   {$ \T_{\!\qe\qz\gamma\delta}^{B,Y} ~= $}
  \put(85,0)  {
  \put(0,15)    {\includepicclax3{42}{45}}
  \put(188,46)    {\pg {\qz}}
  \put(195,197)   {\pg {\qzb}}
  \put(215,141)   {\pg {\qeb}}
  \put(221,115)   {\pg {\qe}}
  \put(159,103.4) {\pl {\phi_\gamma}}
  \put(181,154)   {\pl {\phi_\delta}}
  \put(136,114)   {\pX Y}
  \put(157,187)   {\pX Y}
  \put(183,106)   {\pB B}
  \put(171,168)   {\pB B}
  } } }
Gluing the two three-balls to \erf{glue_tor_inv:45} yields
  \eqpic{cob_4p_df_glue:46} {300} {150} { \put(0,-6){
  \put(0,148)   {$ \M_{\qe\qz\gamma\delta}^{Y,\tau} ~= $}
  \put(75,0)  {  \Includepicclal{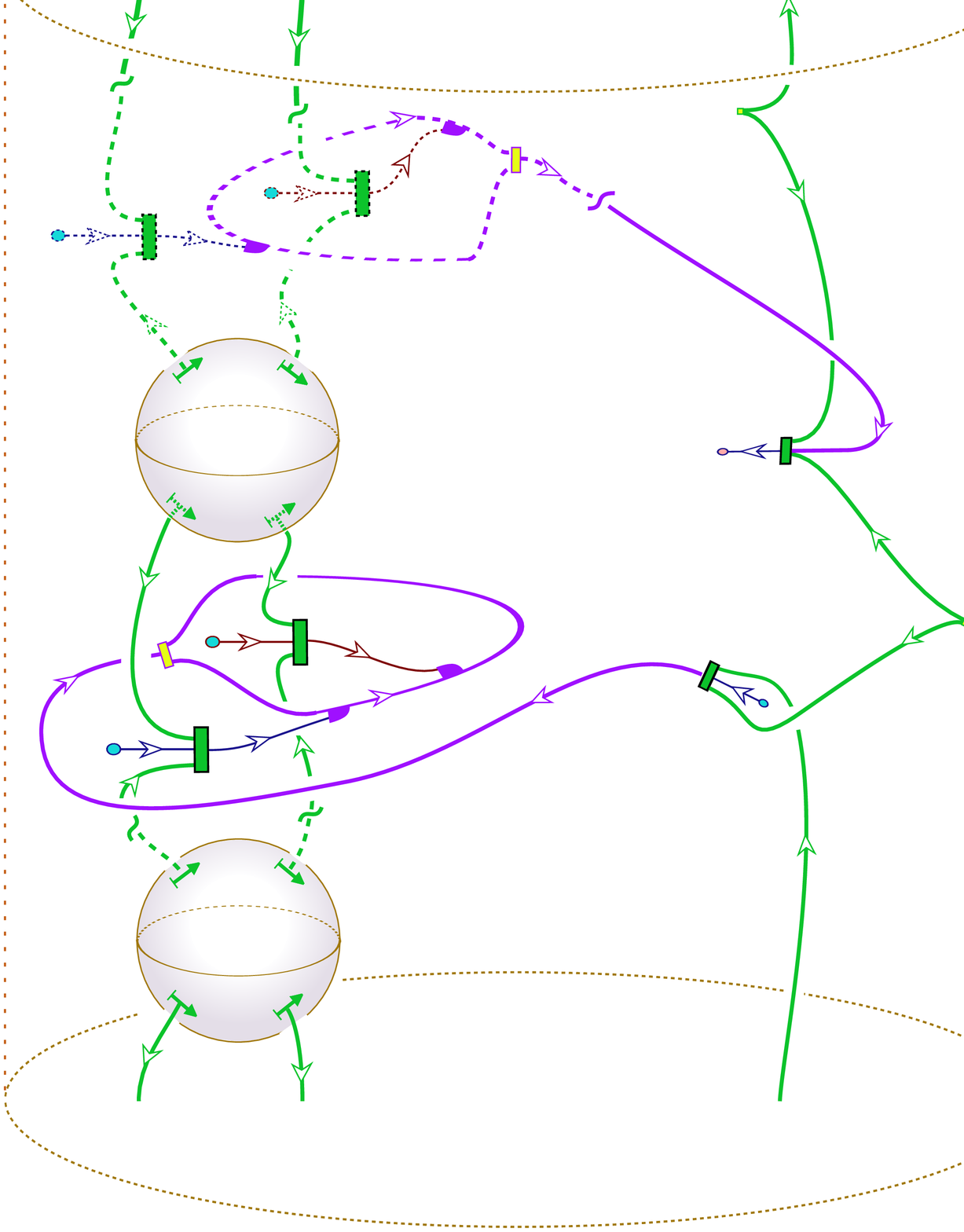}
  \put(215,160)   {\pg {\qeb}}
  \put(220,136)   {\pg {\qe}}
  \put(188,36)    {\pg {\qz}}
  \put(197,226)   {\pg {\qzb}}
  \put(160,122.8) {\pl {\phi_{\!\gamma}^{}}}
  \put(181,174)   {\pl {\phi_{\!\delta}^{}}}
  \put(132,132)   {\pX Y}
  \put(170,228)   {\pX Y}
  \put(42,123)    {\pl{\phi_{\alphe}}}
  \put(67,147.2)  {\pl{\phi_{\alphz}}}
  \put(42,174)    {\pg{\ia_1}}
  \put(43,78.5)   {\pg{\ja_1}}
  \put(59,172)    {\pg{\ia_3}}
  \put(61,78.5)   {\pg{\ja_3}}
  \put(30,244)    {\pl{\phi_{\alphd}}}
  \put(85,235)    {\pl{\phi_{\!\alphv}}}
  \put(43,56)     {\pg{\ja_2}}
  \put(43,197)    {\pg{\ia_2}}
  \put(59,54)     {\pg{\ja_4}}
  \put(62.5,197)  {\pg{\ia_4}}
  \put(97.5,252)  {\pA A}
  \put(90,136)    {\pA A}
  \put(172,187)   {\pB B}
  \put(36,126)    {\pl {\bar\tau}}
  \put(119.5,259) {\pl \tau}
  \put(115,232)   {\pX X}
  \put(101,155)   {\pX X}
  } } }
Note that the boundary of $\M_{\qe\qz\gamma\delta}^{Y,\tau}$ has two
components, each of which is a a two-sphere on which four ribbons start. 

\medskip

To obtain an expression involving the \dtc s, we do not need the whole
correlator \CX, but only its component \cXo\ in the \emph{vacuum channel} 
for both left- and right-movers. \CX\ is an 
element of the vector space $\BCS{U_{\ia_1}, U_{\ia_3},U_{\ia_2},U_{\ia_4}} 
\otic \BCSc{U_{\ja_1}, U_{\ja_3},U_{\ja_2},U_{\ja_4}}$, with $B$ and $B^-$
spaces of conformal four-point blocks on the sphere with outward and
inward orientation, respectively.
For the component \cXo\ to be non-zero it is necessary that
  \be
  \ia_3 = \ib_1\,, \quad \ia_4 = \ib_2\,, \quad \ja_3 = \jb_1
  \quad{\rm and}\quad \ja_4 = \jb_2 \,.
  \labl{ia2ib1}
In this case the component of \CX\ in the vacuum channel
is the projection of \CX\ to the \onedim\ subspace that is spanned by
  \be
  \BB(\ia_1\ib_1\ia_2\ib_2)_0\otimes\BBc(\ja_1\jb_1\ja_2\jb_2)_0 \,,
  \labl{basis_VC}
where $\BB(\ia_1\ib_1\ia_2\ib_2)_0$ is the basis element of
$\BCS{U_{\ia_1}, U_{\ib_1},U_{\ia_3},U_{\ib_3}}$ that corresponds to the
propagation of the subobject $\one$ of $U_{\ia_1} \oti U_{\ib_1}$ in the 
intermediate channel in a chiral factorization of four-point blocks into 
tensor products of three-point blocks, and analogously for 
$\BBc(\ja_1\jb_1\ja_2\jb_2)_0$. The vector $\BB(\ia_1\ib_1\ia_2\ib_2)_0$ 
is the value of the modular functor on the cobordism
  \eqpic{4pbasis:75,4pbasisvac:76} {420} {55} { \setulen 90
  \put(-9,69)  {$ \widehat\BB(\ia_1\ib_1\ia_2\ib_2)_0 ~:= $}
  \put(102,0)  { \includepicclax3{42}{75}
  \put(8,118)  {\pg {\ia_1} }
  \put(29,136) {\pg {\ib_1} }
  \put(68,43)  {\pg {\ib_1} }
  \put(84,65)  {\pg 0 }
  \put(105,135){\pg {\ia_2} }
  \put(126,117){\pg {\ib_2} }
  }
  \put(275,69) {$ \equiv $}
  \put(311,0)  { \includepicclax3{42}{76}
  \put(8,118)  {\pg {\ia_1} }
  \put(29,136) {\pg {\ib_1} }
  \put(105,135){\pg {\ia_2} }
  \put(126,117){\pg {\ib_2} }
  } }
(compare the picture (2.14) of \cite{fuSs}).
Also recall from \cite{fuSs} that there is a canonical projection from 
the space of four-point blocks to its vacuum channel subspace.

\medskip

After restriction to the situation \erf{ia2ib1}, we obtain the component \cXo\
of the correlator \CX\ in the vacuum channel by gluing the basis element dual 
to \erf{basis_VC} to the cobordism \erf{cob_4p_df_glue:46}. This results in 
the following expression for the projection to the vacuum channel:
  \be
  \bearll \cXo \!\!&
  \equiv c(\PHi{\alphz}^{\ib_1\jb_1\,\ccA},\PHi{\alphv}^{\ib_2\jb_2\,\ccA};
  X;\PHi{\alphe}^{\ia_1\ja_1\,\ccB},\PHi{\alphd}^{\ia_2\ja_2\,\ccB})_0^{}
  \nxl2&\dsty
  = \frac1{S_{0,0}^2}\sum_{Y}\sum_{\tau} \sum_{\qe,\qz,\gamma,\delta}
  \dim(U_{\qe})\, \dim(U_{\qz})\, \cdefinv Y{\qe}{\qz}\delta\gamma\,
  Z(\breve\M_{\qe\qz\gamma\delta}^{\,Y,\tau})
  \eear
  \labl{c_long}
with $\breve\M_{\qe\qz\gamma\delta}^{\;Y,\tau}$ the ribbon graph
  \eqpic{cob_4p_df_glue_proj:47} {300} {147} {
  \put(0,-6){
  \put(0,144)   {$ \breve\M_{\qe\qz\gamma\delta}^{\,Y,\tau} ~= $}
  \put(85,0)  {  \Includepicclal{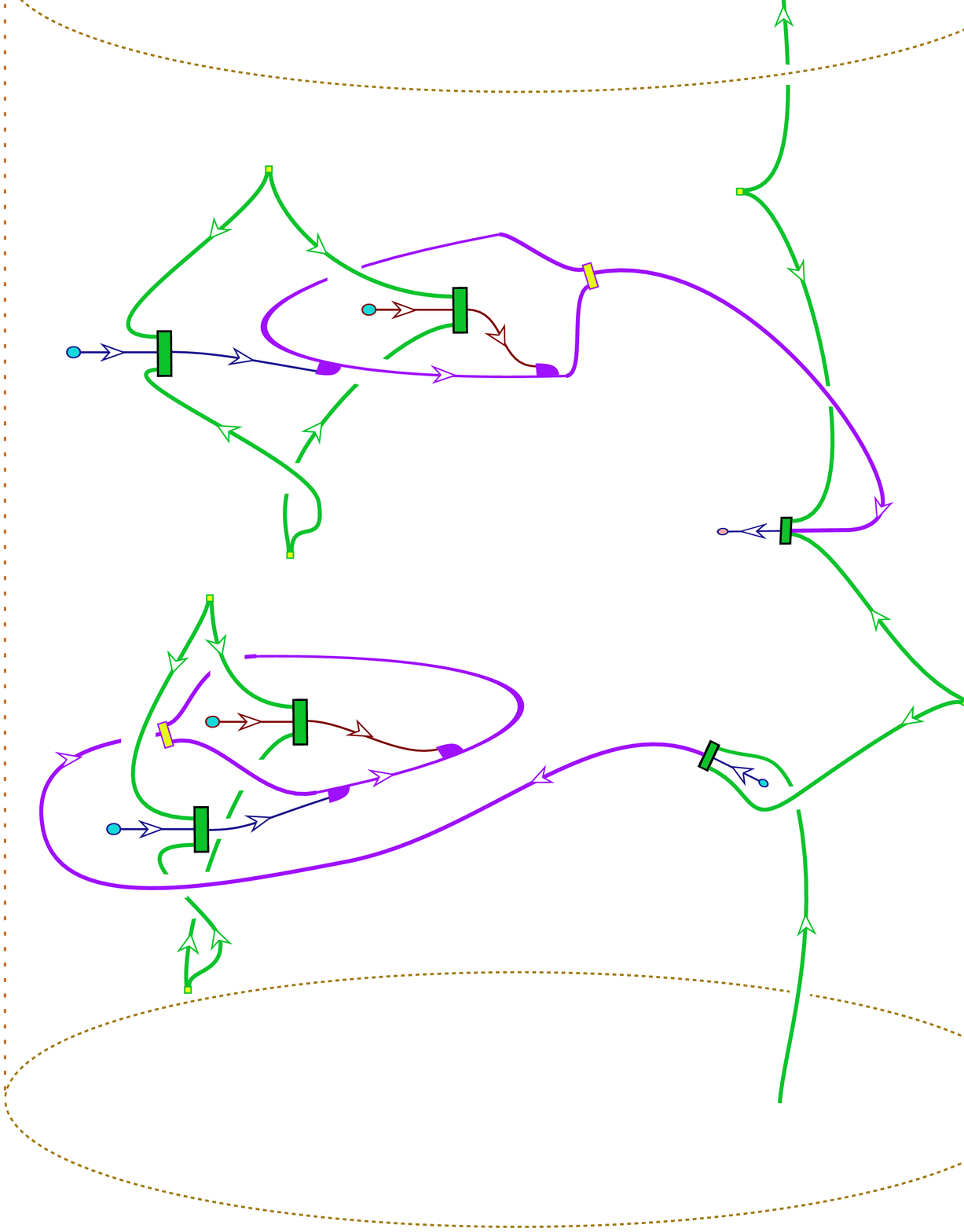}
  \put(187,33)    {\pg {\qz}}
  \put(196,212)   {\pg {\qzb}}
  \put(214,141)   {\pg {\qeb}}
  \put(220,118)   {\pg {\qe}}
  \put(160,103.4) {\pl {\phi_\gamma}}
  \put(181,154.6) {\pl {\phi_\delta}}
  \put(134,113)   {\pX Y}
  \put(171,222)   {\pX Y}
  \put(42,103)    {\pl{\phi_{\alphe}}}
  \put(68,128)    {\pl{\phi_{\alphz}}}
  \put(31,134)    {\pg{\ia_1}}
  \put(52,62)     {\pg{\ja_1}}
  \put(53,144)    {\pg{\ib_1}}
  \put(35,60)     {\pg{\jb_1}}
  \put(35,216)    {\pl{\phi_{\alphd}}}
  \put(105,225)   {\pl{\phi_{\alphv}}}
  \put(39,235)    {\pg{\ia_2}}
  \put(77,168)    {\pg{\ja_2}}
  \put(70,240)    {\pg{\ib_2}}
  \put(58,166)    {\pg{\jb_2}}
  \put(121,209)   {\pA A}
  \put(92,116)    {\pA A}
  \put(173,168)   {\pB B}
  \put(36,107.7)  {\pl {\bar\tau}}
  \put(139,217)   {\pl {\tau}}
  \put(103,236)   {\pX X}
  \put(106,135)   {\pX X}
  } } }
in the closed three-manifold \SzSe.

Next we note that the invariant $Z(\breve\M_{\qe\qz\gamma\delta}^{\;Y,\tau})$
is the trace of an endomorphism of the space of conformal one-point blocks on 
$S^2$. This space is zero unless the chiral field insertion is the identity, and
hence $Z(\breve\M_{\qe\qz\gamma\delta}^{\;Y,\tau})$ can be non-zero only if
$\qe \eq \qz \eq 0$. Moreover, since $Y$ is a simple defect and $B$ is a simple 
Frobenius algebra and thereby simple as a $B$-$B$-defect as well, the spaces 
$\Hombb{U_0\otip B\otim U_0}Y \Cong \Hombb BY$ and $\Hombb Y{U_0\otip B\otim U_0}
\Cong \Hombb YB$ of bimodule morphisms are zero unless $Y \eq B$, in 
which case they are one-dimensional with natural basis given by $\id_B$.
Furthermore, for $Y\eq B$, the morphism $\tau$ in \erf{cob_4p_df_glue_proj:47}
is a basis of the \onedim\ space $\Hombb{X^\vee\Ota X}B$, and thus $\tau$ and 
the dual basis morphism $\bar\tau \iN \Hombb B{X^\vee\Ota X}$ can be written as
  \be
  \tau = \lambda_X\,\delta_X \quad{\rm and}\quad
  \bar\tau = \tilde\lambda_X\tilde \beta_X \qquad{\rm with}\qquad
  \lambda_X\, \tilde\lambda_X = \dim(B)\, / \dim(X) \,,
  \ee
where $\delta_X$ is the evaluation morphism for the right duality and 
$\tilde \beta_X$ the coevaluation morphism for the left duality in the category
of $B$-bimodules. (Their precise definition and the expression for
the product $\lambda_X\, \tilde\lambda_X$ are given in formulas (2.20) and
(3.48), respectively, of \cite{ffrs5}. Note that $\dim(\cdot)$ is the dimension
as an object of \C, rather than as an object of the category of $B$-bimodules.)

This way the sum in the expression \erf{c_long} for \cXo\ reduces to a single 
summand, and up to the explicit prefactors already obtained, this term is just 
the product of two \dtc s as given in \erf{def_b:40,def_op:44}. We thus arrive at
  \be
  \cXo = \frac1{S_{0,0}^2} \dim(B)\,\dim(X)\, \cdefinv B00\nl{\nl\,}\, 
  \dcoef {\ib_1}{\jb_1}{\alphz}{\alphe} X \,
  \dcoef {\ib_2}{\jb_2}{\alphv}{\alphd} X .
  \labl{c=dd.0}

Finally we note that the defect fields that are relevant to the defect field
two-point function $\cdef B00{}{}$ appearing here are actually just ordinary 
bulk fields in phase $B$, so that we have $\cdef B00{}{} \eq S_{0,0}/\dim(B)$ 
by formula (C.14) of \cite{fjfrs}. As a consequence we can rewrite \erf{c=dd.0}
as
  \be
  c(\PHi{\alphz}^{\ib_1\jb_1\,\ccA},\PHi{\alphv}^{\ib_2\jb_2\,\ccA};
  X;\PHi{\alphe}^{\ia_1\ja_1\,\ccB},\PHi{\alphd}^{\ia_2\ja_2\,\ccB})_0^{}
  = \frac1{S_{0,0}^{}}\,\dim(X)\, \dcoef {\ib_1}{\jb_1}{\alphz}{\alphe} X \,  
  \dcoef {\ib_2}{\jb_2}{\alphv}{\alphd} X .
  \labl{c=dd}

~


\section{Double bulk factorization}\label{s4}

The double bulk factorization of the correlator \CX\ is a factorization
along two circles, each of which encircles the two bulk field insertions 
on one of the two hemispheres. In the context of the folding trick, this
corresponds to a single bulk factorization in the folded theory.
We perform both factorizations simultaneously. 

We indicate the two cutting circles $\wsemb_1(S)$ and $\wsemb_2(S)$ as
dashed-dotted lines in the following redrawing (with a slightly different
choice of dual triangulation) of the three-manifold \erf{cob_4p:42}:
  \eqpic{cob_4p_BulkF:48}{420}{111}{ 
  \put(-10,118)   {$ \M_X ~= $}
  \put(40,-2){
  \put(5,-2)   {\includepicclax38{48}}
  \put(71,120)    {\pl{\phi_{\alphe}}}
  \put(114,130)   {\pl{\phi_{\alphd}}}
  \put(197,132)   {\pl{\phi_{\alphz}}}
  \put(234,135)   {\pl{\phi_{\alphv}}}
  \put(211,118)   {\pA A}
  \put(133,103)   {\pB B}
  \put(89,94)     {\begin{rotate}{15} \pl{\wsemb_1(S)}\end{rotate}}
  \put(265,138)   {\begin{rotate}{-36}\pl{\wsemb_2(S)}\end{rotate}}
  \put(95,219)    {\pg {\ia_1}}
  \put(133,219)   {\pg {\ia_2}}
  \put(187,219)   {\pg {\ia_3}}
  \put(242,219)   {\pg {\ia_4}}
  \put(92,26)     {\pg {\ja_1}}
  \put(130,27)    {\pg {\ja_2}}
  \put(186,27)    {\pg {\ja_3}}
  \put(241,27)    {\pg {\ja_4}}
  \put(290,140)   {\pX X}
  } }
Cutting $\M_X$ along the connecting intervals over the two cutting circles 
$\wsemb_1(S)$ and $\wsemb_2(S)$ yields the disconnected sum of three 
three-manifolds with corners. Two of these, to be denoted by \MrmN\ and \MrmS, 
are \emph{nibbled apples}, i.e.\ three-balls with one sticky annulus 
(analogously as in formula (3.6) of \cite{fuSs}). We denote these sticky annuli
by $\rmY_{\!S,A}^1$ and $\rmY_{\!S,B}^1$, respectively, and again indicate 
them in the pictures below by an accentuated shading. The three-balls \MrmN\ 
and \MrmS\ contain the two bulk insertions on the Northern and Southern 
hemisphere, respectively; we depict them as follows:
  \Eqpic{ball_n:49,ball_s:50}{420}{133}{ \setulen 80
  \put(-30,244)  {$ \MrmN_{\ia_3\ja_3,\ia_4\ja_4} ~= $}
  \put(66,140){
  \put(0,0)   {\includepicclax3{04}{49}}
  \put(105,103)  {\pl{\phi_{\alphz}}}
  \put(131,101)  {\pl{\phi_{\alphv}}}
  \put(147,80)   {\pA A}
  \put(90,193)   {\pg {\ia_3}}
  \put(132,185)  {\pg {\ia_4}}
  \put(95,8)     {\pg {\ja_3}}
  \put(137,18)   {\pg {\ja_4}}
  \put(188,80)   {\includepicclax3{04}{lsqarrov}}
  \put(215,83)   {\fbY SA1}
  } 
  \put(239,96)  {$ \MrmS_{\ia_1\ja_1,\ia_2\ja_2} ~= $}
  \put(335,-8){
  \put(0,0)   {\includepicclax3{04}{50}}
  \put(101,102)  {\pl{\phi_{\alphe}}}
  \put(126,101)  {\pl{\phi_{\alphd}}}
  \put(142,81)   {\pB B}
  \put(87,193)   {\pg {\ia_1}}
  \put(129,185)  {\pg {\ia_2}}
  \put(92,8)     {\pg {\ja_1}}
  \put(134,18)   {\pg {\ja_2}}
  \put(188,80)   {\includepicclax3{04}{lsqarrov}}
  \put(215,83)   {\fbY SB1}
  } }
For each of the nibbled apples, a single phase -- either $A$ or $B$ -- is 
relevant.

The third connected component \MrmE, which contains the equatorial region of the
world sheet and thereby in particular the defect line, may be described as a 
\emph{millstone} with two sticky annuli $\rmY_{\!S,A}^2$ and $\rmY_{\!S,B}^2$,
one for each phase $A$ and $B$:
  \eqpic{cyl:51}{225}{91}{ \setulen 70 \put(0,-2){
  \put(0,134) {$ \MrmE_X ~= $}
  \put(83,0){ {\includepicclax2{66}{51}}
  \put(151,137)  {\pA A}
  \put(38,94)    {\pB B}
  \put(148,96)   {\pX X}
  \put(188,80)   {\includepicclax2{66}{lsqarrov}}
  \put(216,82)   {\fbY SB2}
  \put(122,240)  {\includepicclax2{66}{lsqarrov}}
  \put(150,242)  {\fbY SA2}
  } } }
As we are performing \emph{two} bulk factorizations, the sticky parts of the
manifolds \erf{ball_n:49,ball_s:50} and \erf{cyl:51} are to be identified 
pairwise with corresponding sticky annuli $\rmY_{\!\T}$ on \emph{two} \glutori\ 
$\T^A$ and $\T^B$. The latter are of the same type as the one in picture 
\erf{glue_tor_inv:45}, except that the 
$Y$-ribbons are replaced by $A$- and $B$-ribbons, respectively:
  \Eqpic{glue_tor_A:52,glue_tor_B:53}{420}{81}{ \setulen 70
  \put(-40,127)   {$ \T_{\!\!\qe\qz\beta_1\beta_2}^A \,= $}
  \put(55,0){
  \put(0,0)   {\includepicclax2{66}{52}}
  \put(0,-11) {
  \put(220,115)   {\pg {\qe}}
  \put(215,141)   {\pg {\qeb}}
  \put(188,46)    {\pg {\qz}}
  \put(196,197)   {\pg {\qzb}}
  \put(156,101.4) {\pl {\phi_{\!\beta_1}^{}}}
  \put(176,154)   {\pl {\phi_{\!\beta_2}^{}}}
  \put(136,115)   {\pA A}
  \put(157,186.6) {\pA A}
  \put(171,168)   {\pA A}
  \put(2,170)     {\includepicclax2{66}{rsqarrov}}
  \put(-31,183)   {\fbY \T A1}
  \put(2,98)      {\includepicclax2{66}{rsqarrov}}
  \put(-31,95)    {\fbY \T A2}
  } }
  \put(331,127)   {$ \T_{\!\!\qd\qf\beta_3\beta_4}^B \,= $}
  \put(423,0){
  \put(0,0)   {\includepicclax2{66}{53}}
  \put(0,-11) {
  \put(221,115)   {\pg {\qd}}
  \put(213,141)   {\pg {\qdb}}
  \put(188,43)    {\pg {\qf}}
  \put(194,197)   {\pg {\qfb}}
  \put(157,101.7) {\pl {\phi_{\!\beta_3}^{}}}
  \put(178,154)   {\pl {\phi_{\!\beta_4}^{}}}
  \put(135,114)   {\pB B}
  \put(157,187)   {\pB B}
  \put(171,168)   {\pB B}
  \put(2,170)     {\includepicclax2{66}{rsqarrov}}
  \put(-31,183)   {\fbY \T B1}
  \put(2,98)      {\includepicclax2{66}{rsqarrov}}
  \put(-31,95)    {\fbY \T B2}
  } } }

Thus altogether we must perform four identifications of pairs of sticky annuli.
Three of them are straightforward. The manifold (with corners) 
$\MN^{\Alpha;\,\ia_3\ja_3,\ia_4\ja_4} _{\!\qe\qz\beta_1\beta_2}$ obtained by 
gluing the nibbled apple $\MrmN_{\ia_3\ja_3,\ia_4\ja_4}$ to the 
\glutorus\ $\T_{\!\qe\qz\beta_1\beta_2}^A$ (by identifying $\rmY_{\!S,A}^1$ 
with $\rmY_{\!\T,A}^1$) is 
  \Eqpic{glue_tor_A_ball:55,glue_tor_A_ball_b.eps:54}{420}{105}{ \setulen 75
  \put(-38,148)   {$ \MN^{\Alpha;\,\ia_3\ja_3,\ia_4\ja_4}
                    _{\!\qe\qz\beta_1\beta_2} ~= $}
  \put(69,3){
  \put(-121,0)   {\includepicclax2{85}{55}}  
  \put(218,113)   {\pg {\qe}}
  \put(219,141)   {\pg {\qeb}}
  \put(194,35)    {\pg {\qz}}
  \put(200,197)   {\pg {\qzb}}
  \put(160,103.4) {\pl {\phi_{\beta_1}}}
  \put(181,154.5) {\pl {\phi_{\beta_2}}}
  \put(140,117)   {\pA A}
  \put(161,186.8) {\pA A}
  \put(16,211)    {\pg{\ja_3}}
  \put(63.3,208)  {\pg{\ja_4}}
  \put(28,165)    {\pg{\ia_3}}
  \put(75,168)    {\pg{\ia_4}}
  \put(34,194)    {\pl{\phi_{\alphz}}}
  \put(84,204)    {\pl{\phi_{\alphv}}}
  }
  \put(339,148)   {$ = $}
  \put(380,3){
  \put(0,0)   {\includepicclax2{85}{54}}
  \put(189,35)    {\pg {\qz}}
  \put(194.6,198) {\pg {\qzb}}
  \put(214,142)   {\pg {\qeb}}
  \put(221,117)   {\pg {\qe}}
  \put(157,101.8) {\pl {\phi_{\beta_1}}}
  \put(179,154)   {\pl {\phi_{\beta_2}}}
  \put(137,117)   {\pA A}
  \put(157,187)   {\pA A}
  \put(11,216)    {\pg{\ja_3}}
  \put(71,220)    {\pg{\ja_4}}
  \put(22,166)    {\pg{\ia_3}}
  \put(70,170)    {\pg{\ia_4}}
  \put(30,194)    {\pl{\phi_{\alphz}}}
  \put(80,204)    {\pl{\phi_{\alphv}}}
  \put(2,102)     {\includepicclax2{85}{rsqarrov}}
  \put(-29,98)    {\fbY \T A2}
  } }
where the second equality is just a redrawing corresponding to a shift along 
the $S^1$-direction of \SzSe.

Analogously one glues the nibbled apple $\MrmS_{\ia_1\ja_1,\ia_3\ja_3}$ to the 
\glutorus\ $\T_{\!\qd\qf\beta_3\beta_4}^B$ by identifying $\rmY_{\!S,B}^1$ 
with $\rmY_{\!\T,B}^1$. And next $\M^{AB}_X$ is glued to 
$\T_{\!\qd\qf\beta_3\beta_4}^B$ by identifying the two sticky annuli $\rmY_
{\!S,B}^2$ with $\rmY_{\!\T,B}^2$ as well. Together these last two gluings yield
  \Eqpic{glue_tor_B_ball_line_a:56,glue_tor_B_ball_line_b:57}{420}{103}{
  \setulen 75
  \put(-38,148)   {$ \MN^{\Beta;X,\ia_1\ja_1,\ia_2\ja_2}
                   _{\!\qd\qf\beta_3\beta_4} \,= $}
  \put(77,3){
  \put(0,0)   {\includepicclax2{85}{56}}
  \put(203,106)   {\pg {\qd}}
  \put(213,141)   {\pg {\qdb}}
  \put(188,35)    {\pg {\qf}}
  \put(195,195)   {\pg {\qfb}}
  \put(159,101.8) {\pl {\phi_{\!\beta_3}^{}}}
  \put(180,154)   {\pl {\phi_{\!\beta_4}^{}}}
  \put(145,102.6) {\pB B}
  \put(159,187.1) {\pB B}
  \put(12,221)    {\pg{\ja_1}}
  \put(72,221)    {\pg{\ja_2}}
  \put(23,167)    {\pg{\ia_1}}
  \put(70,171)    {\pg{\ia_2}}
  \put(30,194)    {\pl{\phi_{\alphe}}}
  \put(81,204)    {\pl{\phi_{\alphd}}}
  \put(107,125)   {\pX X}
  }
  \put(344,148)   {$ = $}
  \put(380,3){
  \put(0,0)   {\includepicclax2{85}{57}}
  \put(203,106)   {\pg {\qd}}
  \put(213,141)   {\pg {\qdb}}
  \put(188,35)    {\pg {\qf}}
  \put(195,195)   {\pg {\qfb}}
  \put(159,101.8) {\pl {\phi_{\!\beta_3}^{}}}
  \put(180,154)   {\pl {\phi_{\!\beta_4}^{}}}
  \put(145,109)   {\pB B}
  \put(158,187)   {\pB B}
  \put(12,221)    {\pg{\ja_1}}
  \put(72,221)    {\pg{\ja_2}}
  \put(23,167)    {\pg{\ia_1}}
  \put(70,171)    {\pg{\ia_2}}
  \put(30,194)    {\pl{\phi_{\alphe}}}
  \put(81,204)    {\pl{\phi_{\alphd}}}
  \put(145,125)   {\pX X}
  \put(2,102)     {\includepicclax2{85}{rsqarrov}}
  \put(-28,99)    {\fbY SA2}
  } }
Here the first picture is obtained after performing the same deformation that
led to the second picture in
\erf{glue_tor_A_ball:55,glue_tor_A_ball_b.eps:54}, while the second equality 
follows by deforming the $X$-ribbon around the horizontal $S^2$.

Note that as a three-manifold (with corners), both 
$\MN^{\Alpha;\,\ia_3\ja_3,\ia_4\ja_4}_{\!\qe\qz\beta_1\beta_2}$ and
$\MN^{\Beta;X,\ia_1\ja_1,\ia_2\ja_2}_{\!\qd\qf\beta_3\beta_4}$ is \SzSe\ with a
three-ball cut out, and with a sticky annulus on its spherical boundary. It 
remains to identify also the latter two sticky annuli $\rmY_{\!S,A}^2$ and 
$\rmY_{\!\T,A}^2$ with each other. This gluing is a bit harder to describe than
the previous ones.  We relegate the details to appendix \ref{app.glue4} and 
present only the result for the correlator 
$\CX \eq C(\PHi{\alphz}^{\ia_3\ja_3\,\ccA},\PHi{\alphv}^{\ia_4\ja_4\,\ccA}; 
X;\PHi{\alphe}^{\ia_1\ja_1\,\ccB},\PHi{\alphd}^{\ia_2\ja_2\,\ccB})$: 
  \be
  \bearl\dsty
  \CX = \sum_{\qe,\qz,\qd,\qf\in\I}\!
  \dim(U_{\qe}) \dim(U_{\qz}) \dim(U_{\qd}) \dim(U_{\qf})
  \nxl{-3}\dsty \hsp{10}
  \sum_{\beta_1,\beta_2,\beta_3,\beta_4}\! {(\cbulki_{A;\qe,\qz})}
  _{\beta_2\beta_1}\, {(\cbulki_{B;\qd,\qf})}_{\beta_4\beta_3}\,
  \sum_{n\in\I} S_{0,n}\,
  Z(\M_{n;\qe\qz\qd\qf}^{\beta_1\beta_2\beta_3\beta_4})\, 1 \,,
  \eear
  \labl{res4glue}
with $\cbulk_{A;uv}$ the matrix of coefficients of the  
bulk field two-point function on a sphere in phase $A$, and with
$\M_{n;\qe\qz\qd\qf}^{\beta_1\beta_2\beta_3\beta_4}$ the ribbon graph
  \eqpic{fact_MF_S2S1:62}{410}{186}{ \setulen 70
  \put(0,270){$ \dsty\M_{n;\qe\qz\qd\qf}^{\beta_1\beta_2\beta_3\beta_4} ~= $}
  \put(55,0){
  \put(84,-3)   {\includepicclax2{66}{62}}
  \put(213,452)   {\pl{\phi_{\!\alphz}^{}}}
  \put(264,438)   {\pl{\phi_{\!\alphv}^{}}}
  \put(361,410)   {\pl{\phi_{\!\beta_2}^{}}}
  \put(435,320)   {\pl{\phi_{\!\beta_1}^{}}}
  \put(314,436)   {\pA A}
  \put(391,275)   {\pA A}
  \put(284,270)   {\pB B}
  \put(261,195)   {\pX X}
  \put(195,397)   {\pg{\ia_3}}
  \put(248,397)   {\pg{\ia_4}}
  \put(202,353)   {\pg{\ia_1}}
  \put(249,352)   {\pg{\ia_2}}
  \put(410,400)   {\pg{\qeb}}
  \put(444,235)   {\pg{\qz}}
  \put(205,246)   {\pl{\phi_{\!\alphe}^{}}}
  \put(257,256)   {\pl{\phi_{\!\alphd}^{}}}
  \put(303,225)   {\pl{\phi_{\!\beta_4}^{}}}
  \put(284,173)   {\pl{\phi_{\!\beta_3}^{}}}
  \put(331,224)   {\pg{\qdb}}
  \put(192,72)    {\pg{\ja_3}}
  \put(235,77)    {\pg{\ja_4}}
  \put(190,114)   {\pg{\ja_1}}
  \put(235,115)   {\pg{\ja_2}}
  \put(168,200)   {\pg{n}}
  \put(254,133)   {\pg{\qf}}
  } }
in the three-manifold that is obtained by removing two four-punctured 
three-balls from the closed manifold \SzSe.

\medskip

Our next task is to project the correlator \erf{res4glue} to the vacuum channel
spanned by the vector \erf{basis_VC} in the space space $\BCS{U_{\ia_1},U_{\ia
_3},U_{\ia_2},U_{\ia_4}} \otic \BCSc{U_{\ja_1}, U_{\ja_3},U_{\ja_2},U_{\ja_4}}$ 
of conformal blocks. To this end we compose the cobordism \erf{fact_MF_S2S1:62} 
with the basis element dual to \erf{basis_VC}. After some straightforward 
manipulations of the resulting ribbon graph in \SzSe, the details of which 
are given in appendix \ref{app.dubuvacpr}, this yields the coefficient 
$\cXo \eq c(\PHi{\alphz}^{\ib_1\jb_1\,\ccA},\PHi{\alphv}^{\ib_2\jb_2\,\ccA};
X;\PHi{\alphe}^{\ia_1\ja_1\,\ccB},\PHi{\alphd}^{\ia_2\ja_2\,\ccB})_0^{}$
of the vacuum channel as
  \be
  \cXo = \frac1{S_{0,0}^3} \sum_{\qe,\qz\in\I}\! \dim(U_{\qe})^2_{}
  \dim(U_{\qz})^2_{} \!\! \sum_{\beta_1,\beta_2,\beta_3,\beta_4}\!
  {(\cbulki_{A;\qe,\qz})}_{\beta_2\beta_1}\, {(\cbulki_{B;\qeb,\qzb})}
  _{\beta_4\beta_3}\, Z(\M_{0;\qe\qz\qeb\qzb}^{\beta_1\beta_2\beta_3\beta_4})
  \labl{cXo}
with the ribbon graph
  \eqpic{fact_MF_S2S1_res:66}{410}{194}{ \setulen 70
  \put(0,288){$\dsty \M_{0;\qe\qz\qeb\qzb}^{\beta_1\beta_2\beta_3\beta_4} ~= $}
  \put(33,-25){   
  \put(104,27)   {\includepicclax2{66}{66}}
  \put(300,418)   {\pA A}
  \put(208,385)   {\pl{\phi_{\alphz}}}
  \put(259,395)   {\pl{\phi_{\alphv}}}
  \put(338,402)   {\pl{\phi_{\!\beta_2}^{}}}
  \put(472.5,381) {\pl{\phi_{\!\beta_1}^{}}}
  \put(193,360)   {\pg{\ib_1^{}}}
  \put(248,357)   {\pg{\ib_2^{}}}
  \put(198,319)   {\pg{\ia_1^{}}}
  \put(250,321)   {\pg{\ia_2^{}}}
  \put(403,380)   {\pg{\qeb}}
  \put(428.8,91)  {\pg{\qz}}
  \put(332,465)   {\pg{\qzb}}
  \put(291,276)   {\pB B}
  \put(211,247)   {\pl{\phi_{\alphe}}}
  \put(260,256)   {\pl{\phi_{\alphd}}}
  \put(316,258)   {\pl{\phi_{\!\beta_4}^{}}}
  \put(438,439.5) {\pl{\phi_{\!\beta_3}^{}}}
  \put(481,437)   {\pX X}
  \put(456,387)   {\pA A}
  \put(436,454.5) {\pB B}
  \put(375,262)   {\pg{\qe}}
  \put(326,122)   {\pg{\qzb}}
  \put(189.7,93)  {\pg{\ja_1^{}}}
  \put(255.7,93)  {\pg{\ja_2^{}}}
  \put(156,437)   {\pg{\jb_1^{}}}
  \put(208,429)   {\pg{\jb_2^{}}}
  \put(498,342)   {\pg{\qe}}
  \put(510,365)   {\pg{\qeb}}
  \put(475,338)   {\pg{\qz}}
  \put(457,334)   {\pg{\qzb}}
  } }
in \SzSe.

When evaluating the invariant of \erf{fact_MF_S2S1_res:66}, the connected component
of the ribbon graph that contains the ribbon with the defect $X$ just gives a scalar
factor. By comparison with \erf{def_b:40,def_op:44} this number can be written as
  \eqpic{def_op_rew:67}{350}{61}{
  \put(70,-4){
  \put(0,0)   {\Includepicclal{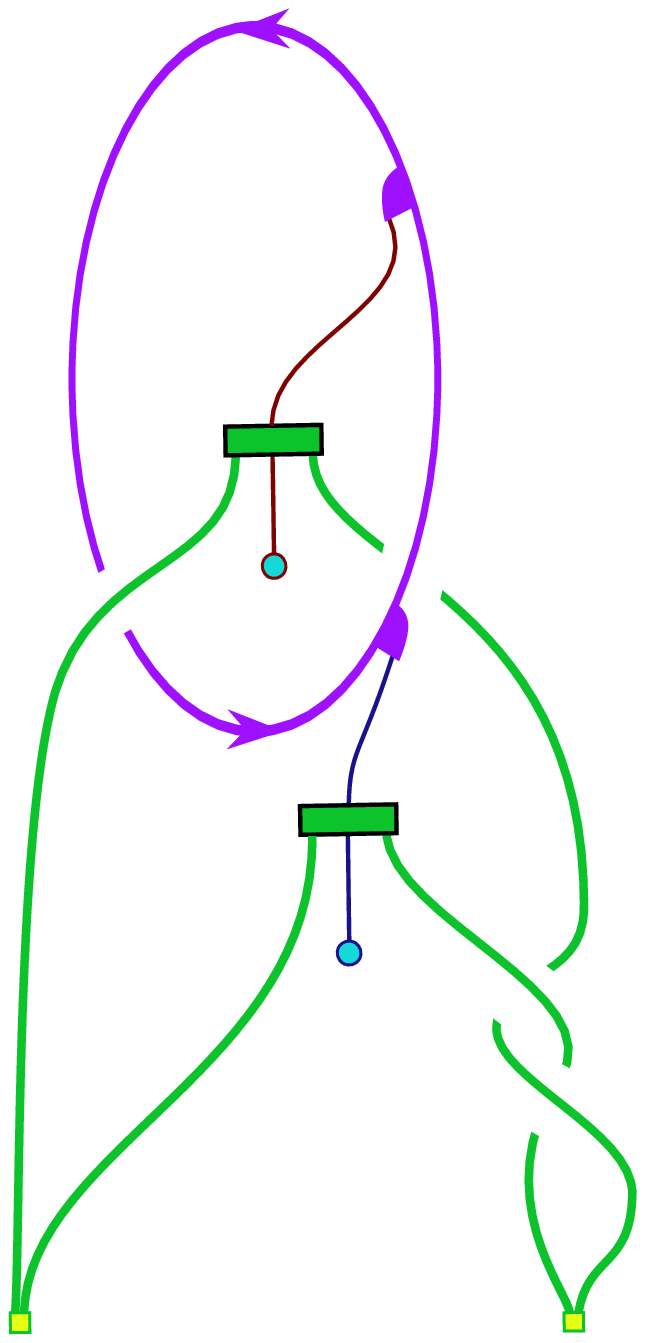}}
  \put(47,138)    {\pX X}
  \put(21,104.8)  {\pl{\phi_{\beta_1}}}
  \put(50.5,56.1) {\pl{\phi_{\beta_3}}}
  \put(39,116)    {\pA A}
  \put(47.4,67)   {\pB B}
  \put(1.2,11)    {\pl{\qe}}
  \put(16,11)     {\pl{\qeb}}
  \put(57,11)     {\pl{\qzb}}
  \put(76.6,11)   {\pl{\qz}}
  }
  \put(163,64)    {$ =~ \dim(X)\, \Rm{\qzb}{\qz}0\nl\nl\,
                     \dcoef{\qe}{\qz}{\beta_1}{\beta_3}X $ }
  }
where ${\sf R}^-$ is an inverse braiding matrix (as defined e.g.\ in (2.42)
of \cite{fuRs4}).

The other component of the ribbon graph in \erf{fact_MF_S2S1_res:66} can be 
slightly simplified further by first rotating the pieces involving bulk fields
in phase $A$ by 180 degrees (and using the bimodule morphism properties of the
bulk fields to rearrange the resulting form of the $A$-ribbons) and then 
deforming the $\qe$- and $\qz$-ribbons in such a way that in particular all the 
braidings involving an $A$-ribbon are removed. This gives rise to a twist of 
the $\ja_1$- and of the $\ja_2$-ribbon, to an inverse twist of the $\qe$-ribbon,
and to a double twist as well as a self-braiding of the $\qz$-ribbon. Note that 
the latter braiding precisely cancels the inverse braiding from 
\erf{def_op_rew:67}, and that the twist values $\theta_{\qe}$ and 
$\theta_{\qz}$ are equal, because $\qe$ and $\qz$ are the
chiral labels of a non-zero bulk field. In addition we can use the
relation between basis three-point couplings and (co)evaluation morphisms
(see formulas (2.34) and (2.35) of of \cite{fuRs4}), which amounts to factors 
of $\dim(U_{\qe})^{-1}_{}$ and $\dim(U_{\qz})^{-1}_{}$. We then end up with
  \be
  \bearl\dsty
  \cXo = \frac1{S_{0,0}^3} \dim(X)\, \theta_{\ja_1}^{}\, \theta_{\ja_2}^{}
  \!\sum_{\qe,\qz\in\I}\! \dim(U_{\qe}) \dim(U_{\qz})\,\theta_{\qz}^{}
  \nxl{-1}\dsty \hsp{11.5}
  \sum_{\beta_1,\beta_2,\beta_3,\beta_4}\! {(\cbulki_{A;\qe,\qz})}
  _{\beta_2\beta_1}\, {(\cbulki_{B;\qeb,\qzb})}_{\beta_4\beta_3}\,
  Z(\K{\ia_1}{\ia_2}{\qe}{\ja_1}{\ja_2}{\qz}{\alphz\alphv\beta_2}
  {\alphe\alphd\beta_4})\, \dcoef{\qe}{\qz}{\beta_1}{\beta_3}X \,
  \eear
  \labl{cXo-fin}
with the ribbon graph
  \eqpic{SCIT:70}{360}{149}{ \setulen 70
  \put(0,222){$\dsty \K{\ia_1}{\ia_2}{\qe}{\ja_1}{\ja_2}{\qz}
               {\alphz\alphv\beta_2}{\alphe\alphd\beta_4} ~= $}
  \put(47,-123){   
  \put(104,110)   {\includepicclax2{66}{70}}
  \put(179,387)   {\pl{\phi_{\!\alphz}^{}}}
  \put(258,389)   {\pl{\phi_{\!\alphv}^{}}}
  \put(309,398)   {\pl{\phi_{\!\beta_2}^{}}}
  \put(197,360)   {\pg{\ib_1^{}}}
  \put(249,360)   {\pg{\ib_2^{}}}
  \put(193,312)   {\pg{\ia_1^{}}}
  \put(251,302)   {\pg{\ia_2^{}}}
  \put(310,313)   {\pg{\qe}}
  \put(308,360)   {\pg{\qeb}}
  \put(210,243)   {\pl{\phi_{\!\alphe}^{}}}
  \put(261,254)   {\pl{\phi_{\!\alphd}^{}}}
  \put(319,256)   {\pl{\phi_{\!\beta_4}^{}}}
  \put(205,182)   {\pg{\ja_1^{}}}
  \put(255.3,182) {\pg{\ja_2^{}}}
  \put(355,180)   {\pg{\qz}}
  \put(180.6,437) {\pg{\jb_1^{}}}
  \put(228,437)   {\pg{\jb_2^{}}}
  \put(336,440)   {\pg{\qzb}}
  \put(294,409.5) {\pA A}
  \put(292,270.5) {\pB B}
  } }
in \SzSe. Note that the invariants $Z(\K{\ia_1}{\ia_2}
{\qe}{\ja_1}{\ja_2}{\qz}{\alphz\alphv\beta_2}{\alphe\alphd\beta_4})$
are traces on spaces of three-point conformal blocks on the sphere.


\section{The classifying algebra}\label{sec:CD}

What we have achieved so far is to express the vacuum channel coefficient \cXo\
of the correlator \erf{CX-equiv} in two different ways: \budefa\
leads according to \erf{c=dd} to a product of two \dtc s, while double bulk 
factorization yields the linear combination \erf{cXo-fin} of \dtc s. We now
compare these two expressions for \cXo\ and thereby find that
  \be
  \dcoef ij\alpha\beta X\, \dcoef kl\gamma\delta X
  = \sum_{p,q\in\I} \sum_{\mu,\nu}
  \clc{ij}\alpha\beta{kl}\gamma\delta{pq}\mu\nu\, \dcoef pq\mu\nu X
  \labl{clc0}
for all $i,j,k,l\iN \I$ and all $\alpha,\beta,\gamma,\delta$ in the relevant
spaces of bimodule morphisms, with coefficients that do not depend on the 
simple defect $X$:
  \be
  \clc{\ia\ja}\alpha\beta{kl}\gamma\delta{pq}\mu\nu
  = \frac1{S_{0,0}^2} \dim(U_p) \dim(U_q)\,\theta_\ja\,\theta_l\,\theta_q
  \sum_{\kappa,\lambda}
  {(\cbulki_{A;pq})}_{\kappa\mu}\, {(\cbulki_{B;\pb\qb})}_{\lambda\nu}\,
  Z(\K \ib\kb p\jb\lb q{\alpha\gamma\kappa}{\beta\delta\lambda}) \,,
  \labl{clc}
where the cobordism $\K \ib\kb p\jb\lb q{\alpha\gamma\kappa}{\beta\delta\lambda}$ 
is given by \erf{SCIT:70}.

Now consider the \findim\ complex vector space
  \be
  \CD = \bigoplus_{p,q\in\I} \Homaa{U_p\otip A\otim U_q}A
  \otic \Hombb{U_\pb\otip B\otim U_{\qb}}B
  \ee
that we introduced in \erf{CD-as-vs}. Recall that this is the space of pairs
$(\phi^{pq}_A,\phi^{\pb\qb}_B)$ of bulk fields in phase $A$ with chiral labels 
$p,q$ and in phase $B$ with labels $\pb,\qb$, respectively. The dimension of the
space of bulk fields $\phi^{ij}_A$ is given by the entry $\Z_{ij}(A)$ of the 
matrix $\Z(A)$ that describes the torus partition function (in phase $A$) in the
standard basis of characters. Thus the dimension of the vector space \CD\ is
  \be
  \dimc(\CD) = \mathrm{tr}\,(\Z(A)\,\Z(B)^{\rm t}) \,,
  \labl{dimCD}
which according to remark 5.19(ii) of \cite{fuRs4} equals the number of
isomorphism classes of simple $A$-$B$-bimodules, i.e.\ the number of types of
simple $A$-$B$-defects.

Choosing a basis $\{\phi^{ij,\alpha}_A\,|\,\alpha \eq1,2,...\,,\Z_{ij}(A)\}$ 
for each space $\Homaa{U_i\otip A\otim U_j}A$ of bulk fields in phase $A$ and
analogously for those in phase $B$, a basis of \CD\ is given by
  \be
  \{ \phi^{pq,\alpha\beta}_{} \}
  = \{ \phi^{pq,\alpha}_A \oti \phi^{\pb\qb,\beta}_B \,|\, p,q\iN\I\,,\,
  \alpha\eq1,2,...\,,\Z_{pq}(A)\,,\, \beta\eq1,2,...\,,\Z_{pq}(B) \} \,.
  \labl{phi-ijab}
We can define a multiplication on \CD\ by using the coefficients \erf{clc}
as structure constants in the basis \erf{phi-ijab}, i.e.\ by setting
  \be
  \phi^{ij,\alpha\beta}_{} \cdot \phi^{kl,\gamma\delta}_{}
  := \sum_{p,q\in\I} \sum_{\mu,\nu} 
  \clc{ij}\alpha\beta{kl}\gamma\delta{pq}\mu\nu\, \phi^{pq,\mu\nu}_{} .
  \labl{def-prod}

This product on \CD\ turns out to behave neatly: we have the following
\nxl3
{\bf Theorem:}
\nxl2
(1)\,~The complex vector space \CD\ endowed with the product \erf{def-prod}
is a semisimple commutative unital associative algebra.
\nxl2
(2)\,~The (\onedim) irreducible \rep s of the algebra \CD\ are in bijection
with the types of simple topological defects separating the phases 
$A$ and $B$, i.e.\ with the isomorphism classes of simple $A$-$B$-bimodules.
Their \rep\ matrices are furnished by the \dtc s.

\medskip

Let us first have a look at the specialization of the classifying algebra
\CD\ to the Cardy case. The rest of this section will then be devoted to 
the proof of the theorem in the general case.

\subsubsection*{The Cardy case}

In the Cardy case, i.e.\ for $A\eq B$ being (Morita equivalent to) the tensor 
unit $\one$, the invariant of
$\K{\ia_1}{\ia_2}\qe{\ja_1}{\ja_2}\qz{\alphz\alphv\beta_2}{\alphe\alphd\beta_4}$
in \erf{SCIT:70} is non-zero only if $\ja_1 \eq \ib_1$, $\ja_2 \eq \ib_2$ and
$\qz \eq \qeb$, with all labels $\alphe,...\,,\beta_4$ taking only a single 
value $\nl$. And in this case the invariant reduces, after straightening out the
ribbons, to the trace over the identity morphism of the object $U_{\ia_1}\oti 
U_{\ia_2}\oti U_\qe$ which, in turn, is given by the fusion rule coefficient 
$\N{\ia_1}{\ia_2}\qeb$. The structure constants \erf{clc} then take the form
  \be
  \clc{\ia\ib}\nl\nl{\ja\jb}\nl\nl{p\pb}\nl\nl
  =  \frac{\theta_\ia^{}\,\theta_\ja^{}}{\theta_p^{}}\,
  \frac{{\dim(U_p)}^2} {{\dim(U_\ia)}^2\, {\dim(U_\ja)}^2}\, \N\ia\ja p \,. 
  \labl{CD-caca}
Thus we recognize \CD\ as the \emph{fusion algebra} that describes the fusion 
product of the chiral sectors of the theory (or in other words, as the 
complexified Grothendieck ring of the category \C\ of chiral sectors), albeit 
expressed in a non-standard basis related to the standard one by a rescaling 
with $\theta_\ia\,{/}\!\dim(U_\ia)^2$.

The first part of the theorem is now a well-known statement about fusion 
algebras of rational conformal field theories, while the second part reduces to 
the fact that the inequivalent \onedim\ irreducible \rep s $R_x$, $x\iN\I$, of 
the fusion algebra are given by the generalized quantum dimensions $(S_{i,x}
{/}\!S_{0,x})^{}_{i\in\I}$. And indeed according to \erf{cardy-case} in the 
Cardy case the \dtc s are nothing but rescaled generalized quantum dimensions.

\subsubsection*{Commutativity}

Note that the fusion algebra of defects is, in general, \emph{not}
commutative, as the tensor category of defects is not braided. Nevertheless
the classifying algebra \CD\ \emph{is} commutative.

That \CD\ is commutative and has a unit is easy to see. For commutativity, 
the basic observation is that by simple properties of the braiding and of
bimodule morphisms (of braided-induced bimodules, with prescribed
choice of over- or underbraiding) one has
  \Eqpic{comm1..comm6:74}{440}{44}{ \setulen 90
  \put(-17,10){\includepicclax3{42}{74a}
  \put(-2,-9)  {\pg i}   
  \put(18,-9)  {\pg j}   
  \put(32,-9)  {\pl C}
  \put(28.2,41){\pl C}
  \put(32.8,104) {\pl C}
  \put(50,-9)  {\pg k}   
  \put(70,-9)  {\pg l}   
  \put(42,25.6){\pl{\phi_\bet^{}}}
  \put(42,75.6){\pl{\phi_{\!\alpha}^{}}}
  \put(78,35)  {$=$}
  }
  \put(78,10){\includepicclax3{42}{74b}
  \put(-2,-9)  {\pg i}   
  \put(18,-9)  {\pg j}   
  \put(33,-9)  {\pl C}
  \put(32.9,104) {\pl C}
  \put(50,-9)  {\pg k}   
  \put(70,-9)  {\pg l}   
  \put(76,35)  {$=$}
  }
  \put(171,10){\includepicclax3{42}{74c}
  \put(-2,-9)  {\pg i}   
  \put(18,-9)  {\pg j}   
  \put(32,-9)  {\pl C}
  \put(31.1,48){\pl C}
  \put(33.2,104) {\pl C}
  \put(50,-9)  {\pg k}   
  \put(70,-9)  {\pg l}
  \put(78,35)  {$=$}
  }
  \put(266,10){\includepicclax3{42}{74d}
  \put(-2,-9)  {\pg i}   
  \put(18,-9)  {\pg j}   
  \put(32.2,-9){\pl C}
  \put(34.4,106) {\pl C}
  \put(50,-9)  {\pg k}   
  \put(70.5,-9)  {\pg l}   
  \put(78,35)  {$=$}
  }
  \put(363,10){\includepicclax3{42}{74e}
  \put(-2,-9)  {\pg i}   
  \put(18,-9)  {\pg j}   
  \put(32.5,-9){\pl C}
  \put(31.5,110) {\pl C}
  \put(50,-9)  {\pg k}   
  \put(70.5,-9){\pg l}   
  \put(77,35)  {$=$}
  }
  \put(456,10){\includepicclax3{42}{74f}
  \put(-2,-9)  {\pg i}   
  \put(18,-9)  {\pg j}   
  \put(32.4,-9){\pl C}
  \put(28.2,31){\pl C}
  \put(33.4,104) {\pl C}
  \put(50,-9)  {\pg k}   
  \put(70.5,-9){\pg l}   
  \put(40.1,29.1){\pl{\phi_{\!\alpha}^{}}}
  \put(42,75.8){\pl{\phi_\bet^{}}}
  } }
for any $\phi_\alpha \iN \HomCC(U_i\otip C\otim U_l,C)$ and $\phi_\beta \iN 
\HomCC(U_j\otip C\otim U_k,C)$, as well as a similar identity with the 
over-braiding $c_{U_i,U_j}^{}$ replaced by the under-braiding $c_{U_j,U_i}^{-1}$
and the under-braiding $c_{U_l,U_k}^{-1}$ by the over-braiding $c_{U_k,U_l}^{}$,
  \eqpic{comm1,comm7:74}{430}{43}{ \setulen 90
  \put(150,10){\includepicclax3{42}{74a}
  \put(-2,-9)  {\pg i}   
  \put(18,-9)  {\pg j}   
  \put(32.3,-9){\pl C}
  \put(28.3,41){\pl C}
  \put(33.4,103.5) {\pl C}
  \put(50,-9)  {\pg k}   
  \put(70,-9)  {\pg l}   
  \put(42,25.6){\pl{\phi_\bet^{}}}
  \put(42,75.6){\pl{\phi_{\!\alpha}^{}}}
  \put(96,35)  {$=$}
  }
  \put(276,10){\includepicclax3{42}{74g}
  \put(-2,-9)  {\pg i}   
  \put(18,-9)  {\pg j}   
  \put(32.3,-9){\pl C}
  \put(28.2,31){\pl C}
  \put(33.1,103.5) {\pl C}
  \put(50,-9)  {\pg k}   
  \put(70,-9)  {\pg l}   
  \put(40.1,29.1){\pl{\phi_{\!\alpha}^{}}}
  \put(42,75.8){\pl{\phi_\bet^{}}}
  } }
Applying the identity \erf{comm1..comm6:74} to the bulk fields $\PHi{\alphz}$ 
and $\PHi{\alphv}$ in the ribbon graph \erf{SCIT:70}, and the identity 
\erf{comm1,comm7:74} to the bulk fields $\PHi{\alphe}$ and $\PHi{\alphd}$, one 
sees that the invariant of that ribbon graph is symmetric under simultaneous 
exchange of the corresponding quadruples of labels,
  \be
  Z(\K ikpjlq{\alpha\gamma\kappa}{\beta\delta\lambda})
  = Z(\K kipljq{\gamma\alpha\kappa}{\delta\beta\lambda}) \,.
  \labl{Ksymm}
Applying \erf{comm1..comm6:74} and \erf{comm1,comm7:74} also to other pairs of 
fields one shows likewise that the invariant is even \emph{totally} symmetric 
in the three quadruples $(ij\alpha\beta)$, $(kl\gamma\delta)$ and 
$(pq\kappa\lambda)$. 

The symmetry \erf{Ksymm} induces a symmetry of the structure constants \erf{clc}:
  \be
  \clc{ij}\alpha\beta{kl}\gamma\delta{pq}\mu\nu 
  = \clc{kl}\gamma\delta{ij}\alpha\beta{pq}\mu\nu \,.
  \ee
Thus the product \erf{def-prod} is commutative,
  \be
  \phi^{ij,\alpha\beta}_{} \cdot \phi^{kl,\gamma\delta}_{}
  = \phi^{kl,\gamma\delta}_{} \cdot \phi^{ij,\alpha\beta}_{} .
  \ee

\subsubsection*{Unitality}

Using dominance one easily shows that if one of the bulk fields in each phase
is an identity field, then the invariant of \erf{SCIT:70} is essentially
a product of two two-point functions on the sphere, i.e.
  \be 
  Z(\K \ia k0\ja l0{\beta\delta\nl}{\alpha\gamma\nl})
  = \frac{S_{0,0}^2} {\dim(U_{\ia}) \dim(U_{\ja})\,\theta_{\ia}\,\theta_{\ja}}\,
  \delta_{\ib,k}\, \,\delta_{\jb,l}\,
  {(\cbulk_{B;\ia,\ja})}_{\gamma\alpha}\, {(\cbulk_{A;\ib,\jb})}_{\beta\delta}
  \ee
and analogously for $\ia\eq\ja\eq 0$ and for $k\eq l\eq 0$ (recall that the 
invariant is totally symmetric). This implies that
  \be
  \clc{00}\nl\nl{ij}\alpha\beta{pq}\mu\nu
  = \delta_{i,p}\, \delta_{j,q}\, \delta_{\alpha,\mu}\, \delta_{\beta,\nu}\,
  = \clc{ij}\alpha\beta{00}\nl\nl{pq}\mu\nu \,,
  \ee
and thus the basis element $\phi^{00,\nl\nl}$ is a unit for the product
\erf{def-prod}.

Likewise it follows that the map from $\CD\oti\CD$ to \complex\ defined by
  \be
  \phi^{ij,\alpha\beta}_{} \oti \phi^{kl,\gamma\delta}_{}
  \mapsto \clc{ij}\alpha\beta{kl}\gamma\delta{00}\nl\nl
  \ee
is a non-degenerate bilinear form on \CD.

\subsubsection*{Associativity}

For establishing associativity it proves to be convenient to express
the structure constants \erf{clc} in terms of the coefficients that appear 
in the identity (compare section A.2 of \cite{fuSs})
  \eqpic{IV_67_for_A:05}{380}{48}{
  \put(0,0){     {\Includepicclal{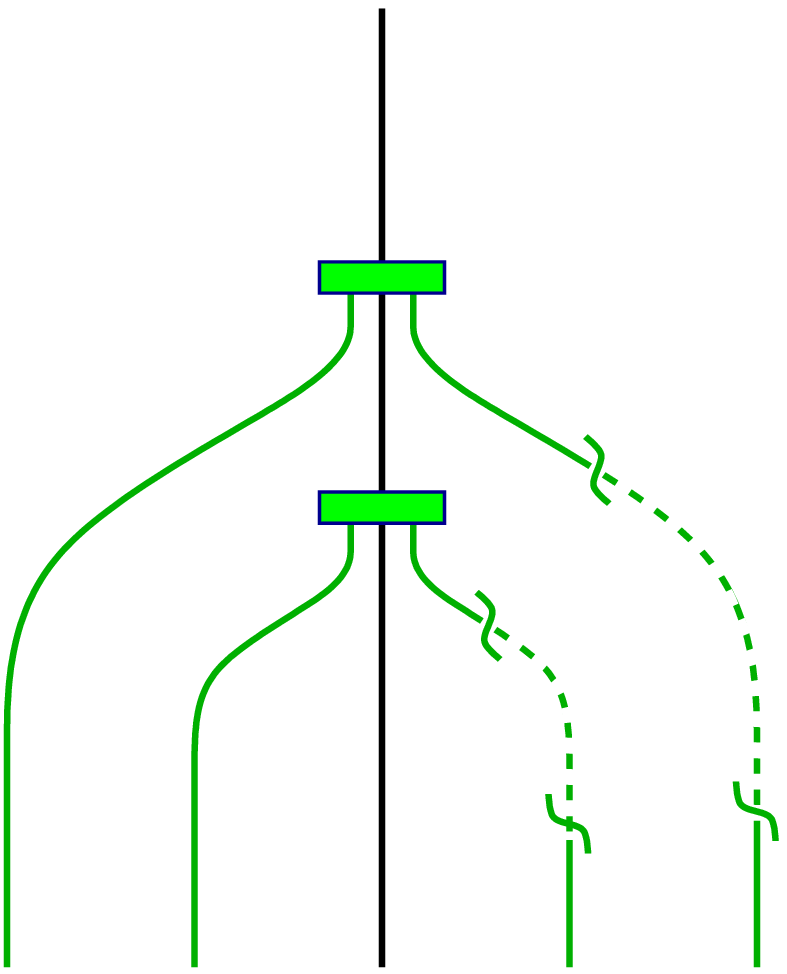}}
  \put(-0.9,-8.8){\pg i}
  \put(18.8,-8.8){\pg j}
  \put(38.2,-8.8){\pl C}
  \put(34.1,57.3){\pl C}
  \put(38.8,110.6){\pl C}
  \put(49.8,49.6){\tiny$\alpha $}
  \put(49.8,74.6){\tiny$\beta $}
  \put(59.9,-8.8){\pg k}
  \put(81.6,-8.8){\pg l}
  }
  \put(110,47)   {$ =~ \dsty\sum_{q,q'\in\I} \sum_{\gamma=1}^{\Z_{qq'}}
                 \sum_{\delta=1}^{\N ijq}\sum_{\delta'=1}^{\N ij{q'}}\,
                 \FC ijkl\alpha\beta\gamma {qq'}{\delta\delta'} $}
  \put(284,0){   {\Includepicclal{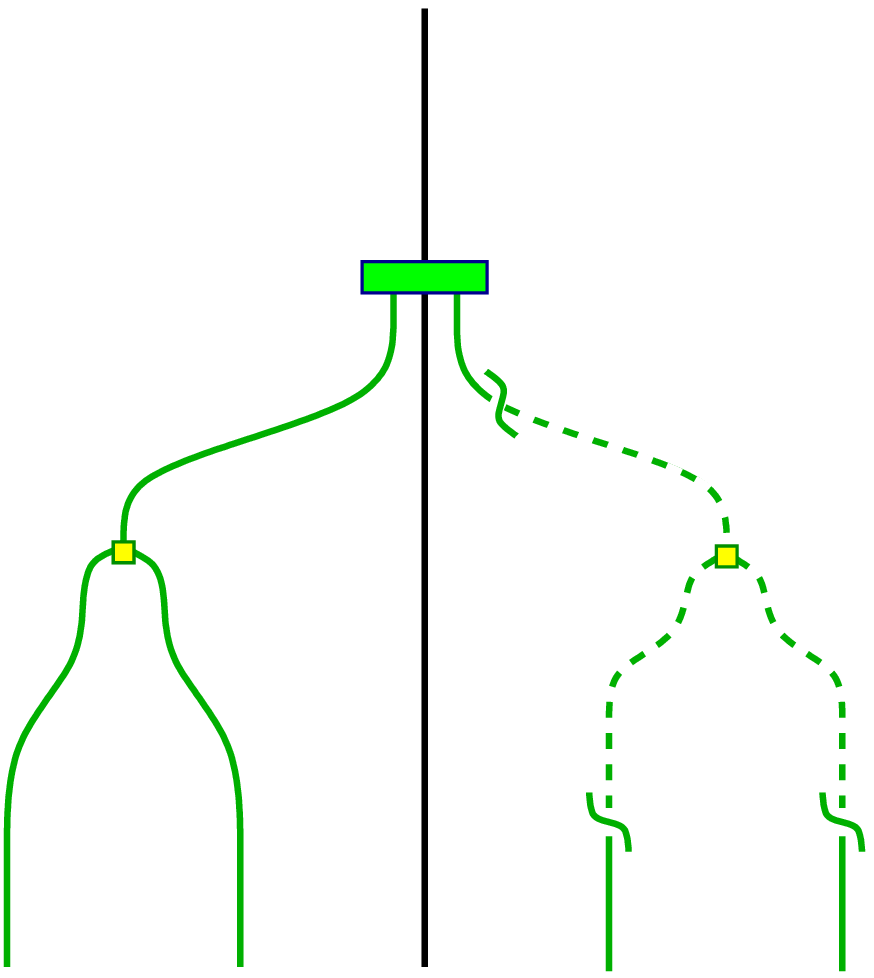}}
  \put(-0.8,-8.8){\pg i}
  \put(24.0,-8.8){\pg j}
  \put(24.1,63.2){\pg q}
  \put(43.1,-8.8){\pl C}
  \put(43.8,110.6){\pl C}
  \put(54.8,75.9){\scriptsize$ \gamma $}
  \put(62.4,63.2){\pg {q'}}
  \put(64.6,-8.8){\pg k}
  \put(15.2,49.1){\tiny$\delta $}
  \put(81.9,47.3){\tiny$\delta' $}
  \put(91.3,-8.8){\pg l}
  } }
between bimodule morphisms. By a four-fold application of this identity we obtain
  \be
  \hspace*{-.8em}\bearl\dsty
  Z(\K \ia kp\ja lq{\alpha\gamma\kappa}{\beta\delta\lambda})
  = \sum_{m_1,m_2,m_3,m_4}\,\sum_{\mu_1,\mu_2,\nu_1,\nu_2,\nu_3,\nu_4}
  \FA{\ib}{\kb}{\lb}{\jb}\gamma\alpha{\mu_1}{m_1m_2}{\nu_1\nu_2}\,
  \FB{k}{\ia}{\ja}{l}\beta\delta{\mu_2}{m_3m_4}{\nu_3\nu_4}
  \Nxl2\hsp{6.8}\dsty
  \sum_{n_1,n_2,n_3,n_4}\,\sum_{\mu_3,\mu_4,\sigma_1,\sigma_2,\sigma_3,\sigma_4}
  \FA{m_1}{\pb}{\qb}{m_2}{\kappa}{\mu_1}{\mu_3}{n_1n_2}{\sigma_1\sigma_2}\,
  \FB{p}{m_3}{m_4}{q}{\mu_2}{\lambda}{\mu_4}{n_3n_4}{\sigma_3\sigma_4}\,
  Z(\SCIT{\ia}{k}{p}{\ja}{l}{q}{\alpha\gamma\kappa}{\beta\delta\lambda})
  \eear
  \labl{FFFF}
with
  \eqpic{SCIT_simp:71}{320}{151}{ \setulen 70
  \put(0,230){$\dsty \SCIT{\ia}{k}{p}{\ja}{l}{q}
              {\alpha\gamma\kappa}{\beta\delta\lambda} ~= $}
  \put(27,-116){
  \put(104,110)   {\includepicclax2{66}{71}}
  \put(178,408)   {\pl{\phi_{\!\mu_3}^{}}}
  \put(279,353)   {\pg{\kb}}
  \put(206,342)   {\pg{\ib}}
  \put(194,311)   {\pg{\ia}}
  \put(241,310)   {\pg k }
  \put(296,311)   {\pg p }
  \put(327,357)   {\pg{\pb}}
  \put(316,276)   {\pl{\phi_{\!\mu_4}^{}}}
  \put(206.5,182) {\pg{\ja}}
  \put(260,182)   {\pg l }
  \put(346,182)   {\pg q }
  \put(195,437)   {\pg{\jb}}
  \put(286,432)   {\pg{\lb}}
  \put(338,437)   {\pg{\qb}}
  \put(264,420)   {\pg{\scriptstyle \nu_{\!2}^{}}}
  \put(233,400)   {\pg{m_2}}
  \put(217,417)   {\pg{\scriptstyle \sigma_{\!2}}}
  \put(197,414)   {\pg{n_2}}
  \put(202,382)   {\pg{n_1}}
  \put(232,378)   {\pg{\scriptstyle \sigma_{\!1}^{}}}
  \put(245,371)   {\pg{m_1}}
  \put(266,377)   {\pg{\scriptstyle \nu_{\!1}^{}}}
  \put(299,279)   {\pg{n_3}}
  \put(281,276)   {\pg{\scriptstyle \sigma_{\!3}^{}}}
  \put(253,281)   {\pg{m_3}}
  \put(237,293)   {\pg{\scriptstyle \nu_{\!3}^{}}}
  \put(312,253)   {\pg{n_4}}
  \put(292,253)   {\pg{\scriptstyle \sigma_{\!4}}}
  \put(252,243)   {\pg{m_4}}
  \put(239,239)   {\pg{\scriptstyle \nu_{\!4}^{}}}
  } }
Since the space of one-point blocks with insertion $U_i$ on the sphere vanishes
unless $i\eq0$, only $n_1 \eq n_2 \eq n_3 \eq n_4 \eq 0$ gives a non-zero 
contribution in the summations in \erf{FFFF}. In the resulting expression we
use the identities
  \be
  {S_{0,0}^{}}\, {(\cbulk_{C;\qe\qz})}_{\beta\gamma} 
  = {\dim(C)}\, \FC{\qe}{\qeb}{\qzb}{\qz}\gamma\beta\nl{00}{\nl\nl}
  = {S_{0,0}^{}}\, \R{\qe}{\qeb}0\nl\nl\, \Rm{\qz}{\qzb}0\nl\nl \,
  {(\cbulk_{C;\qeb\qzb})}_{\gamma\beta}
  \ee
which follow by performing some elementary fusing and braiding moves on the 
result (C.14) of \cite{fjfrs} for the bulk two-point function; they allow us 
to cancel the factors $\cbulki_{A;pq}$ and $\cbulki_{B;\pb\qb}$ in \erf{clc}. 
Doing so, further summations over morphism spaces become trivial. We then end 
up with the expression
  \be
  \clc{\ia\ja}\alpha\beta{kl}\gamma\delta{pq}\mu\nu
  = \dim(U_p) \dim(U_q)\,\theta_\ja\,\theta_l\,\theta_q\,
  \R p\pb 0\nl\nl\, \Rm q\qb 0\nl\nl
  \!\sum_{\kappa,\lambda,\rho,\sigma}\!
  \FA \ia k l\ja \gamma\alpha\mu {pq} {\kappa\lambda}\,
  \FB \kb\ib\jb\lb \beta\delta\nu {\pb\qb} {\rho\sigma}\,
  \SCmorph p\kb\ib {\kappa}{\rho}\, \SCmorphb \jb\lb q {\lambda}{\sigma}
  \labl{clc2}
for the structure constants, where we introduced the morphisms
  \eqpic{SCmorph:72,SCmorphb:73}{410}{50}{
  \put(20,50)  {$\SCmorph ijk \kappa\lambda ~:= $}
  \put(73,5){ {\Includepicclal{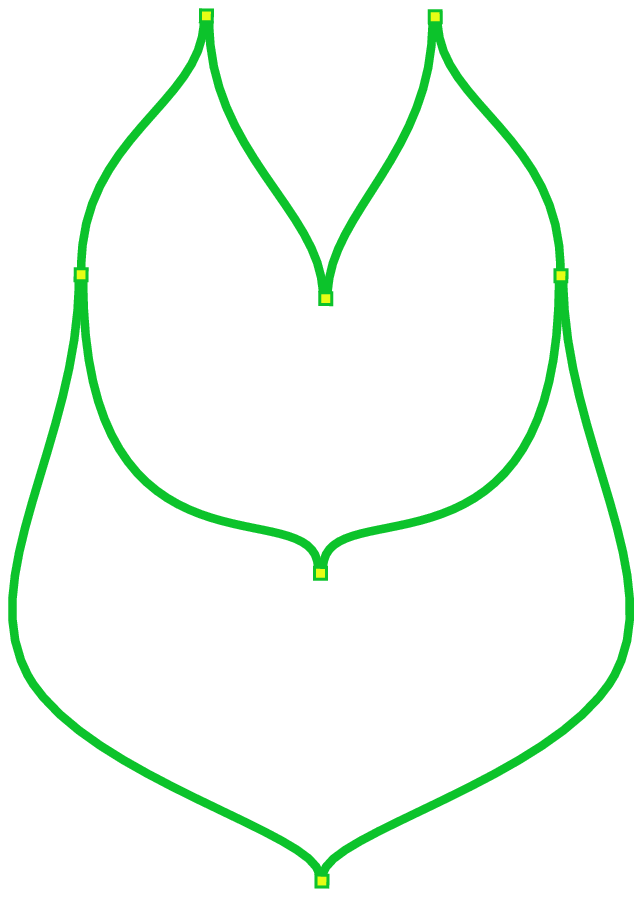}}
  \put(43,75)   {\pg i }
  \put(43,44)   {\pg j }
  \put(43,12)   {\pg k }
  \put(10.2,66.4){\pg{\kappa}}
  \put(63.2,66.4){\pg{\lambda}}
  }
  \put(182,50)  {and $\qquad \SCmorphb ijk \kappa\lambda ~:= $}
  \put(282,0){ {\Includepicclal{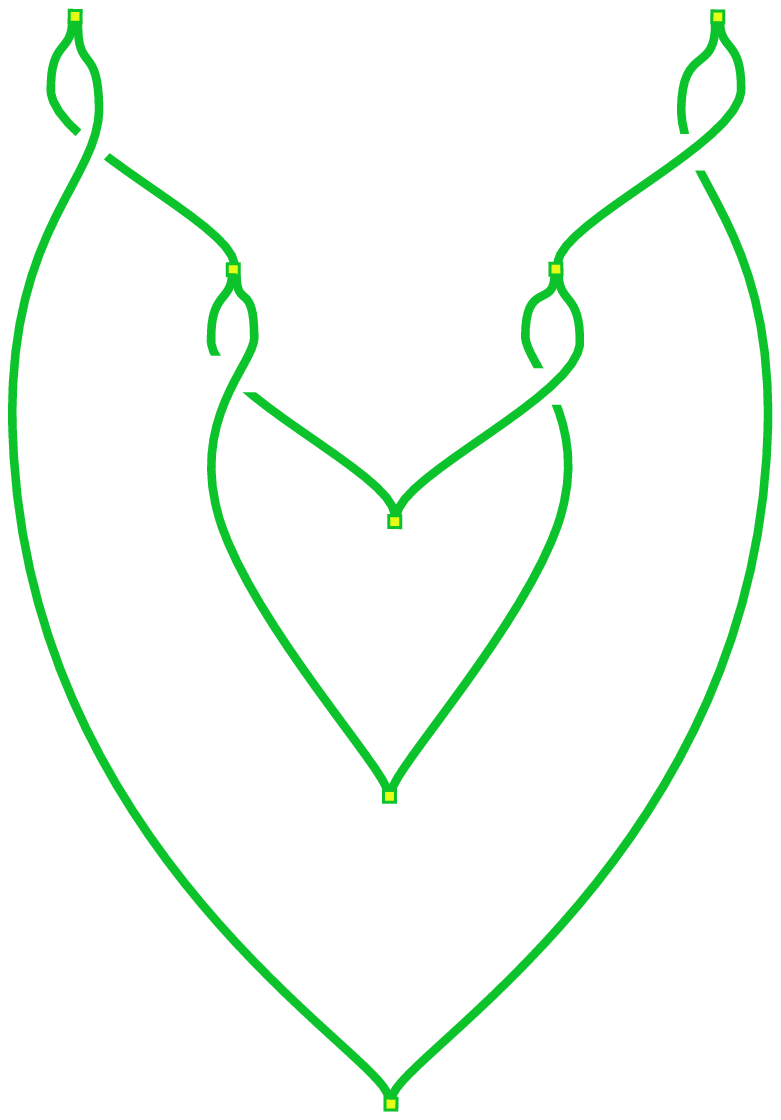}}
  \put(48,76)   {\pg i }
  \put(57,51)   {\pg j }
  \put(71,27)   {\pg k }
  \put(27.1,91.9){\pg{\kappa}}
  \put(53.9,91.9){\pg{\lambda}}
  } }
in $\End(\one) \,{\cong}\, \complex$.

\medskip

Next we observe that in the same way as we obtained an ordinary binary
product on \CD, we can endow \CD\ also with an $n$-ary product for any integer 
$n\,{\ge}\,2$. To define the structure constants of these products in the basis
\erf{phi-ijab}, we use the same expression as in \erf{clc}, with the only 
modification that the ribbon graph \erf{SCIT:70} is replaced by another
ribbon graph $\K{\ia_1\ia_2...}{\ia_n}{\qe}{\ja_1\ja_2...}{\ja_n}{\qz}
{\alpha_2^{}\alpha_4^{}...\alpha_{2n}^{}\beta_2}{\alpha_1^{}\alpha_3^{}...
\alpha_{2n-1}^{}\beta_4}$ in \SzSe\ in which for each additional basis 
element in the argument of the product there is an extra ribbon along the 
$S^1$-direction and two extra bimodule morphisms:
  \eqpic{SCIT_n:77}{410}{161}{ \put(0,-8){ \setulen 75
  \put(0,223){$\dsty \K{\ia_1...}{\ia_n}{\qe}{\ja_1...}{\ja_n}{\qz} {\alpha_2^{}
             ...\alpha_{2n}^{}\beta_2}{\alpha_1^{}...\alpha_{2n-1}^{}\beta_4} ~=$}
  \put(180,-2)   {\includepicclax2{85}{77}}
  \put(176,-11)   {
  \put(34,286)   {\pl{\phi_{\!\alpha_2}^{}}}
  \put(113,288)  {\pl{\phi_{\!\alpha_4}^{}}}
  \put(201.3,289){\pl{\phi_{\!\alpha_{2n}}^{}}}
  \put(246,297)  {\pl{\phi_{\!\beta_2}^{}}}
  \put(71.5,272) {\pg{\ib_1}}
  \put(129,266)  {\pg{\ib_2}}
  \put(200,260)  {\pg{\ib_n}}
  \put(261,272)  {\pg{\qeb}}
  \put(47,214)   {\pg{\ia_1}}
  \put(101,216)  {\pg{\ia_2}}
  \put(195.6,214){\pg{\ia_n}}
  \put(252,217)  {\pg{\qe}}
  \put(58,142)   {\pl{\phi_{\!\alpha_1}^{}}}
  \put(108,152)  {\pl{\phi_{\!\alpha_3}^{}}}
  \put(201,152)  {\pl{\phi_{\!\alpha_{2n\!-\!1}}^{}}}
  \put(257.5,154){\pl{\phi_{\!\beta_4}^{}}}
  \put(59.4,81)  {\pg{\ja_1}}
  \put(110.7,81) {\pg{\ja_2}}
  \put(205,81)   {\pg{\ja_n}}
  \put(291,81)   {\pg{\qz}}
  \put(50.5,342) {\pg{\jb_1}}
  \put(101.5,342){\pg{\jb_2}}
  \put(194,342)  {\pg{\jb_n}}
  \put(274,341)  {\pg{\qzb}}
  \put(71.5,301) {\pA A }
  \put(218,169)  {\pB B }
  } } }

By analogous arguments as for the binary product one checks that all the
$n$-ary products are totally commutative, i.e.\ invariant under any 
permutation of their arguments, and one obtains expressions for the
structure constants that generalize those for the binary product in an
obvious manner. In particular, the structure constants of the ternary
product can be written as
  \be
  \bearl\dsty \hspace*{-1.6em}
  \clc{\ia\ja}\alpha\beta{kl}\gamma{\delta;mn\kappa\lambda}{pq}\mu\nu
  = \! \frac1{S_{0,0}^{}} \dim(U_p) \dim(U_q)\,
  \theta_\ja\,\theta_l\,\theta_n\,\theta_q\, \R p\pb 0\nl\nl\, \Rm q\qb 0\nl\nl
  \Nxl1\hsp4\dsty
  \sum_{r,s\in\I} \sum_{\rho_1,\rho_2,
  \sigma_1,\sigma_2,\sigma_3,\sigma_4,\tau_1,\tau_2,\tau_3,\tau_4} \!\! 
  \dim(U_r) \dim(U_s)\,(\theta_s)^2\, \Rm{\bar r}r0\nl\nl\, \R{\bar s}s0\nl\nl\,
  \Nxl3\hsp5\dsty 
  \FA \ia k l \ja \gamma\alpha {\rho_1} {\bar r\bar s} {\sigma_1\sigma_2}\,
  \FB \kb\ib\jb\lb \beta\delta {\rho_2} {rs} {\sigma_3\sigma_4}\,
  \FA {\bar r}mn{\bar s} \lambda{\rho_1}\mu {pq} {\tau_1\tau_2}\,
  \FB {\bar m}rs{\bar n} {\rho_2}\lambda\nu {\bar p\bar q} {\tau_3\tau_4}\,
  \SCmorph {\bar r}\kb\ib {\sigma_1}{\sigma_3}\, 
  \SCmorphb \jb\lb{\bar s} {\sigma_2}{\sigma_4} \,
  \SCmorph p{\bar m}r{\tau_1}{\tau_3}\, \SCmorphb s{\bar n}q{\tau_2}{\tau_4}
  \nxl{-2.9}~
  \eear
  \labl{clc3}
with the factors of $\omega$ and $\varpi$ as defined in 
\erf{SCmorph:72,SCmorphb:73}.

By comparison with the expression \erf{clc2} for the structure constants of the
binary product it is the easy to see (using also the identity
$\Rm u{\bar u}0\nl\nl \eq \theta_u^2\, \R u{\bar u}0\nl\nl$) that the structure
constants of the ternary product can be expressed through those of the binary
one as
  \be
  \clc{\ia\ja}\alpha\beta{kl}\gamma{\delta;mn\kappa\lambda}{pq}\mu\nu
  = \sum_{r,s\in\I} \sum_{\rho,\sigma}
  \clc{\ia\ja}\alpha\beta{kl}\gamma\delta{rs}\rho\sigma \,
  \clc{rs}\rho\sigma{mn}\kappa\lambda{pq}\mu\nu .
  \ee
This proves associativity: it is an elementary fact that a commutative algebra 
endowed with a totally commutative ternary product is associative if at least 
one bracketing of a twofold binary product equals the ternary product.

\subsubsection*{Semisimplicity}

Obviously, the formula \erf{clc0} expresses the fact that the \dtc s for a
given simple defect $X$ constitute the ($1\Times1$) \rep\ matrices for
a \onedim\ irreducible \CD-\rep.

We now invoke theorem 4.2 of \cite{ffrs5}, according to which the 
$\dimc(\CD)\,{\times}\dimc(\CD)$\,-matrix furnished by the \dtc s (with rows and 
columns labeled by simple $A$-$B$-bimodules and by pairs of bulk fields,
respectively) is non-degenerate. This means that non-iso\-morp\-hic simple 
bimodules give rise to inequivalent \onedim\ representations of \CD\ and thus
implies that the number $n_{\text{simp}}(\CD)$ of inequivalent irreducible 
representations of \CD\ is at least as large as its dimension,
  \be
  n_{\text{simp}}(\CD) \ge \dimc(\CD) \,.
  \labl{ss1}

On the other hand, as any \findim\ associative algebra, \CD\ is isomorphic, 
as a module over itself, to the direct sum over all inequivalent indecomposable 
projective \CD-modules, each one occurring with a multiplicity given by the 
dimension of the corresponding irreducible module (see e.g.\ Satz G.10
of \cite{JAsc}), so that in particular
  \be
  \dimc(\CD) \ge n_{\text{simp}}(\CD) \,.
  \labl{ss2}
Thus in fact the number $n_{\text{simp}}(\CD)$ of inequivalent irreducible 
representations equals the dimension $\dimc(\CD)$. This is only possible if
every irreducible \rep\ is \onedim\ and projective, which in turn implies that 
\CD\ is semisimple.

\subsubsection*{Representation matrices}

Besides semisimplicity the previous arguments show at the same time that the 
$\dimc(\CD)$ irreducible modules obtained this way already exhaust all 
irreducible modules of \CD. Together with the fact that $\dimc(\CD)$ equals 
the number of isomorphism classes of simple $A$-$B$-bimodules (see \erf{dimCD}),
this establishes part (2) of the theorem.

\subsubsection*{Comparison with the algebra {\boldmath$\mathscr D_{\!A|A}^{\rm PZ}$}}

As already mentioned in section \ref{syno}, in the special case $A \eq B$ an
algebra $\mathscr D_{\!A|A}^{\rm PZ}$ over \complex\ that has the same 
dimension as $\mathscr D_{\!A|A}$ and also shares other properties of 
$\mathscr D_{\!A|A}$, including in particular commutativity, has been obtained 
in \cite[Sect.\,7.4]{pezu6}. According to \cite[(7.15)]{pezu6}, there is a 
Verlinde-like relation for the structure constants of the fusion algebra of 
defects in the basis of simple defect types, in which 
instead of the entries of the modular $S$-matrix the numbers
  \be
  \gcoef ij\alpha\beta X 
  := \sqrt{\dim(U_i)\,\dim(U_j)}\, \dim(X)\, \dcoef ij\alpha\beta X
  \labl{gcoef}
appear and the summation is over the set $\mathcal J$ of labels 
$(ij\alpha\beta)$ of the basis \erf{phi-ijab}. One may then define another 
algebra $\mathscr D_{\!A|A}^{\rm PZ}$, `dual' to the defect fusion algebra, 
by requiring its structure constants in a basis labeled by $\mathcal J$
to be given by another Verlinde-like relation featuring again the numbers
\erf{gcoef}, but now summed over the types $X$ of simple defects. 

Note that $|\mathcal J|$ equals the number of simple defect types, i.e.\ the 
matrix $\gc \eq ( \gcoef ij\alpha\beta X )$ formed by the numbers \erf{gcoef} is
a square matrix. Assuming, as in \cite{pezu6}, that this matrix $\gc$ is unitary, 
and also invoking that according to \cite{pezu6} 
the mapping $\gcoef ij\alpha\beta X \,{\mapsto}\, {(\gcoef ij\beta\alpha X)}^{\!*}$ of the
numbers \erf{gcoef} corresponds to an involution on the set $\mathcal J$, 
it follows that $\mathscr D_{\!A|A}^{\rm PZ}$ is commutative. Conversely, when
making these two assumptions one can show that the matrices of structure 
constants of the classifying algebra $\mathscr D_{\!A|A}$ can be 
simultaneously diagonalized, i.e.\ are given by a Verlinde-like formula. This 
in turn implies that $\mathscr D_{\!A|A}^{}$ and $\mathscr D_{\!A|A}^{\rm PZ}$ 
are in fact isomorphic as algebras over \complex\ and also that their 
irreducible \rep s furnish the \dtc s,
so that the algebra $\mathscr D_{\!A|B}$ can be seen as a
generalization of $\mathscr D_{\!A|A}^{\rm PZ}$ to $B \,{\ne}\, A$.
However, our approach, which works over any algebraically closed field of 
characteristic zero, cannot be used to prove that in the case of the complex 
numbers the matrix $\gc$ is indeed unitary or that complex conjugation of its 
entries gives rise to an involution on $\mathcal J$. (On the other hand, as 
observed in \cite{pezu6}, these properties are indeed satisfied e.g.\ for the 
particularly important case of the $\mathfrak{sl}(2)$ WZW models.)

\subsubsection*{Classification of boundary conditions}

Since the categories of left $A$-modules and of $A$-$\one$-bimodules are
equivalent, the classifying algebra $\mathscr D_{\!A|\one}$ should be isomorphic
to the classifying algebra for boundary conditions obtained in \cite{fuSs}. It 
is straightforward to show that this is indeed the case. Even more, in
the renormalized basis $\big\{\theta_p^{-1}S_{0,0}\dim(U_p)
\,\phi^{p\pb,\alpha\nl}_{} \big\}$ of $\mathscr D_{\!A|\one}$
the structure constants coincide with the structure constants of the 
classifying algebra for boundary conditions as given in \cite{fuSs}.


\section{The defect partition function}

In a rational CFT the classifying algebra \CA\ for boundary conditions governs 
in particular also the annulus partition functions. Denote by 
$\mathrm A_{k,M}^{~~~N}$ the expansion coefficients, in the basis of characters,
of the partition function for an annulus in phase $A$ with elementary boundary 
conditions $M$ and $N$ on its two boundary circles. One finds that the annulus
coefficients $\mathrm A_{k,M}^{~~~N}$ can be naturally expressed in terms of 
products of reflection coefficients \cite{prss3,bppz2},
i.e.\ of irreducible \CA-\rep s. Specifically \cite{stig5},
  \be
  \mathrm A_{k,M}^{~~~N} = \dim(M) \dim(N) \sum_{q\in\I} S_{k,q}\, \theta_{q}
  \sum_{\gamma,\delta=1}^{\Z_{q\qb}} (\cbulki_{A;q\qb})^{}_{\delta\gamma}\,
  \bcoef q\gamma N\,\bcoef {\qb}\delta M . 
  \ee
The purpose of this section is to demonstrate that information about the 
partition functions on a torus with parallel defect lines is encoded in an 
analogous manner in the \dtc s, and hence in the \rep\ theory of the classifying
algebra \CD\ of defect lines.

Consider the partition function $Z\TXY$ of a torus $\torus$ with two circular 
defect lines labeled by an $A$-$B$-defect $X$ and a $B$-$A$-defect $Y$ and 
running parallel to a non-contractible cycle that represents a basis element 
of the first homology of $\torus$. Such partition functions were introduced in 
\cite{pezu5}, where they are termed `generalized twisted partition functions'.
In the framework of the TFT construction the partition functions $Z\TXY$
were already studied in \cite[Sect.\,5.10]{fuRs4}; in that framework
$Z\TXY$ is described as the invariant of the ribbon graph
  \eqpic{manT2d_op:78}{450}{91}{
  \put(115,86){$\dsty \M\TXY ~=$}
  \put(185,-5)   {
  \put(0,-2)   {\Includepicclal{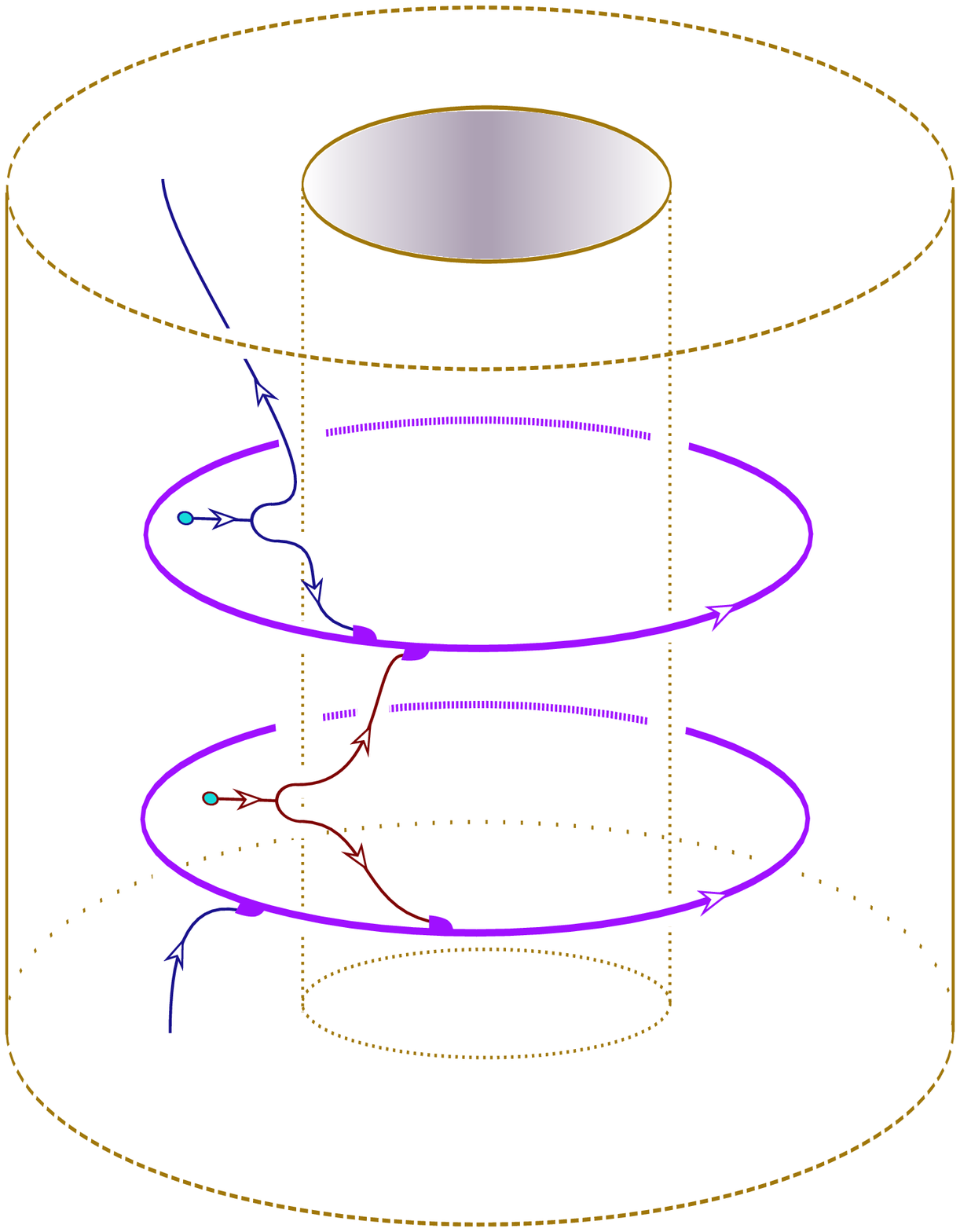}}
  \put(133.1,63){\pX X}
  \put(134,111) {\pX Y}
  \put(28.5,32) {\pB B}
  \put(54.5,69) {\pA A}
  } }
in the connecting three-manifold for the torus, i.e.\ in $(S^1\Times S^1)\Times
[-1,1]$, which is drawn here as a horizontal annulus $S^1\Times [-1,1]$ times a
vertically running circle (thus top and bottom of the picture are to be 
identified).

\medskip

Our aim is to express $Z\TXY$ in terms of the \dtc s for the two defects $X$ and
$Y$. To this end we perform a double bulk factorization of \erf{manT2d_op:78}, 
with the two cutting circles running parallel to the defects lines (and thus 
lying in horizontal planes in the picture \erf{manT2d_op:78}) in such a way that
each of the two full tori (with corners) that result from the cutting contains 
one of the defect lines. This way we obtain a description of $Z\TXY$ as a sum
  \be
  \bearl\dsty
  Z\TXY = \sum_{\qe,\qz,\qd,\qf\in\I}
  \dim(U_{\qe}) \dim(U_{\qz}) \dim(U_{\qd}) \dim(U_{\qf})
  \nxl{-2}\dsty \hsp{12}
  \sum_{\beta_1,\beta_2,\beta_3,\beta_4}
  {(\cbulki_{A;\qe,\qz})}_{\beta_2\beta_1}\,
  {(\cbulki_{B;\qd,\qf})}_{\beta_4\beta_3}\,
  Z(\Tfact XY{\qe\qz}{\qd\qf}{\beta_1\beta_2}{\beta_3\beta_4}) \, 1 \,,
  \eear
  \ee
where in each summand $\Tfact XY{\qe\qz}{\qd\qf}{\beta_1\beta_2}{\beta_3\beta_4}$
is a three-manifold with embedded ribbon graph that can be described as follows.
$\Tfact XY{\qe\qz}{\qd\qf}{\beta_1\beta_2}{\beta_3\beta_4}$ is obtained from 
four pieces $\M_{\torus}^X$, $\M_{\torus}^Y$, $\T_{\torus}^A$ and $\T_{\torus}
^B$ (which are three-manifolds with corners) by pairwise identification of 
sticky annuli that are located on their boundaries, in much the same way as the 
three-manifold \erf{cob_4p_BulkF:48} can be obtained by pairwise identification 
of sticky annuli on the pieces \erf{ball_n:49,ball_s:50}, \erf{cyl:51} and 
\erf{glue_tor_A:52,glue_tor_B:53}. Indeed, two of the four pieces -- 
$\T_{\torus}^A$ and $\T_{\torus}^B$ -- are precisely given by full tori with 
corners of the form displayed in \erf{glue_tor_A:52,glue_tor_B:53}, while the 
other two pieces, which are factorization tori (and thus full tori with 
corners as well), look as
  \Eqpic{manT2d_cutX_op:79,manT2d_cutY_op:80}{420}{71}{ \put(0,-4){
  \put(-27,75){$\dsty \M_{\torus}^X\;=$}
  \put(21,0)   {
  \put(0,0)   {\Includepicclal{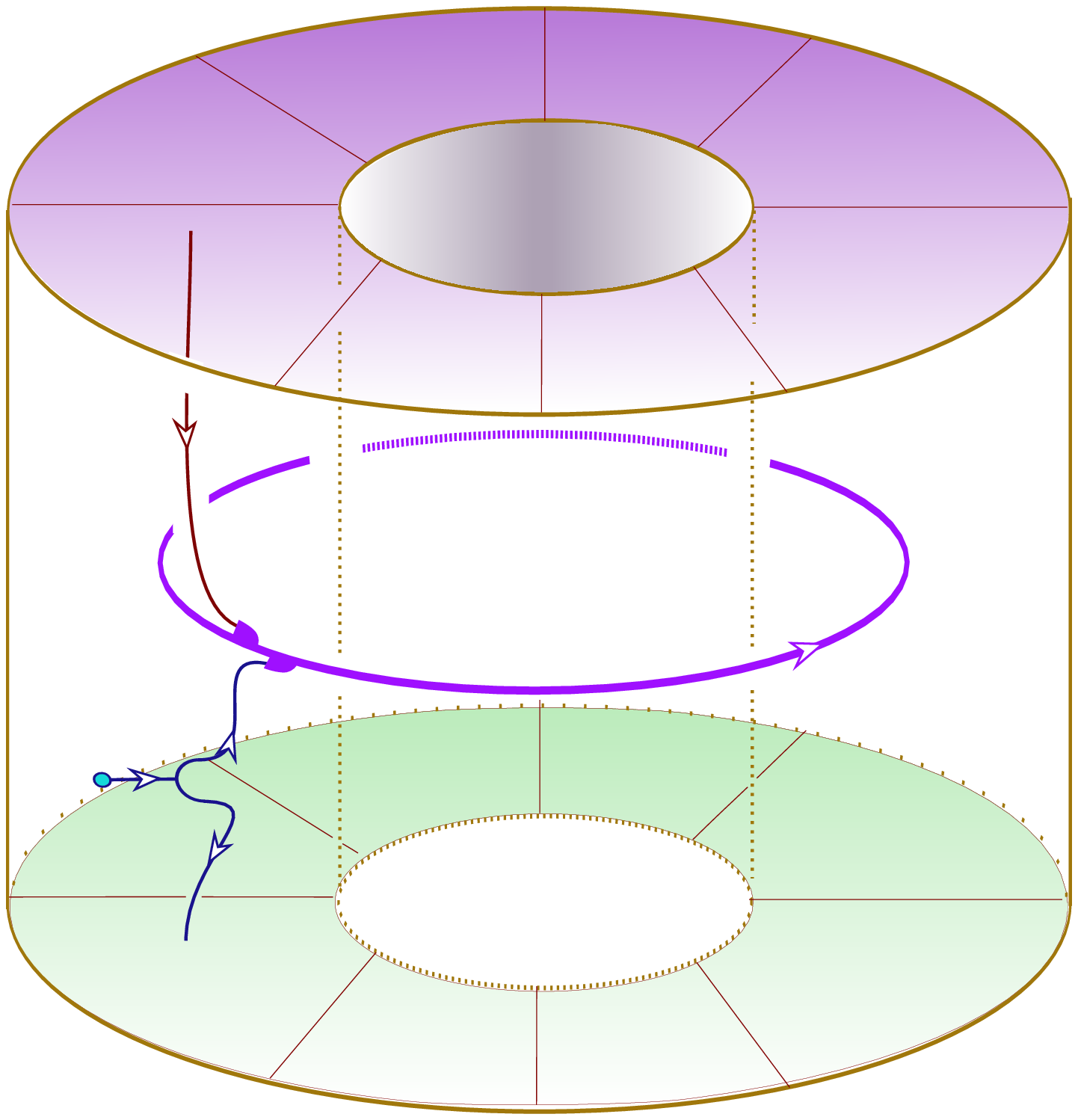}}
  \put(134,78)  {\pX X}
  \put(33,36)   {\pB B}
  \put(18,92)   {\pA A}
  \put(138,138) {\Includepicclal{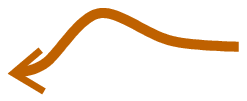}}
  \put(165,141) {$\fbY SA2$}
  \put(138,17)  {\Includepicclal{lsqarrov}}
  \put(165,20)  {$\fbY SB1$}
  }
  \put(229,75){$\dsty \M_{\torus}^Y\;=$}
  \put(278,0)   {
  \put(0,0)   {\Includepicclal{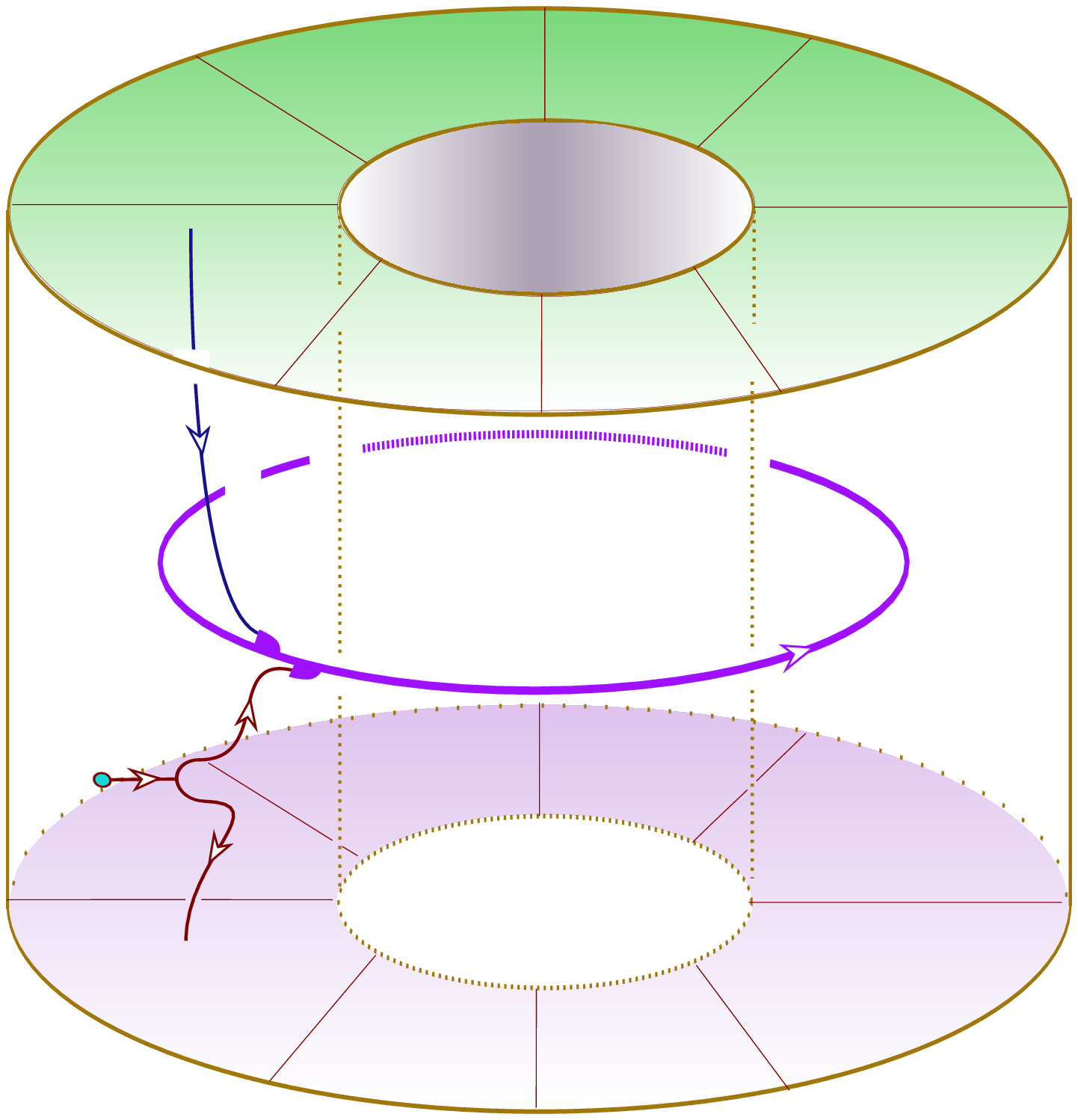}}
  \put(134,80)  {\pX Y}
  \put(33.4,37) {\pA A}
  \put(19,101)  {\pB B}
  \put(138,138) {\Includepicclal{lsqarrov}}
  \put(165,141) {$\fbY SB2$}
  \put(138,17)  {\Includepicclal{lsqarrov}}
  \put(165,20)  {$\fbY SA1$}
  } } }
(Here and below, the sticky annuli are denoted by $\mathrm Y_{\!S;C}^\ell$
with $\ell\iN\{1,2\}$ and $C\iN\{A,B\}$, and are accentuated by their shading,
analogously as we already did in pictures like \erf{ball_n:49,ball_s:50}.
The corresponding sticky annuli of the other two pieces $\T_{\torus}^A$ and 
$\T_{\torus}^B$ will be denoted by $\mathrm Y_{\!\T;C}^\ell$, as in
\erf{glue_tor_A:52,glue_tor_B:53}.)

For dealing with the manifolds $\M_{\torus}^X$ and $\M_{\torus}^Y$ more 
easily when performing the various identifications, we
redraw them in a slightly deformed manner, such that they look as follows:
  \Eqpic{manT2d_cutX_def_op:81,manT2d_cutY_def_op:82}{420}{75}{ \put(0,-13){
  \put(-28,75){$\dsty \M_{\torus}^X ~=$}
  \put(22,0)   {
  \put(0,0)   {\Includepicclal{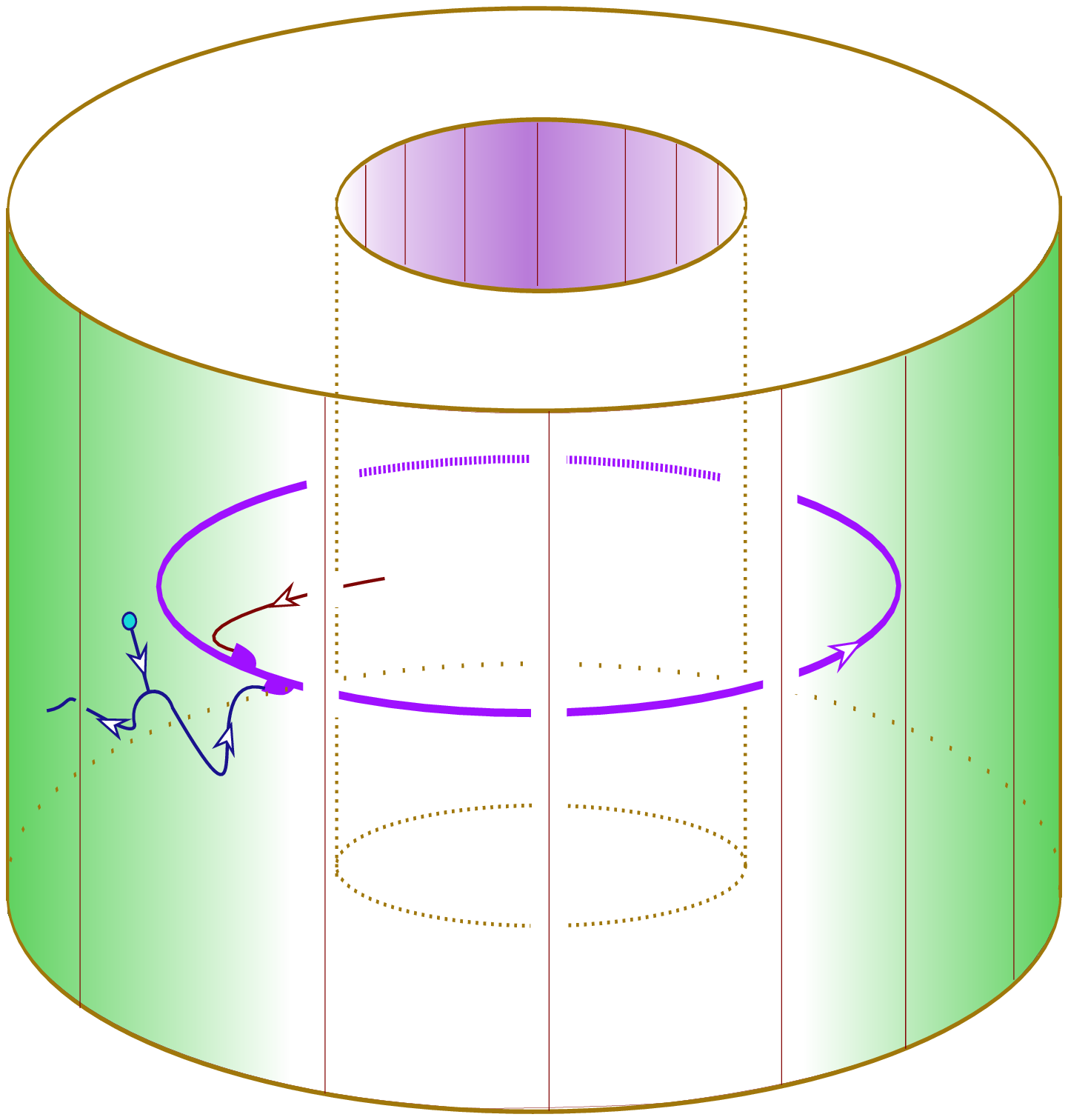}}
  \put(122,92)  {\pX X}
  \put(20,49)   {\pB B}
  \put(31,77)   {\pA A}
  \put(101,134) {\begin{rotate}{45}\Includepicclal{lsqarrov}\end{rotate}}
  \put(116,157) {\begin{rotate}{45}$\fbY SA2$\end{rotate}}
  \put(138,28)  {\Includepicclal{lsqarrov}}
  \put(165,31)  {$\fbY SB1$}
  }
  \put(227,75){$\dsty \M_{\torus}^Y ~=$}
  \put(277,0)   {
  \put(0,0)   {\Includepicclal{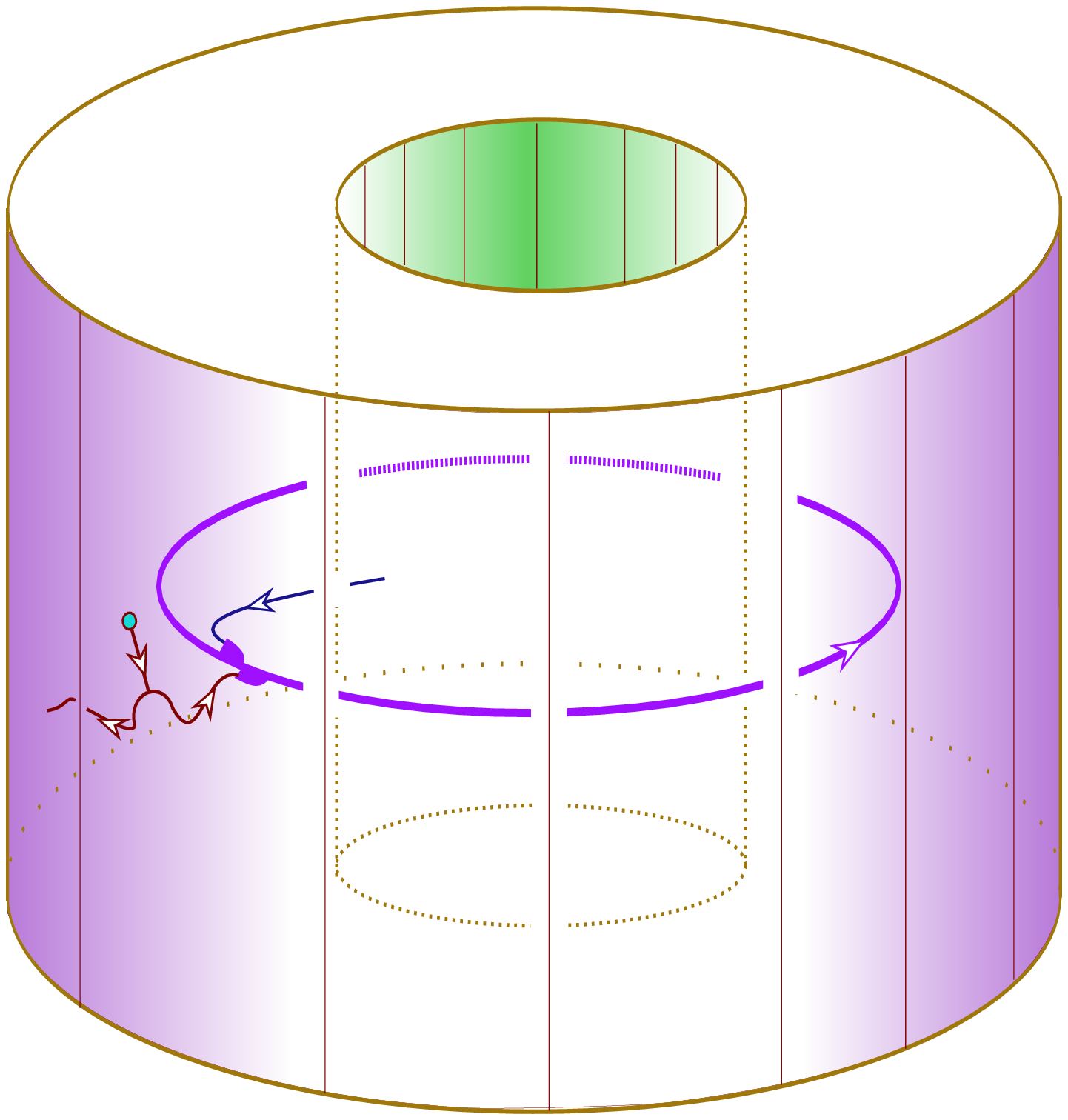}}
  \put(122,92)  {\pX Y}
  \put(19,53)   {\pA A}
  \put(29,76.6) {\pB B}
  \put(101,134) {\begin{rotate}{45}\Includepicclal{lsqarrov}\end{rotate}}
  \put(116,157) {\begin{rotate}{45}$\fbY SB2$\end{rotate}}
  \put(138,28)  {\Includepicclal{lsqarrov}}
  \put(165,31)  {$\fbY SA1$}
  } } }
Gluing $\M_{\torus}^X$ to $\T_{\torus}^A$, respectively $\M_{\torus}^Y$
to $\T_{\torus}^B$, yields the two manifolds
  \Eqpic{ST_A_glue1:83,ST_B_glue1_op:84}{420}{126}{ \put(0,-5){
  \put(-27,259) {$ \STg{\qe}{\qz}{\beta_1}{\beta_2}AX \;:= $}
  \put(0,-3){ \setulen84
  \put(24,-2) {\includepicclax3{19}{83} }
  \put(124,-2){
  \put(-8,26)   {\pg {\qz}}
  \put(12,215)  {\pg {\qzb}}
  \put(8.4,171) {\pg {\qeb}}
  \put(33,130)  {\pg {\qe}}
  \put(-18,74.6){\pl {\phi_{\!\beta_1}^{}}}
  \put(-10,179.6){\pl {\phi_{\!\beta_2}^{}}}
  \put(-19,84.5){\pA A}
  \put(-47.4,65.6){\pB B}
  \put(-6,101)  {\pX X}
  \put(-27,205) {\pA A}
  \put(-108,216){\Includepicclal{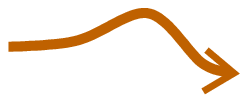}}
  \put(-138,221){$\fbY \T A1$}
  \put(-108,110){\Includepicclal{rsqarrov}}
  \put(-138,115){$\fbY SB1$}
  } }
  \put(180,259)   {$ \STg{\qd}{\qf}{\beta_3}{\beta_4}BY \;:= $}
  \put(224,-3)  {\Includepicclal{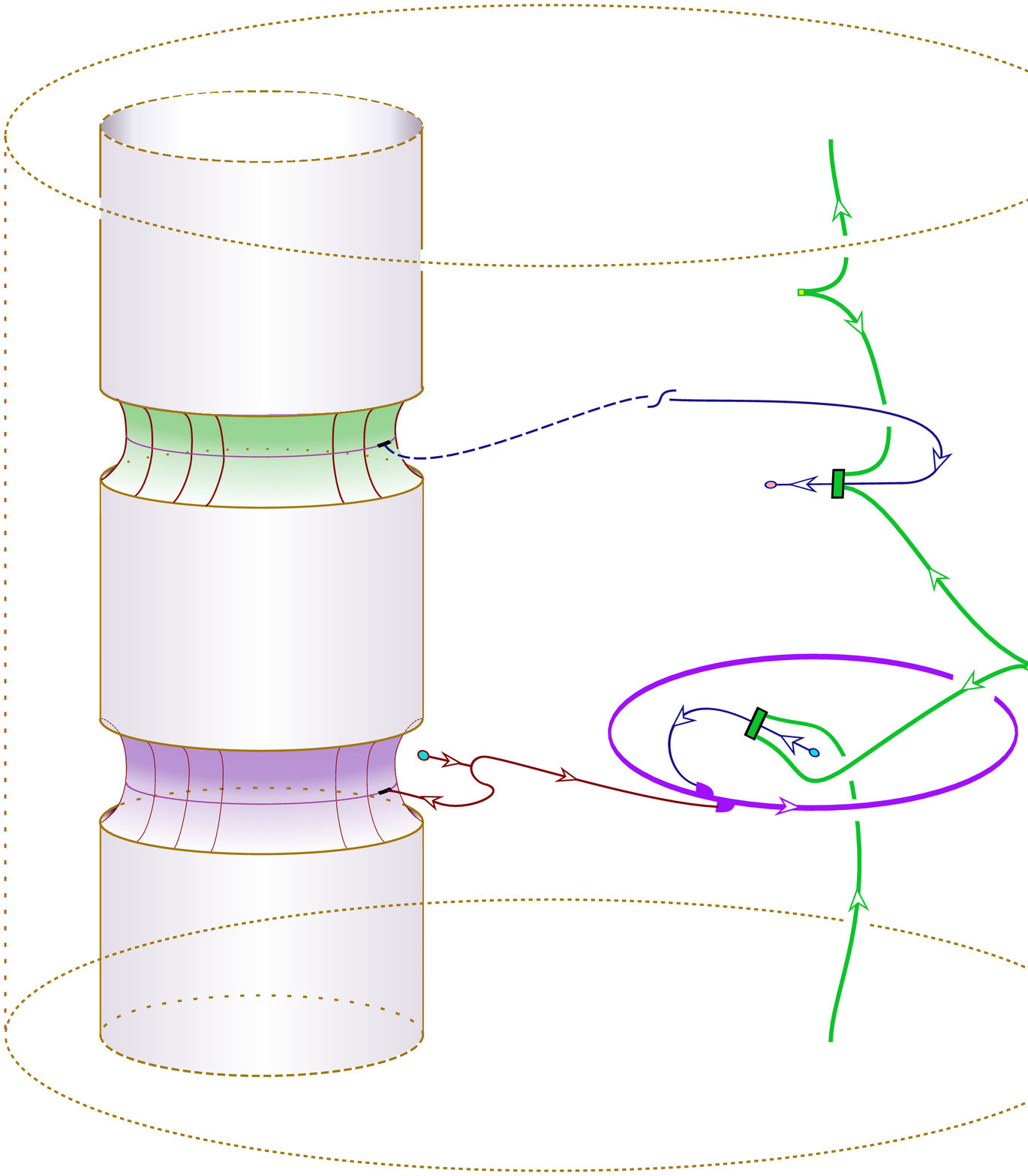}}
  \put(224,-14) {
  \put(186.7,42){\pg {\qf}}
  \put(195,194) {\pg {\qfb}}
  \put(214,138) {\pg {\qdb}}
  \put(203,106) {\pg {\qd}}
  \put(165,117) {\pl {\phi_{\beta_3}}}
  \put(178,156) {\pl {\phi_{\!\beta_4}^{}}}
  \put(149.7,103.9) {\pB B}
  \put(172,166.7) {\pB B}
  \put(159,185) {\pB B}
  \put(1,174)   {\Includepicclal{rsqarrov}}
  \put(-22,179) {$\fbY \T B1$}
  \put(1,98)    {\Includepicclal{rsqarrov}}
  \put(-22,103) {$\fbY SA1$}
  \put(114,104) {\pA A}
  \put(226.4,108) {\pX Y}
  } } }
(the second of these is drawn in the same way as we did in 
\erf{glue_tor_A:52,glue_tor_B:53}, i.e.\ with the full torus turned inside out,
while the first shows the full torus directly). The remaining two pairwise 
identifications of sticky annuli are then immediate, leading to
  \eqpic{M_tor:85}{370}{124}{
  \put(0,128)   {$\Tfact XY{\qe\qz}{\qd\qf}{\beta_1\beta_2}{\beta_3\beta_4}~=$}
  \put(104,-5)  {\Includepicclal{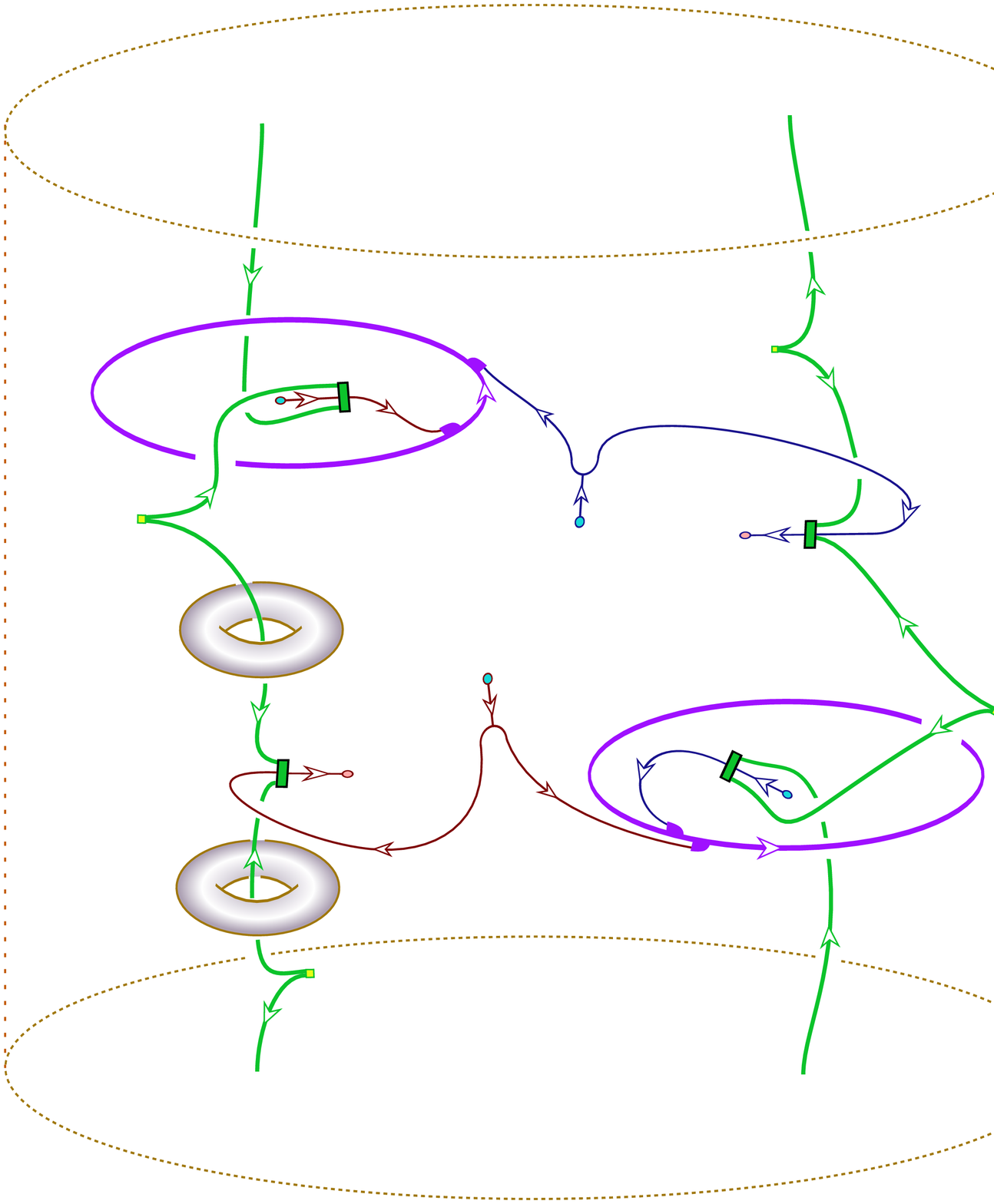}}
  \put(104,-16) {
  \put(186,43)  {\pg {\qf}}
  \put(194,194) {\pg {\qfb}}
  \put(203,106) {\pg {\qd}}
  \put(213,140) {\pg {\qdb}}
  \put(165.8,117.5) {\pl {\phi_{\!\beta_3}^{}}}
  \put(179,156) {\pl {\phi_{\!\beta_4}^{}}}
  \put(95,191)  {\pA A}
  \put(148,104) {\pB B}
  \put(171,167) {\pB B}
  \put(157,191) {\pB B}
  \put(119,115) {\pA A}
  \put(207,88)  {\pX Y}
  \put(60.2,43) {\pg {\qz}}
  \put(59,219)  {\pg {\qzb}}
  \put(50,171)  {\pg {\qe}}
  \put(42,155)  {\pg {\qeb}}
  \put(62.8,115.5){\pl {\phi_{\!\beta_2}^{}}}
  \put(74,203.2){\pl {\phi_{\!\beta_1}^{}}}
  \put(89,175)  {\pX X}
  } }
which is a ribbon graph in \SzSe\ with two full tori cut out.

\medskip

We now extract the coefficients of the partition function $Z\TXY$ with respect 
to a basis of the space of conformal blocks for the boundary of $\M\TXY$. This 
boundary is the complex double $\torus\,{\sqcup}\,{-}\torus$ of the torus. 
Every way to represent $\torus$ as the boundary of a full torus gives rise to 
a basis of the space of conformal blocks on $\torus$. The basis element 
$|\chii_i;\torus\rangle$ is given as the invariant of a ribbon graph 
consisting of an annular ribbon that runs along the core of the torus and is 
labeled by the simple object $U_i$ (see e.g.\ \cite[Sect.\,5.2]{fuRs4}). To 
compute the coefficients in the corresponding basis, we compose the cobordism
\erf{M_tor:85} with the cobordisms for the elements of the dual basis.

The interpretation as a partition function requires that there is a basis in 
which the coefficients are non-negative integers. The coefficients $Z\TXYij$ 
in this basis are obtained by gluing the solid tori to \erf{M_tor:85}
precisely in the way suggested by the pictorial description of the boundary
$\partial \Tfact XY{\qe\qz}{\qd\qf}{\beta_1\beta_2}{\beta_3\beta_4} \Cong
\torus\,{\sqcup}\,{-}\torus$ in \erf{M_tor:85}, in particular without applying
any large diffeomorphism. That this is the correct gluing is e.g.\ seen by
noticing that in order for $Z\TXY$ to have a proper interpretation as a
partition function, the `time' direction of $\torus$ must be taken along the
defect lines, and keeping track of this prescription for the time direction
while performing the manipulations that lead from \erf{manT2d_op:78} to 
\erf{M_tor:85}.\,%
 \footnote{~This is in agreement with the result (5.151) of \cite{fuRs4}
 for the coefficients of $Z\TXY\!$. Note, however, that in the conventions of
 \cite{fuRs4} the two defect lines have opposite orientation, while here we
 have chosen the same orientation for both of them.}
This way we find
  \be
  \bearl\dsty
  Z\TXYij = \sum_{\qe,\qz,\qd,\qf\in\I}
  \dim(U_{\qe}) \dim(U_{\qz}) \dim(U_{\qd}) \dim(U_{\qf})
  \nxl{-3}\dsty \hsp{12}
  \sum_{\beta_1,\beta_2,\beta_3,\beta_4}
  {(\cbulki_{A;\qe,\qz})}_{\beta_2\beta_1}\,
  {(\cbulki_{B;\qd,\qf})}_{\beta_4\beta_3}\,
  Z(\Tfact X{Y;ij}{\qe\qz}{\qd\qf}{\beta_1\beta_2}{\beta_3\beta_4}) \, 1 
  \eear
  \ee
with
  \eqpic{M_tor_proj_op:88}{420}{143}{  \setlength\unitlength{1.2pt}
  \put(-10,121)   {$ \Tfact X{Y;ij}{\qe\qz}{\qd\qf}
                   {\beta_1\beta_2}{\beta_3\beta_4} ~:= $}
  \put(89,-4){ {\includepicclax4{56}{88}}
  \put(0,-9) {
  \put(187,43)   {\pg {\qf}}
  \put(194,197)  {\pg {\qfb}}
  \put(203,106)  {\pg {\qd}}
  \put(213,141)  {\pg {\qdb}}
  \put(165,117.4){\pl {\phi_{\bet_3}^{}}}
  \put(178.1,157){\pl {\phi_{\bet_4}^{}}}
  \put(95,184)   {\pA A}
  \put(119,115)  {\pA A}
  \put(150,104)  {\pB B}
  \put(172,167)  {\pB B}
  \put(163,190)  {\pB B}
  \put(207,88)   {\pX Y}
  \put(62,44)    {\pg {\qz}}
  \put(52,180)   {\pg {\qe}}
  \put(44,153)   {\pg {\qeb}}
  \put(62,115.9) {\pl {\phi_{\bet_2}^{}}}
  \put(76,195)   {\pl {\phi_{\bet_1}^{}}}
  \put(91,170)   {\pX X}
  \put(69,133)   {\pg i}
  \put(69,75)    {\pg j}
  } } }
Again the invariant of this ribbon graph can be evaluated with the help of
dominance, which we apply to both $\id_{U_\qe} \oti \id_{U_\qdb}$ and
$\id_{U_\qz} \oti \id_{U_\qfb}$. In both cases only the tensor unit can give a 
non-zero contribution to the invariant $Z(\Tfact X{Y;ij}{\qe\qz}{\qd\qf}
{\beta_1\beta_2}{\beta_3\beta_4})$, so that we need $\qd\eq\qe$ and $\qf\eq\qz$,
and in this case the invariant is equal to the one of the ribbon graph
  \eqpic{M_tor_dom_op:97}{160}{113}{ \setulen 70
  \put(0,-8)   {\includepicclax2{66}{97}}
  \put(0,-19) {
  \put(215,120)   {\pg {\qe}}
  \put(67,186)    {\pg {\qe}}
  \put(95,220)    {\pg {\qe}}
  \put(59,137)    {\pg {\qeb}}
  \put(196,163)   {\pg {\qeb}}
  \put(163,204)   {\pg {\qeb}}
  \put(152,69)    {\pg {\qz}}
  \put(67.5,290)  {\pg {\qz}}
  \put(94,65)     {\pg {\qzb}}
  \put(220,252)   {\pg {\qzb}}
  \put(167,117)   {\pl {\phi_{\bet_3}^{}}}
  \put(194,243)   {\pl {\phi_{\bet_4}^{}}}
  \put(170,129)   {\pB B}
  \put(170,250)   {\pB B}
  \put(127,117)   {\pA A}
  \put(223,102)   {\pX Y}
  \put(72,127)    {\pl {\phi_{\bet_2}^{}}}
  \put(95.8,270.5){\pl {\phi_{\bet_1}^{}}}
  \put(108,240)   {\pX X}
  \put(79,146)    {\pg i}
  \put(80,88)     {\pg j}
  } }
in \SzSe. This invariant, in turn, reduces to the product of the two morphisms 
in $\End(\one)\,{\cong}\,\complex$ that are given by the two components of the 
ribbon graph. Each of these two morphisms can be simplified with the help of 
elementary fusing and braiding moves, whereby they are seen to be essentially
given by \dtc s. We suppress the details; the final result is
  \be
  Z\TXYij = \frac{\dim(X)\dim(Y)}{S_{0,0}^2}
  \sum_{\qe,\qz\in\I} S_{i,\qe}^{}\, S_{j,\qz}^{*} \!
  \sum_{\beta_1,\beta_2,\beta_3,\beta_4} \!
  {(\cbulki_{A;\qe,\qz})}_{\beta_2\beta_1}\,
  {(\cbulki_{B;\qe,\qz})}_{\beta_4\beta_3}\,
  \dcoef \qe\qz{\beta_1}{\beta_4}X\, \dcoef\qe\qz{\beta_3}{\beta_2}Y \,.
  \labl{ZTXYij}

{}From this result we can deduce various properties of the numbers $Z\TXYij$.
We mention two of them. First, for $X \eq Y \eq A \eq B$ we recover the ordinary
torus partition function $Z_\torus(A)$, in agreement with theorem 5.23(ii) of 
\cite{fuRs4}: The identity
  \be
  \dcoef pq\alpha\beta A
  = \frac{S_{0,0}}{\dim(A)}\, {(\cbulk_{A;p,q})}_{\alpha,\beta}
  \ee
together with $\dimc\HomAA(U_\qeb\otip A\otim U_\qzb,A) \eq \Z(A)_{\qeb\qzb}$
implies that
  \be
  Z_{\torus\,,ij}^{A|A} = \sum_{\qe,\qz\in\I} S_{i,\qe}^{}\, S_{j,\qz}^{*}
  \sum_{\beta_2=1}^{\Z(A)_{\qeb\qzb}} 1
  = \sum_{\qe,\qz\in\I} S_{i,\qeb}^{-1}\, \Z(A)_{\qeb\qzb}\, S_{\qzb,j}^{}
  = \Z(A)_{ij} \,,
  \ee
where the last equality holds by modular invariance of $Z_\torus(A)$. (More 
generally, when $X$ and $Y$ are arbitrary simple topological defects, but still
$A$ equals $B$, \erf{ZTXYij} is equivalent to formula (1.4) of \cite{petk4}.)

Second, we observe that the \dtc s for defect lines with opposite orientation
are related by
  \be
  \dcoef \ia\ja\alpha\beta{X^\vee_{}}
  = \R \ia\ib0\nl\nl\, \Rm \ja\jb0\nl\nl\, \dcoef \ib\jb\beta\alpha X .
  \ee
When combined with the identity
  \be
  \R \ia\ib0\nl\nl\, \Rm \ja\jb0\nl\nl\, {(\cbulki_{A;\ia,\ja})}_{\alpha,\beta}
  = {(\cbulki_{A;\ib,\jb})}_{\beta,\alpha}
  \ee
for the bulk field two-point functions on the sphere, it follows that
  \be
  Z_{\torus\,,\ib\jb}^{X^\vee_{}|Y^\vee_{}} = Z_{\torus\,,\ia\ja}^{Y|X} .
  \ee
This reproduces the result of theorem 5.23(iii) of \cite{fuRs4}.

\newpage
\appendix

\section{The final gluing in the double bulk factorization}\label{app.glue4}

In this appendix we present the derivation of the result \erf{res4glue} and
\erf{fact_MF_S2S1:62} for the gluing of the two three-manifolds (with corners)
\erf{glue_tor_A_ball:55,glue_tor_A_ball_b.eps:54} and
\erf{glue_tor_B_ball_line_a:56,glue_tor_B_ball_line_b:57}, which is the
last of the four pairwise identifications of sticky annuli that result from
the double bulk factorization. According to the discussion leading to the 
expressions \erf{glue_tor_A_ball:55,glue_tor_A_ball_b.eps:54} and
\erf{glue_tor_B_ball_line_a:56,glue_tor_B_ball_line_b:57} for the 
three-manifolds $\M^\Alpha_{} \,{\equiv}\, \MN^{\Alpha;\,\ia_3\ja_3,\ia_4\ja_4}
_{\!\qe\qz\beta_1\beta_2}$ and $\MN^{\Beta;X}_{} \,{\equiv}\,
\MN^{\Beta;X,\ia_1\ja_1,\ia_2\ja_2}_{\!\qd\qf\beta_3\beta_4}$, the correlator
$\CX \,{\equiv}\, C(\PHi{\alphz},\PHi{\alphv};X;\PHi{\alphe},\PHi{\alphd})$
can be written as
  \be
  \bearl\dsty
  \CX = \sum_{\qe,\qz,\qd,\qf\in\I}\sum_{\beta_1,\beta_2,\beta_3,\beta_4}\!
  \dim(U_{\qe}) \dim(U_{\qz}) \dim(U_{\qd}) \dim(U_{\qf})\,
  \nxl{-3} \hsp{16}
  (\cbulki_{A;\qe,\qz})_{\beta_2\beta_1}^{}\,
  (\cbulki_{B;\qd,\qf})_{\beta_4\beta_3}^{}\,
  Z(\M_{\qe\qz\qd\qf}^{\beta_1\beta_2\beta_3\beta_4})\, 1 \,,
  \eear
  \ee
where $\M_{\qe\qz\qd\qf}^{\beta_1\beta_2\beta_3\beta_4}$ is the three-manifold 
obtained by identifying the two sticky annuli $\rmY_{\!S,A}^2$ and $\rmY
_{\!\T,A}^2$ which are subsets of the boundary of $\M^\Alpha_{}$ and of 
$\MN^{\Beta;X}_{}$, respectively, and where $\cbulk_{A;uv}$ is the matrix of 
coefficients of the two-point function of bulk fields (in phase $A$) on the 
sphere in a standard basis of blocks. (For more details about the bulk field 
two-point function see appendix A.6 of \cite{fuSs}. According to theorem 2.13 
of \cite{fjfrs}, each bulk factorization gives rise to a factor of 
$\cbulki_{A;uv}\!$.)

To properly display the manifold $\M_{\qe\qz\qd\qf}^{\beta_1\beta_2\beta_3
\beta_4}$ including its embedded ribbon graph requires some care. We proceed as 
follows. First we give an alternative description of the three-manifolds 
$\M^\Alpha_{}$ and $\MN^{\Beta;X}_{}$, and thereby of 
$\M_{\qe\qz\qd\qf}^{\beta_1\beta_2\beta_3\beta_4}$, by cutting each of them 
along a two-sphere, such that they are written as the compositions
  \be
  \MN^\Alpha_{} = \widehat\MN^\Alpha_{} \circ \mathcal S^\Alpha_{}
  \qquad{\rm and}\qquad
  \MN^{\Beta;X}_{} = \widehat\MN^{\Beta;X}_{}  \circ \mathcal S^\Beta_{} \,.
  \labl{MMSMMS}
Here each of $\widehat\MN^\Alpha_{} \,{\equiv}\,
\widehat\MN^{\Alpha;\, \ia_3\ja_3,\ia_4\ja_4}_{\!\qe\qz\beta_1\beta_2}$ and
$\widehat\MN^{\Beta;X}_{} \,{\equiv}\,
\widehat\MN^{\Beta; X;\ia_1\ja_1,\ia_2\ja_2}_{\!\qd\qf\beta_3\beta_4}$ is
topologically an \SzSe\ with a three-ball cut out, while each of 
$\mathcal S^\Alpha_{}$ and $\mathcal S^\Beta_{}$ is a three-manifold with
corners, namely a spherical shell \SzI\ that has a sticky annulus (namely 
$\rmY_{\!S,A}^2$ and $\rmY_{\!\T,A}^2$, respectively) as part of its `inner' 
boundary component. (Even though $\mathcal S^\Alpha_{}$ and 
$\mathcal S^\Beta_{}$ are thus not cobordisms themselves, the compositions in 
\erf{MMSMMS} are in the sense of cobordisms, since the relevant identifications 
do not involve the sticky parts $\rmY_{\!S,A}^2$ and $\rmY_{\!\T,A}^2$ of the 
boundaries of $\mathcal S^\Alpha_{}$ and $\mathcal S^\Beta_{}$.)
Including their ribbon graphs, $\widehat\MN^\Alpha_{}$ and
$\widehat\MN^{\Beta;X}_{}$ are given by
  \Eqpic{glue_tor_A_cut:58,glue_tor_B_cut:59}{440}{102}{ \setulen 75
  \put(-23,144)   {$\widehat\MN^\Alpha_{} ~= $}
  \put(38,0){
  \put(0,0)   {\includepicclax2{85}{58}}
  \put(188,35)    {\pg {\qz}}
  \put(195,196)   {\pg {\qzb}}
  \put(207,108)   {\pg {\qe}}
  \put(218,139)   {\pg {\qeb}}
  \put(157,103.1) {\pl {\phi_{\bet_1}^{}}}
  \put(177,155.5) {\pl {\phi_{\bet_2}^{}}}
  \put(145,120)   {\pA A}
  \put(160,187)   {\pA A}
  \put(10,216)    {\pg{\ja_3}}
  \put(72.7,222)  {\pg{\ja_4}}
  \put(29.1,165)  {\pg{\ia_3}}
  \put(70.5,171)  {\pg{\ia_4}}
  \put(29.5,194.6){\pl{\phi_{\alphz}}}
  \put(80.2,204)  {\pl{\phi_{\alphv}}}
  }
  \put(318,144)   {$ \widehat\MN^{\Beta;X}_{} ~= $}
  \put(393,0){
  \put(0,0)   {\includepicclax2{85}{59}}
  \put(187,35)    {\pg {\qf}}
  \put(195,194)   {\pg {\qfb}}
  \put(204,106)   {\pg {\qd}}
  \put(215,139)   {\pg {\qdb}}
  \put(157.7,103.4) {\pl {\phi_{\bet_3}^{}}}
  \put(178,154.5) {\pl {\phi_{\bet_4}^{}}}
  \put(145,107.5) {\pB B}
  \put(159,187)   {\pB B}
  \put(10,216)    {\pg{\ja_1}}
  \put(73.2,222)  {\pg{\ja_2}}
  \put(28.7,166.5){\pg{\ia_1}}
  \put(70.5,171)  {\pg{\ia_2}}
  \put(29.5,194.6){\pl{\phi_{\alphe}}}
  \put(80.2,204)  {\pl{\phi_{\alphd}}}
  \put(143,125)   {\pX X}
  } }

The resulting description of the three-manifold
$\M_{\qe\qz\qd\qf}^{\beta_1\beta_2\beta_3\beta_4}$ is as the composition
  \be
  \M_{\qe\qz\qd\qf}^{\beta_1\beta_2\beta_3\beta_4}
  = \mathcal S \circ (\widehat\MN^\Alpha_{} \sqcup \widehat\MN^{\Beta;X}_{}) 
  \labl{McircMcupM}
of cobordisms, where $\mathcal S$ is the manifold (without corners) that is 
obtained by gluing the spherical shells $\mathcal S^\Alpha_{}$ and 
$\mathcal S^\Beta_{}$ along the two sticky annuli on their respective inner 
boundary components. It is easy to see that as a result of this identification 
$\mathcal S$ is topologically a three-ball with three three-balls cut out. But 
to be able to visualize this gluing, we need to change the embedding of one 
of the two spherical shells into $\reals^3$ (or rather, $S^3$) by
``turning it inside out'', such that the sticky annulus is now located on the 
outer rather than the inner boundary component. After this manipulation the 
identification of the sticky annuli on $\mathcal S^\Alpha_{}$ and 
$\mathcal S^\Beta_{}$ is straightforward, yielding
  \eqpic{cut_ball:60}{250}{88}{ \setlength\unitlength{1.2pt}
  \put(0,80)   {$ \mathcal S ~= $}
  \put(50,0){ {\includepicclax4{56}{60}}
  \put(60.1,98.8) {\pg{\ia_1}}
  \put(91.8,98.8) {\pg{\ia_2}}
  \put(38,142)    {\pg{\ia_3}}
  \put(102,150)   {\pg{\ia_4}}
  \put(58.4,63)   {\pg{\ja_1}}
  \put(91.1,63)   {\pg{\ja_2}}
  \put(45.6,12.2) {\pg{\ja_3}}
  \put(103.3,9.5) {\pg{\ja_4}}
  \put(105,86)    {\pA A}
  } }

According to the composition in \erf{McircMcupM}, the boundary spheres of the 
three cobordisms $\mathcal S$, $\widehat\MN^\Alpha_{}$ and 
$\widehat\MN^{\Beta;X}_{}$ are to be identified, in the obvious manner that is 
already apparent from the labeling of the arcs on their respective boundaries. 
Again one the these two gluings, say of $\mathcal S$ to $\widehat\MN^\Alpha_{}$,
is easy. To achieve the other gluing in a convenient way, we first perform 
surgery on $\widehat\MN^{\Beta;X}_{}$ along a torus which is a tubular 
neighbourhood of a surgery link. This yields a sum over ribbon graphs in $S^3$ 
with one three-ball cut out, where the summation is over the different labelings
by elements of $\I$ of the resulting surgery link. Furthermore, $S^3$ with a 
three-ball cut out is topologically again a three-ball, and we may present it 
as the interior, rather than the exterior, of a two-sphere in $S^3$. This way 
we arrive at
  \eqpic{glue_tor_B_S3_inv:61}{420}{142}{ 
  \put(-8,136)   {$ \widehat\MN^{\Beta;X}_{} ~=~ \dsty\sum_{{\color{DarkGreen}n}
                  \in\I} S_{0,{\color{DarkGreen}n}} $}
  \put(115,0){
  \put(-15,-10)   {\includepicclax38{61}}
  \put(98.7,159.7){\pl{\phi_{\alphe}}}
  \put(148.8,170.3) {\pl{\phi_{\alphd}}}
  \put(96,286)    {\pg{\ia_1}}
  \put(95,1.5)    {\pg{\ja_1}}
  \put(135,287)   {\pg{\ia_2}}
  \put(152.3,-2)  {\pg{\ja_2}}
  \put(65,119.5)  {\pg{{\color{DarkGreen}n}}}
  \put(263,100)   {\pA A}
  \put(172,168.2) {\pB B} 
  \put(177.4,73)  {\pl{\phi_{\bet_3}^{}}}
  \put(197,139)   {\pl{\phi_{\bet_4}^{}}}
  \put(212,83)    {\pg{\qd}}
  \put(160,34.4)  {\pg{\qf}}
  \put(223,55)    {\pX X}
  } }
Here the surgery link is the annular ribbon labeled by $n\iN\I$.

\medskip

With the so obtained description of $\widehat\MN^{\Beta;X}_{}$ the gluing 
with $\mathcal S$ is readily performed, yielding the resulted quoted in 
\erf{res4glue} and \erf{fact_MF_S2S1:62}.
\\~


\section{Projection of \erf{res4glue} to the vacuum channel}\label{app.dubuvacpr}

Here we explain how to obtain the vacuum channel component $\cXo \eq
c(\PHi{\alphz},\PHi{\alphv};X;\PHi{\alphe},\PHi{\alphd})_0$ of the correlator 
\CX\ as given by \erf{res4glue}, by composing the cobordism 
\erf{fact_MF_S2S1:62} with the basis element dual to \erf{basis_VC}. This 
composition yields a ribbon graph in \SzSe. After slightly deforming this 
graph so as to reduce the number of braidings, we use dominance for
$\End(U_{\qf}\oti U_{\qz}\oti U_{\qe}\oti U_{\qd})$, which results in an
additional summation over $u\iN\I$ and over three-point couplings 
$\gamma\iN\hom(U_{\qf}\oti U_{\qz},U_u)$ and 
$\delta\iN\hom(U_{\qe}\oti U_{\qd},U_{\bar u})$. Hereby we obtain 
  \be
  \bearl\dsty
  \cXo = \frac1{S_{0,0}^2} \sum_{\qe,\qz,\qd,\qf\in\I}\!
  \dim(U_{\qe}) \dim(U_{\qz}) \dim(U_{\qd}) \dim(U_{\qf})
  \nxl{-2}\dsty \hsp{7.2}
  \sum_{\beta_1,\beta_2,\beta_3,\beta_4}\!
  {(\cbulki_{A;\qe,\qz})}_{\beta_2\beta_1}\,
  {(\cbulki_{B;\qd,\qf})}_{\beta_4\beta_3}\,
  \sum_{n\in\I} S_{0,n}\, \sum_{u\in\I}\sum_{\gamma,\delta} Z(\widetilde\M
  _{n,u;\qe\qz\qd\qf}^{\gamma\delta,\beta_1\beta_2\beta_3\beta_4}) \,,
  \eear
  \labl{cXo-B}
where the ribbon graph in $\widetilde\M_{n,p;\qe\qz\qd\qf}^{\gamma\delta,
\beta_1\beta_2\beta_3\beta_4}$ consists of two connected components, of which 
the one containing the $X$-ribbon is `localized' in the $S^1$-direction:
  \eqpic{fact_MF_S2S1_proj_dom_def:64}{405}{198}{ \setulen 70
  \put(0,290)      {$\dsty \widetilde\M_{n,u;\qe\qz\qd\qf}
                     ^{\gamma\delta,\beta_1\beta_2\beta_3\beta_4} ~= $}
  \put(35,-29){
  \put(104,27)   {\includepicclax2{66}{64}}
  \put(202,385)   {\pl{\phi_{\alphz}}}
  \put(254,396)   {\pl{\phi_{\alphv}}}
  \put(334,401)   {\pl{\phi_{\bet_2}^{}}}
  \put(474.8,381) {\pl{\phi_{\bet_1}^{}}}
  \put(297,402)   {\pA A}
  \put(194,360)   {\pg{\ib_1}}
  \put(248,357)   {\pg{\ib_2}}
  \put(193,300)   {\pg{\ia_1}}
  \put(254,304)   {\pg{\ia_2}}
  \put(406,376)   {\pg{\qeb}}
  \put(429,93)    {\pg{\qz}}
  \put(211,249)   {\pl{\phi_{\alphe}}}
  \put(254,257.6) {\pl{\phi_{\alphd}}}
  \put(317,260.8) {\pl{\phi_{\bet_4}^{}}}
  \put(436.5,441) {\pl{\phi_{\bet_3}^{}}}
  \put(293,275)   {\pB B}
  \put(377,260)   {\pg{\qdb}}
  \put(335,127)   {\pg{\qf}}
  \put(332,465)   {\pg{\qzb}}
  \put(191,95)    {\pg{\ja_1}}
  \put(257,95)    {\pg{\ja_2}}
  \put(155,437)   {\pg{\jb_1}}
  \put(208,429)   {\pg{\jb_2}}
  \put(276,180)   {\pg n}
  \put(424,165)   {\pg u} 
  \put(437,177)   {\pg{\bar u}}
  \put(443,142)   {\pg \gamma} 
  \put(460,158)   {\pg \delta} 
  \put(476,444)   {\pX X} 
  \put(490,360)   {\pg{\qe}}
  \put(474,340)   {\pg{\qz}}
  \put(510,366)   {\pg{\qd}}
  \put(456,340)   {\pg{\qf}}
  \put(497,304)   {\pg u} 
  \put(515,314)   {\pg{\bar u}}
  \put(471,316)   {\pg{\bar\gamma}}
  \put(517.8,340) {\pg{\bar\delta}}
  } }

Next we use dominance once more, this time for the space $\End(U_{\ib_1}\oti 
U_{\ib_2}\oti U_{\qeb})$. In the resulting summation over intermediate simple 
objects only the contribution from the tensor unit gives a non-zero 
contribution, so that
  \be
  \widetilde\M_{n,u;\qe\qz\qd\qf}^{\gamma\delta,\beta_1\beta_2\beta_3\beta_4}
  = \sum_{\kappa} \widetilde{\widetilde\M}{}_{n,u;\qe\qz\qd\qf}
  ^{\kappa,\gamma\delta,\beta_1\beta_2\beta_3\beta_4}
  \ee
with $\kappa\iN\hom(U_{\ib_1}\oti U_{\ib_2},U_{\qe})$ and
  \eqpic{fact_MF_S2S1_proj_3dom_simp:68}{405}{194}{ \setulen 70
  \put(0,292)   {$\dsty \widetilde{\widetilde\M}{}_{n,u;\qe\qz\qd\qf}
                  ^{\kappa,\gamma\delta,\beta_1\beta_2\beta_3\beta_4} ~= $}
  \put(39,-25){
  \put(104,27)   {\includepicclax2{66}{68}}
  \put(206,408)   {\pl{\phi_{\alphz}}}
  \put(257,418.5) {\pl{\phi_{\alphv}}}
  \put(330,404)   {\pl{\phi_{\bet_2}^{}}}
  \put(479.3,376) {\pl{\phi_{\bet_1}^{}}}
  \put(300,401.5) {\pA A}
  \put(184,382)   {\pg{\ib_1}}
  \put(235,384)   {\pg{\ib_2}}
  \put(177,260)   {\pg{\ia_1}}
  \put(248,278.5) {\pg{\ia_2}}
  \put(211,249)   {\pl{\phi_{\alphe}}}
  \put(254,257.6) {\pl{\phi_{\alphd}}}
  \put(317,260.8) {\pl{\phi_{\bet_4}^{}}}
  \put(440.5,435.5) {\pl{\phi_{\bet_3}^{}}}
  \put(293,275)   {\pB B}
  \put(381,257)   {\pg{\qzb}}
  \put(331,464)   {\pg{\qzb}}
  \put(228,312)   {\pg{\qe}}
  \put(229,348)   {\pg{\qe}}
  \put(407,282)   {\pg{\qeb}}
  \put(373.4,404) {\pg{\qeb}}
  \put(336,119)   {\pg{\qf}}
  \put(429,93)    {\pg{\qz}}
  \put(191,95)    {\pg{\ja_1}}
  \put(257,95)    {\pg{\ja_2}}
  \put(155,434)   {\pg{\jb_1}}
  \put(211,429)   {\pg{\jb_2}}
  \put(282,176)   {\pg n} 
  \put(423,154)   {\pg u}
  \put(436,169)   {\pg{\bar u}}
  \put(441,130)   {\pg \gamma}
  \put(460,158)   {\pg \delta}
  \put(211,294)   {\pg{\kappa}}
  \put(207,374)   {\pg{\bar\kappa}}
  \put(481,436)   {\pX X}
  \put(494,355)   {\pg{\qe}}
  \put(480,335)   {\pg{\qz}}
  \put(515,362)   {\pg{\qd}}
  \put(461,334)   {\pg{\qf}}
  \put(502,299)   {\pg u}
  \put(519.9,307) {\pg{\bar u}}
  \put(475,312)   {\pg{\bar\gamma}}
  \put(523.5,335) {\pg{\bar\delta}}
  } }

At this point we can make use of the fact that a horizontal section of the 
three-manifold is the two-sphere $S^2$: we can deform the $U_n$-ribbon around
the horizontal two-sphere in such a way that it just encircles the $U_u$-ribbon.
As a consequence, we can trade the $U_n$-ribbon for a scalar factor 
${S_{u,n}}/{S_{u,0}}$. Owing to the unitarity of the $S$-matrix we can then 
perform the $n$-summation according to $\sum_{n\in\I}S_{0,n}S_{u,n}\eq 
\delta_{0,u}$. As a result only $u\eq 0$ contributes in the summation over $u$, 
and accordingly also the $\gamma$- and $\delta$-summations consist of only a 
single term; moreover, for this contribution to be non-zero we need 
$\qd\eq \qeb$ and $\qf\eq \qzb$. Inserting this information 
into the formula \erf{cXo-B} for \cXo\ gives the result quoted in \erf{cXo}.

\vskip 2.5em

\noindent{\sc Acknowledgments:}
We thank Jens Fjelstad for helpful discussions.
We are grateful to Valya Petkova for insisting that the TFT approach should
allow one to put classifying algebras on a firm mathematical ground, and to her
and Jean-Bernard Zuber for reminding us about the relevance of their earlier work.
JF is partially supported by VR under project no.\ 621-2009-3993.
CSc is partially supported by the Collaborative Research Centre 676 ``Particles,
Strings and the Early Universe - the Structure of Matter and Space-Time'' and
by the DFG Priority Programme 1388 ``Representation Theory''.

\newpage

 \newcommand\wb{\,\linebreak[0]} \def\wB {$\,$\wb}
 \newcommand\Bi[1]    {\bibitem{#1}}
 \newcommand\Erra[3]  {\,[{\em ibid.}\ {#1} ({#2}) {#3}, {\em Erratum}]}
 \newcommand\J[5]     {{\em #5}, {#1} {#2} ({#3}) {#4} }
 \renewcommand\K[6]   {{\em #6}, {#1} {#2} ({#3}) {#4} {\tt[#5]} }
 \newcommand\Prep[2]  {{\em #2}, preprint {\tt #1} }
 \newcommand\PhD[2]   {{\em #2}, Ph.D.\ thesis (#1)}
 \newcommand\BOOK[4]  {{\em #1\/} ({#2}, {#3} {#4})}
 \newcommand\inBO[7]  {{\em #7}, in:\ {\em #1}, {#2}\ ({#3}, {#4} {#5}), p.\ {#6}}
 \def\adma  {Adv.\wb in Math.}
 \def\bacp  {Ba\-nach\wB Cen\-ter\wB Publ.}
 \def\cntp  {Com\-mun.\wB Number\wB Theory\wB Phys.}
 \def\comp  {Com\-mun.\wb Math.\wb Phys.}
 \def\jgap  {J.\wb Geom.\wB and\wB Phys.}
 \def\jhep  {J.\wb High\wB Energy\wB Phys.}
 \def\jopa  {J.\wb Phys.\ A} 
 \def\nupb  {Nucl.\wb Phys.\ B}
 \def\phlb  {Phys.\wb Lett.\ B}
 \def\phrd  {Phys.\wb Rev.\ D}
 \def\phrl  {Phys.\wb Rev.\wb Lett.}
 \def\taac  {Theo\-ry\wB and\wB Appl.\wb Cat.}
 \newcommand\ecmiii[4] {\inBO{Third European Congress of Mathematics%
            {\rm, Vol.\,#1}} {C.\ Casucuberta, R.M.\ Mir\'o-Roig, J.\ Verdera,
            and S.\ Xamb\'o-Descamps, eds.} {Birk\-h\"au\-ser}{Basel}{2001}
            {{#2}}{{#4}} {\tt[#3]}}

\end{document}